\documentclass[times,twocolumn,final]{elsarticle}
\usepackage{framed,multirow}

\usepackage{amssymb}
\usepackage{latexsym}
\usepackage{amsmath}

\usepackage{url}
\usepackage{xcolor}

\usepackage{hyperref}

\definecolor{newcolor}{rgb}{.8,.349,.1}

\begin{document}

%\verso{Luyi Han \textit{et~al.}}

%\begin{frontmatter}

\title{To deform or not: treatment-aware longitudinal registration for breast DCE-MRI during neoadjuvant chemotherapy via unsupervised keypoints detection
%Type the title of your paper, only capitalize first word and proper nouns\tnoteref{tnote1}
}%
%\tnotetext[tnote1]{This is an example for title footnote coding.}

\author[1,2]{Luyi Han}

\author[2,4]{Tao Tan\corref{cor1}}%
%\fntext[fn1]{This is author footnote for second author.}
\cortext[cor1]{Corresponding author: Tao Tan;}
\ead{taotanjs@gmail.com}
%  Tel.: +0-000-000-0000;  
%  fax: +0-000-000-0000;}
\author[2,3]{Tianyu Zhang}
%% Third author's email
%\ead{author3@author.com}
\author[2,3]{Yuan Gao}
\author[2,3]{Xin Wang}
\author[6]{Valentina Longo}
\author[7]{Sof\'ia Ventura-D\'iaz}
\author[8]{Anna D'Angelo}
\author[5]{Jonas Teuwen}
\author[1,2]{Ritse Mann}

\address[1]{Department of Radiology and Nuclear Medicine, Radboud University Medical Centre, Geert Grooteplein 10, 6525 GA, Nijmegen, The Netherlands}
\address[2]{Department of Radiology, The Netherlands Cancer Institute, Plesmanlaan 121, 1066 CX, Amsterdam, The Netherlands}
\address[3]{GROW School for Oncology and Developmental Biology, Maastricht University Medical Centre, P. Debyelaan 25, 6202 AZ, Maastricht, The Netherlands}
\address[4]{Faculty of Applied Sciences, Macao Polytechnic University, 999078, Macao SAR, China}
\address[5]{Department of Radiation Oncology, The Netherlands Cancer Institute, Plesmanlaan 121, 1066 CX, Amsterdam, The Netherlands}
\address[6]{Institute of Radiology, Catholic University of the Sacred Heart, Largo A. Gemelli 8, 00168, Rome, Italy}
\address[7]{Department of Pathology, Hospital Universitario Ram\'on y Cajal, Madrid, Spain}
\address[8]{Dipartimento di diagnostica per immagini, Radioterapia, Oncologia ed ematologia, Fondazione Universitaria A. Gemelli, IRCCS Roma, Roma, Italy}

%\received{1 May 2013}
%\finalform{10 May 2013}
%\accepted{13 May 2013}
%\availableonline{15 May 2013}
%\communicated{S. Sarkar}

\begin{abstract}
%%%
Clinicians compare breast DCE-MRI after neoadjuvant chemotherapy (NAC) with pre-treatment scans to evaluate the response to NAC.
Clinical evidence supports that accurate longitudinal deformable registration without deforming treated tumor regions is key to quantifying tumor changes.
We propose a conditional pyramid registration network based on unsupervised keypoint detection and selective volume-preserving to quantify changes over time. 
In this approach, we extract the structural and the abnormal keypoints from DCE-MRI, apply the structural keypoints for the registration algorithm to restrict large deformation, and employ volume-preserving loss based on abnormal keypoints to keep the volume of the tumor unchanged after registration.
We use a clinical dataset with 1630 MRI scans from 314 patients treated with NAC.
The results demonstrate that our method registers with better performance and better volume preservation of the tumors.
Furthermore, a local-global-combining biomarker based on the proposed method achieves high accuracy in pathological complete response (pCR) prediction, indicating that predictive information exists outside tumor regions. The biomarkers could potentially be used to avoid unnecessary surgeries for certain patients.
It may be valuable for clinicians and/or computer systems to conduct follow-up tumor segmentation and response prediction on images registered by our method.
Our code is available on \url{https://github.com/fiy2W/Treatment-aware-Longitudinal-Registration}.
%%%%
\end{abstract}

%\begin{keyword}
%% Keywords
%\KWD Breast DCE-MRI\sep Deformable Registration\sep Neoadjuvant Chemotherapy\sep Unsupervised Keypoint Detection
%\end{keyword}

%\end{frontmatter}
\maketitle
%\linenumbers

\section{Introduction}\label{sec:introduction}

\begin{figure*}[!htbp]
\centering
\begin{minipage}{0.1\linewidth}
        \centerline{Baseline}
    \end{minipage}
    \begin{minipage}{0.28\linewidth}
        \centerline{\includegraphics[width=\textwidth]{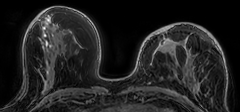}}
    \end{minipage}
    \begin{minipage}{0.28\linewidth}
        \centerline{\includegraphics[width=\textwidth]{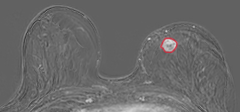}}
    \end{minipage}
    \begin{minipage}{0.28\linewidth}
        \centerline{\includegraphics[width=\textwidth]{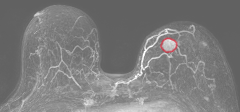}}
    \end{minipage}
    
    \vspace{3pt}
    
    \begin{minipage}{0.1\linewidth}
        \centerline{Follow-up}
    \end{minipage}
    \begin{minipage}{0.28\linewidth}
        \centerline{\includegraphics[width=\textwidth]{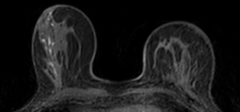}}
    \end{minipage}
    \begin{minipage}{0.28\linewidth}
        \centerline{\includegraphics[width=\textwidth]{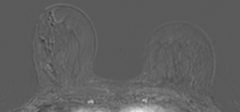}}
    \end{minipage}
    \begin{minipage}{0.28\linewidth}
        \centerline{\includegraphics[width=\textwidth]{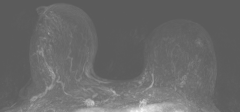}}
    \end{minipage}
    
    \vspace{3pt}
    
    \begin{minipage}{0.1\linewidth}
        \centerline{}
    \end{minipage}
    \begin{minipage}{0.28\linewidth}
        \centerline{T1w pre-contrast}
    \end{minipage}
    \begin{minipage}{0.28\linewidth}
        \centerline{Wash-in}
    \end{minipage}
    \begin{minipage}{0.28\linewidth}
        \centerline{MIP}
    \end{minipage}
    
\caption{Examples of breast DCE-MRIs for one patient during different visits. The first line is acquired before NAC, and the second is scanned after NAC and before surgery. The columns show the T1-weighted (T1w) pre-contrast image, the wash-in image, and a maximum intensity projection (MIP) of the wash-in image. The breast tumor is circled with the red contour in the wash-in image and MIP.} \label{fig:introduction}
\end{figure*}

Breast cancer has surpassed lung cancer, becoming the most common global cancer, with an estimated 2.3 million new female diagnosed cases worldwide in 2020~\citep{sung2021global}.
Neoadjuvant chemotherapy (NAC) is one of the standard treatments for locally advanced breast cancer~\citep{curigliano2017escalating}.
NAC aims at reducing the size of the primary tumor to increase the likelihood of breast conservation and suppressing occult systemic metastasis to improve overall survival~\citep{bonadonna1998primary,cho2013oncologic,earl2015neoadjuvant}.
To monitor tumor changes during NAC and predict the pathological response, clinicians evaluate longitudinal changes in breast dynamic contrast-enhanced magnetic resonance imaging (DCE-MRI).
DCE-MRI is usually acquired before and after NAC and commonly also halfway. The total therapy time lasts three to six months. An optimal response in which no residual invasive tumor is left, is referred to as a pathological complete response (pCR).
In practice, two volume-of-interest (VOI)-based measures, tumor volume and histogram analysis, can quantify the tumor response among baseline and follow-up scans but are not sensitive to heterogeneous intra-tumoral changes~\citep{thakran2022impact}.
In order to analyze heterogeneous changes in various subregions within the same tumors, areas from different scans need to be aligned first.
It is shown that a better deformable registration for the multi-visit MRI scans provides a better comparison of tissue morphological features and a more precise quantification of lesion changes, and further a better assessment of short/long-term treatment response prediction~\citep{ou2015deformable,thakran2022impact}.
Thus, it is clinically extremely important to improve the registration performance on longitudinal breast DCE-MRIs.

Various proposed methods for medical image registration can be divided into (1) conventional registration methods~\citep{avants2008symmetric,klein2009elastix,vercauteren2009diffeomorphic,modat2010fast,ou2011dramms,heinrich2013mrf} and (2) deep learning-based methods.
%Conventional methods are inefficient in the clinical setting due to expensive computation costs and difficulty in parameter adaptation.
Conventional methods are time-inefficient for inferencing the deformation field in the clinical setting due to the need for iterative searches for optimal parameters.
Some researchers~\citep{ou2015deformable,mehrabian2018deformable,thakran2022impact} compared different conventional methods and attempted to identify the optimal input parameter set for breast DCE-MRI.
However, it is still time-consuming that each pair of samples needs to be optimized iteratively during the registration approach.
Recently, deep learning-based methods can resolve the deformable registration task faster due to better generalization ability.
Learning-based registration methods can be classified into two categories according to the usage of manual ground truth: (1) fully supervised and (2) weakly supervised/unsupervised.
Some models~\citep{cao2018deformable,eppenhof2018pulmonary} learned supervised patterns by utilizing deformation fields simulated from conventional registration methods as labels.
In contrast, other models~\citep{hu2018weakly,balakrishnan2019voxelmorph,de2019deep} are learned in a weakly supervised or unsupervised way by optimizing label-level or intensity-level similarity between warped images and fixed images.
Although deep learning methods have been proven efficient and robust in medical image registration, previous researches seldom discuss an abnormal-to-normal or abnormal-to-abnormal registration pattern.
Different than the normal-to-normal pattern, as illustrated in Fig.~\ref{fig:introduction}, longitudinal breast DCE-MRI registration has the following challenges: (1) large deformation occurs between two visits due to different positioning of the patient, different forces to fix the breast, and weight loss/gain for patients, etc.; (2) the changes of tumors during NAC tend to mislead the algorithm producing incorrect registration results.

\subsection{Longitudinal deformable registration}
The intensity-level and/or distribution-level similarity can not be used for aligning abnormal tissue with normal or changed abnormal tissue. The abnormal area can be squeezed to disappear due to the absence of a similar matched structure in the corresponding normal area.
Some researchers consider avoiding negative effects caused by abnormal areas by performing abnormal-to-normal registration.
One of the most direct resolutions is to exclude pathological regions from tumor images to apply for registration only on normal regions~\citep{brett2001spatial,andersen2010cost}.
%These methods need segmentation of abnormal regions and cannot force the tumor unchanged during registration.
Building a sensible mapping of abnormal regions is another choice.
\citet{gooya2010deformable,gooya2012glistr} utilize a glioma growth model to synthesize a tumor in a normal atlas for subsequent registration.
In contrast, other researches~\citep{liu2014low,liu2015low,tang2017groupwise,tang2018new} try to map tumor images to the corresponding normal-appearance counterpart with low-rank plus sparse matrix decomposition.
Another method is to utilize implicit constraints.
\citet{estienne2020deep} propose a dual network that addresses tumor segmentation and registration at the same time. The method can couple two tasks and relax the registration constraints in the tumor region based on the segmentation path.
\citet{rohlfing2003volume,gigengack2011motion} employ Jacobian determinant to restrict deformation on target tissues.
\citet{wodzinski2021semi} introduce volume penalty into the registration process to guarantee full tumor resection.
However, these methods need segmentation masks for tumors, which are time-consuming for clinicians to create during clinical settings. The manual annotation also has limited precision at boundaries and is a major driver of inter and intra-clinician variability.

\subsection{Unsupervised keypoint detection}
Apart from tumor masks, other keypoints can assist registration but are harder to annotate in 3D due to the lack of obvious anatomical structures in breast tissue.
\citet{yan2020self} propose a self-supervised anatomical embedding approach to discover the intrinsic structure from unlabeled images.
Based on pixel-level contrastive learning and simple nearest neighbor searching, the corresponding point can be located in the target image when any point of interest is annotated on an atlas image.
However, it needs laborious handcraft annotation of keypoints for the template image.
Therefore, several studies have developed the research on learning keypoints without manual supervision.
The key to these methods is to build up an appearance consistency for an object with different poses.
Most notably, \citep{jakab2018unsupervised,zhang2018unsupervised} encode keypoint coordinates with a conditional autoencoder to capture the meaningful structure and semantic representation.
To synthesize the target pose, the decoder is fed with the concatenation of the keypoint heatmaps and the feature embeddings.
\citet{kulkarni2019unsupervised} refine a transported feature embedding by exchanging features on keypoint locations from different poses.
However, these methods focus on general structures unaffected by deformation in 2D images and videos. 
In contrast, it is more difficult to detect keypoints automatically from 3D DCE-MRIs than from natural images. 
Moreover, these methods are incapable of detecting individual variant keypoints that focus on tumor areas that contribute to preserving tumor volume during registration.

\subsection{Contributions}
For facilitating longitudinal treatment analysis, we propose a treatment-aware registration scheme that learns the structure representation and captures abnormal areas from different keypoints, respectively.
We can obtain better registration performance on longitudinal breast DCE-MRIs by restricting large deformation and tumor changes based on the detected keypoints.
The proposed registration framework based on unsupervised keypoints detection contributes in three folds:
\begin{itemize}
	\item[1.] Based on unsupervised learning methods, we automatically extract the structural and the abnormal keypoints from paired breast DCE-MRI.
	\item[2.] We restrict large deformation on structural keypoints, and prevent the volume change of the tumors on abnormal keypoints.
	\item[3.] To evaluate the performance of the registration methods, we propose local and global registration-based biomarkers to assess the pixel-level breast tissue response between multi-visit MRIs, which can be used for pathological response prediction. The proposed biomarker is composed of the responses in both the intratumoral area and extraneoplastic area.
\end{itemize}

\section{Methods}
\subsection{Overview}

\begin{figure*}[!htbp]
	\centering
	\includegraphics[width=0.85\textwidth]{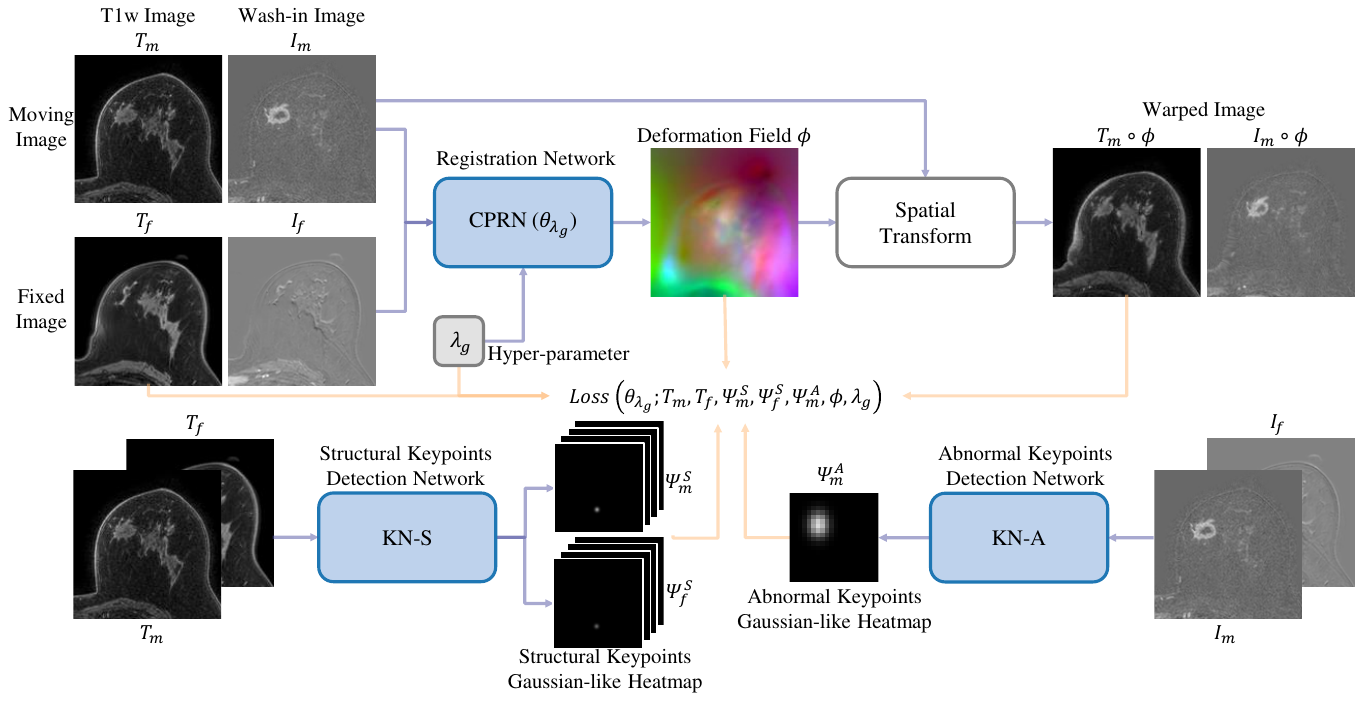}
	\caption{The overview of the proposed framework. We first extract the structural and the abnormal keypoints by KN-S and KN-A, respectively. Then we employ the heatmaps of keypoints to restrict the CPRN.} \label{fig:overview}
\end{figure*}

DCE-MRI involves multiple T1-weighted (T1w) sequences acquired once pre-contrast injection and at least twice post-contrast. Wash-in images are calculated by subtraction between the first post-contrast (1.5min after contrast injection) and the pre-contrast images. We leverage T1w pre-contrast images for registration due to a similar intensity histogram before injection. We also input the wash-in images, which show the contrast enhancement, to assist in aligning tumor regions.

Denoting T1w pre-contrast images and wash-in images by $T$ and $I$. The purpose of the framework is to learn a deformable transform function $\mathbf{\Phi}: (T_m, I_m, T_f, I_f)\rightarrow\phi$, where $T_m$ and $I_m$ refer to moving images, and $T_f$ and $I_f$ are fixed images. The moving and fixed images are paired data of baseline and follow-up images for one patient.

The framework consists of three networks -- the Structure Keypoint Network (KN-S), the Abnormal Keypoint Network (KN-A), and the Conditional Pyramid Registration Network (CPRN). KN-S is utilized to capture the patient-invariant breast structures, referred to as structural keypoints. KN-A is capable of detecting patient-specific abnormal areas between moving and fixed images, producing differential abnormal keypoints. CPRN is a registration network fitting the deformable transform function $\mathbf{\Phi}$.
As illustrated in Fig.~\ref{fig:overview}, we first detect structural keypoints and abnormal keypoints with KN-S and KN-A based on unsupervised learning methods, and then employ these keypoints into CPRN as constraint loss function.

\subsection{Structural keypoints detection}
\begin{figure}[!htbp]
	\centering
	\includegraphics[width=0.4\textwidth]{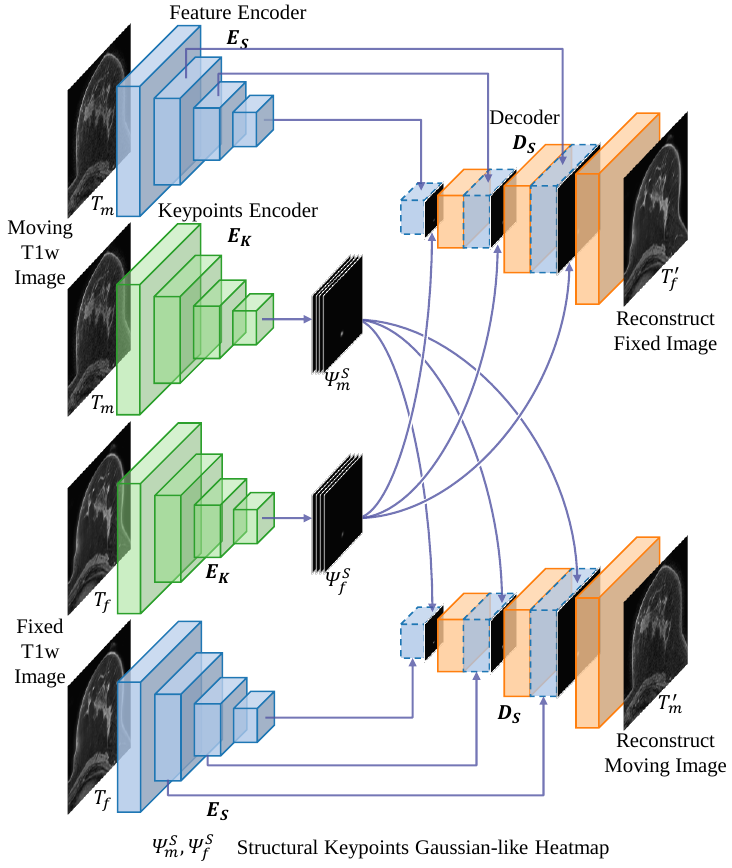}
	\caption{The architecture of the proposed KN-S.} \label{fig:keynetS}
\end{figure}

It is difficult to find a golden standard for breast structure keypoints in clinical practice due to individual variance in breast tissue.
Based on identical features within all the breast MRIs, it is possible to abstract an invariant representation of the breast structure as structural keypoints.
Registration based on structural keypoints contributes to matching at the structure level and reduces the influence of intensity bias between the baseline and follow-up images.
We assume that structural keypoints with a superior representation of breast structure are capable of recovering the images from features and controlling their spatial transformation.
To leverage the structure consistency between the baseline image and follow-up image, we proposed the KN-S for structural keypoints detection in Fig.~\ref{fig:keynetS}.
We extend the model proposed by \citet{jakab2018unsupervised} from 2D to 3D to handle 3D breast DCE-MRI and also introduce bilateral networks to extract structural keypoints from both moving and fixed images to control the opposite images, preserving better structural consistency.

Wash-in images are created by subtracting the first post-contrast T1 weighted acquisition from the pre-contrast T1 acquisition. Considering that wash-in images thus show areas with blood flow or high vessel permeability, e.g., vessels and tumors, only T1w pre-contrast images $T$ are utilized to extract structural keypoints to avoid the influence from the contrast.
The KN-S includes the feature encoder $\mathbf{E}_S$, the keypoint encoder $\mathbf{E}_K$, and the decoder $\mathbf{D}_S$. $\mathbf{E}_S$ is utilized to extract feature maps from $T_m$ and $T_f$, $\mathbf{E}_K$ aims to encode $K_S$-channel heatmaps from corresponding images.
%$K_C$ refers to the number of structural keypoints.
Referring to \citep{jakab2018unsupervised}, the heatmaps $\mathbf{E}_K(T_m)$ and $\mathbf{E}_K(T_f)$ are softmax-normalized and condensed to structural keypoints $\mu_m^S,\mu_f^S\in\mathbb{R}^{K_S\times3}$ and further generated as Gaussian-like heatmaps $\Psi_m^S$ and $\Psi_f^S$ with $K_S$ channels and the same size as feature maps $\mathbf{E}_S(T_m)$ and $\mathbf{E}_S(T_f)$.
$\mathbf{D}_S$ is utilized to generate reconstruction images by inputting feature maps and Gaussian-like heatmaps with concatenation.
%by computing the center of gravity on each channel.
%Then, Gaussian-like heatmaps are generated from $\mu_C$ to guide the feature maps, $\mathbf{\Psi}$
In order to learn the structure consistency, we exchange $\Psi_m^S$ and $\Psi_f^S$ and force $\mathbf{D}_S$ to generate images corresponding to heatmaps.
Considering both intensity similarity and perceptual similarity, the cross-reconstruction loss is defined as,
\begin{equation}
	%\mathcal{L}^C_{cross} = \left\|T_m-\mathbf{D}_S(\mathbf{E}_S(T_f),\Psi_m^S)\right\|_1+\left\|T_f-\mathbf{D}_S(\mathbf{E}_S(T_m),\Psi_f^S)\right\|_1
	\begin{aligned}
    	\mathcal{L}^S_{cross} &= \left\|T_m-T'_{f\rightarrow m}\right\|_1 + \left\|T_f-T'_{m\rightarrow f}\right\|_1 \\
    	&+ \mathcal{L}_{p}(T_m, T'_{f\rightarrow m}) + \mathcal{L}_{p}(T_f, T'_{m\rightarrow f})
	\end{aligned}
\end{equation}
where $\mathcal{L}_{p}$ refers to perceptual loss by utilizing the pre-trained VGG16 model. $T'_{f\rightarrow m}$ and $T'_{m\rightarrow f}$ are cross reconstruction prediction, and are formulated as
\begin{equation}
    \begin{aligned}
    	T'_{f\rightarrow m} = \mathbf{D}_S(\mathbf{E}_S(T_f),\Psi_m^S) \\
    	T'_{m\rightarrow f} = \mathbf{D}_S(\mathbf{E}_S(T_m),\Psi_f^S)
	\end{aligned}
\end{equation}

In the training phase of KN-S, $\Psi_m^S$, $\Psi_f^S$ are Gaussian-like heatmaps generated based on $\mu_m^S$, $\mu_f^S$ with a standard deviation of 0.1, respectively.
The number of structural keypoints $K_S$ is set to 64.

\subsection{Abnormal keypoints detection}
\begin{figure*}[!htbp]
	\centering
	\includegraphics[width=0.85\textwidth]{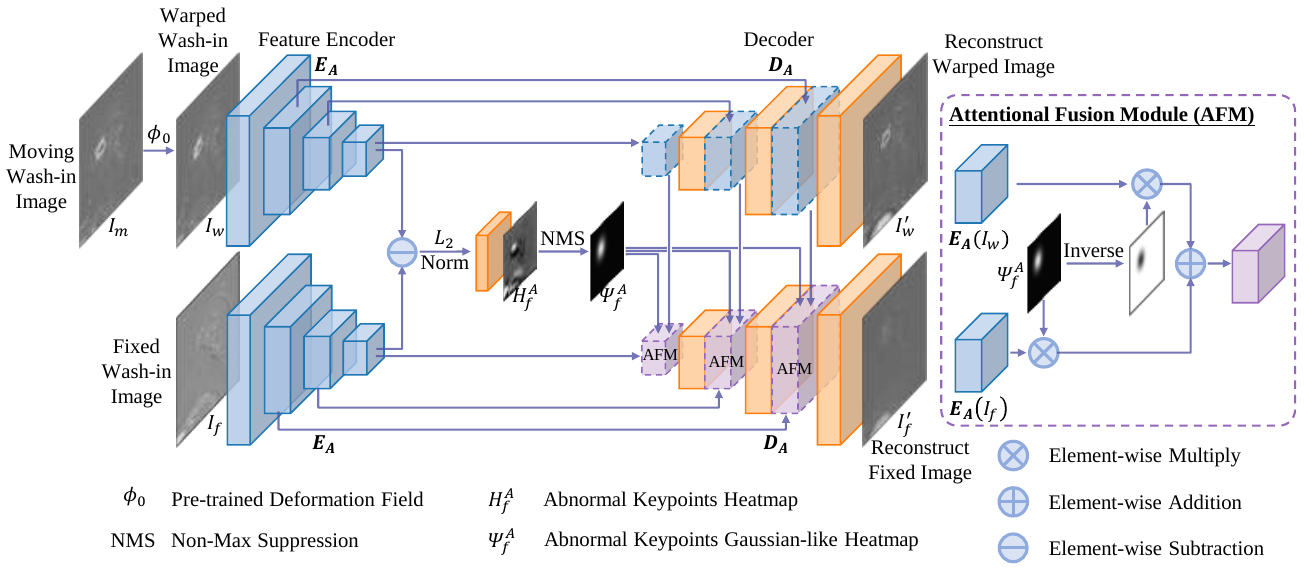}
	\caption{The architecture of the proposed KN-A.} \label{fig:keynetA}
\end{figure*}

The size, shape, and density of breast tumors sometimes are quite different between baseline and follow-up images due to different responses to NAC. To eliminate negative effects caused by these abnormal areas during registration and also be aware of changes in abnormal regions, we proposed the KN-A for abnormal keypoints detection as shown in Fig.~\ref{fig:keynetA}.

The wash-in images $I$, which highlight the tumor areas and are less susceptible to interference from other normal tissue intensities than $T$, are utilized to extract abnormal keypoints.
%The wash-in images $I$, which highlight the tumor areas and have less disturbing factors than $T$, are utilized to extract abnormal keypoints.
The KN-A is composed of a feature encoder $\mathbf{E}_A$ and a decoder $\mathbf{D}_A$. To reduce the detection bias, we first employ a pre-trained deformation field $\phi_0$ generated by models like VoxelMorph~\citep{balakrishnan2019voxelmorph} to transform $I_m$ as warped image $I_w$. Then $\mathbf{E}_A$ is utilized to extract feature maps from $I_w$ and $I_f$. To represent the abnormal area from feature maps, $L_2$ normalization is calculated on $\mathbf{E}_A(I_w)$ and $\mathbf{E}_A(I_f)$, and followed with a convolutional layer fusing the error maps as a one-channel heatmap $H^A_f$.
High responses in $H^A_f$ correspond to abnormal locations. To reduce overlapping and obtain well-performed keypoints, we apply non-max suppression (NMS) on $H^A_f$ and locate peaks as abnormal keypoints $\mu^A_f\in\mathbb{R}^{K_A\times3}$. Similarly, a Gaussian-like heatmap $\Psi^A_f$ is generated~\citep{jakab2018unsupervised} to assist subsequent image reconstruction. $\mathbf{D}_A$ rebuilds images by inputting from the feature maps.
%By replacing abnormal areas from $\mathbf{E}_A(I_w)$ to $\mathbf{E}_A(I_f)$ with the attentional fusion module (AFM), it is believed to reconstruct an image with abnormal areas removed.
Then we employ the attentional fusion module (AFM) for exchanging regions between the feature maps $E_A(I_w)$ and $E_A(I_f)$, which correspond to the abnormal keypoints. We can reconstruct the image with the abnormal regions replaced successfully when the abnormal keypoints are accurately located.
Based on this, two consistency loss functions are defined as follows,

\begin{equation}
	%\mathcal{L}^A_{rec} = \left\|I_w-\mathbf{D}_A(\mathbf{E}_A(I_w))\right\|_1
	\mathcal{L}^A_{rec} = \left\|I_w-I'_w\right\|_1 + \mathcal{L}_{p}(I_w, I'_w)
\end{equation}
\begin{equation}
	%\mathcal{L}^A_{cross} = \left\|I_f-\mathbf{D}_A(\mathbf{E}_A(I_w)\cdot(1-\max\limits_{K_A}(\Psi_A))+\mathbf{E}_A(I_f)\cdot\max\limits_{K_A}(\Psi_A))\right\|_1
	\mathcal{L}^A_{cross} = \left\|I_f-I'_{w\rightarrow f}\right\|_1 + \mathcal{L}_{p}(I_f, I'_{w\rightarrow f})
\end{equation}
where $I'_w$ refers to the reconstruction prediction of $I_w$, and $I'_{w\rightarrow f}$ indicates the cross reconstruction prediction. $I'_w$ and $I'_{w\rightarrow f}$ are formulated as
\begin{equation}
    \begin{aligned}
    	I'_w &= \mathbf{D}_A(\mathbf{E}_A(I_w)) \\
    	I'_{w\rightarrow f} &= \mathbf{D}_A(\mathbf{E}_A(I_w)\cdot(1-\Psi^A_f)+\mathbf{E}_A(I_f)\cdot\Psi^A_f)
	\end{aligned}
\end{equation}
%$\max\limits_{K_A}(\Psi_A)$ indicates the element-wise maximum of $\Psi_A$.

For moving images, abnormal keypoints $\mu^A_m$ and corresponding Gaussian-like heatmaps $\Psi^A_m$ are similarly formulated by moving image $I_m$ and inversed warped image $I_f\circ\phi^{-1}_0$. In the training process of KN-A, $\Psi_m^A$, $\Psi_f^A$ are Gaussian-like heatmaps generated based on $\mu_m^A$, $\mu_f^A$ with a standard deviation of 0.2, respectively.
The number of abnormal keypoints $K_A$ is set to 1.

\subsection{keypoint-based registration}
Different from deformable registration between images of other body parts (particularly the brain), registration for DCE-MRI images faces large deformation of breasts and changes of tumors between multiple scans, which can influence the effect of the registration algorithm. The large deformation comes from the deformation of the breast during positioning and also because of changes induced by treatment.
An important key to predicting the deformation field is the balance between the similarity term and regularization term during the optimization of the network.
Hence, improvements in three aspects are applied to the longitudinal registration of the DCE-MRI series:
(1) we propose a conditional deformable pyramid architecture to balance the similarity term and regularization term automatically;
(2) we restrict the similarity of breast structure by structural keypoints to fit the large deformation; and
(3) we apply volume-preservation based on abnormal keypoints to keep the tumor unchanged.

\subsubsection{Network architecture}\label{sec:regnet}
\begin{figure}[!htbp]
	\centering
	\includegraphics[width=0.4\textwidth]{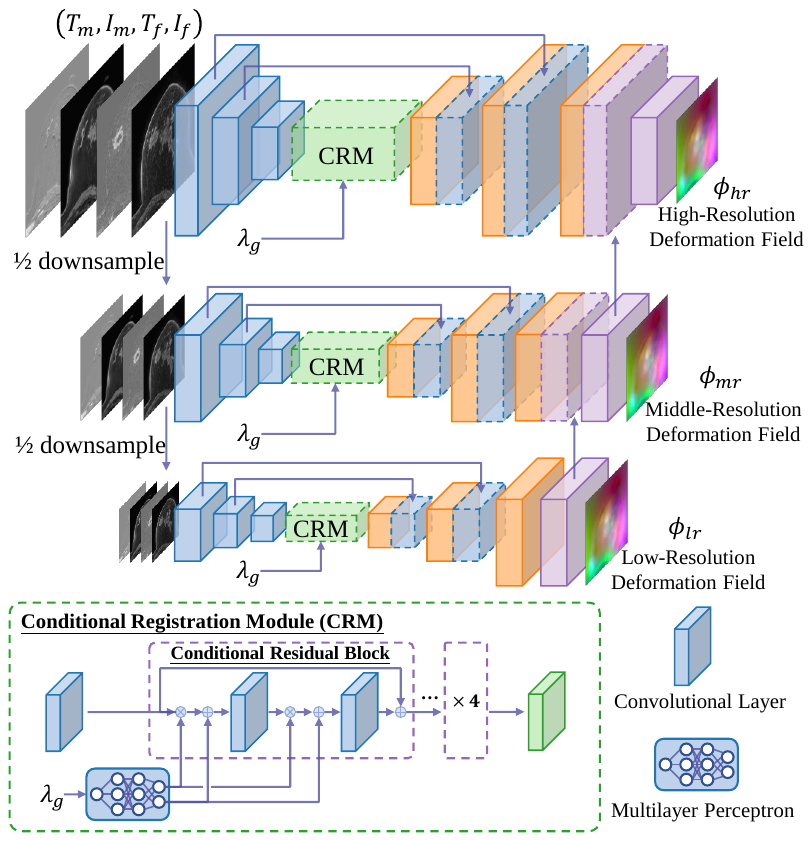}
	\caption{The architecture of the proposed CPRN.} \label{fig:regnet}
\end{figure}

As illustrated in Fig.~\ref{fig:regnet}, the proposed registration model CPRN involves a three-level pyramid coarse-to-fine architecture with conditional registration modules.
The input image pyramid is generated by downsampling input T1w and wash-in images with trilinear interpolation. We obtain paired moving and fixed images from the input image pyramid with low, middle, and high resolution.
Different from using the residual connection between the deformation field in multi-resolution and training the model step by step~\citep{mok2021conditional}, the proposed CPRN inputs the original moving images with different resolutions for training simultaneously and concatenates the deformation fields of different resolutions to avoid the accumulation of wrong deformations.

A U-Net-like CNN-based subnetwork is involved in outputting the deformation field in the corresponding resolution for each resolution level.
The subnetwork can capture the large deformation and avoid leading toward local minimum convergence for the coarsest resolution level. Then the output features are 2-times upsampled to the middle resolution and predicted as the 3-channel deformation field $\phi_{lr}$.
Local details can be extracted from the subnetworks for the middle- and high-resolution levels. Combining with the features embedded from the previous pyramid level, the coarse-to-fine fusion features can learn complex non-linear deformation in a higher resolution, such as $\phi_{mr}$ and $\phi_{hr}$.
Note that, we downsample the fusion features in the highest resolution level and output with the deformation field $\phi_{hr}$ in order to achieve higher registration accuracy and fewer folding areas~\citep{han2021deformable}.

The U-Net-like subnetwork consists of an encoding path, a conditional registration module (CRM), and a decoding path.
The encoding path is comprised of six 3D convolutional layers with the kernel size of 3$\times$3$\times$3, in which the third and fifth layers are with the stride of 2 in order to downsample the feature maps.
Referring to \citet{mok2021conditional}, CRM involves five conditional residual blocks, and each block is similar to the residual block by replacing the instance normalization with conditional instance normalization. The given hyper-parameter $\lambda_g$ can be mapped to a latent code by multi-layer perceptron to control mean and variation in the conditional instance normalization.
The utilization of a conditional or hyperparameter network, as described in previous works~\citep{hoopes2021hypermorph,mok2021conditional}, facilitates the association of $\lambda_g$ with the regularization constraints of the model. By doing so, the model can explore diverse weight configurations of the regularization constraints within a single training process, thereby enabling the manipulation of the smoothness of the resulting deformation field during inference via searching the hyper-parameter $\lambda_g$.
The decoding path includes two upsample layers and four convolutional layers with the size of 3$\times$3$\times$3. Similar to the U-Net, each 2-times upsampled layer in the decoding path is followed by two convolutional layers, and the output features are concatenated with the skip-connected features from the encoding path.

\subsubsection{Pyramid similarity losses}
%To avoid the negative effects of abnormal areas, we utilize abnormal landmark heatmaps $\Psi_A$ extracted by KN-A to suppress difference on abnormal areas from images before calculating losses.
Fitting the registration mapping with intensity-level loss function in a single resolution can lead toward local minimum convergence. Multi-scale information contributes to the network to predict complex non-linear deformation fields in global and local areas. Hybrid losses~\citep{han2021deformable} can restrict image-matching from different aspects, such as intensity, statistical distribution, and boundary levels.
Hereby, we employ pyramid similarity losses to optimize the pyramid registration model CPRN on both full-resolution images $T^{hr}_m$, $T^{hr}_f$ and halved resolution images $T^{mr}_m$, $T^{mr}_f$.
Note that, wash-in images are not used for the loss function to avoid negative effects from tumor changes.
%where $\phi'$ is the half size deformation field, $\mathbf{Down}(\cdot)$ is a downsampling layer with factor of 2. $\Psi'_A$ is upsampled from $\Psi_A$ to the same size as $\phi'$.
%$I'_f$ and $I'_w$ are similarly formulated.
%To restrict the similarity at intensity, statistic, and boundary level on reformed images $T'_w$, $I'_w$, $T'_f$, and $I'_f$, we employ hybrid losses on halved resolution.
\begin{equation}
	\mathcal{L}^R_i(\phi_i) = \left\|T^{mr}_f-T^{mr}_m\circ\phi_i\right\|^2_2 + \left\|T^{hr}_f-T^{hr}_m\circ\phi'_i\right\|^2_2
\end{equation}
\begin{equation}
	\mathcal{L}^R_s(\phi_i) = \mathcal{L}_{mi}(T^{mr}_f,T^{mr}_m\circ\phi_i) + \mathcal{L}_{mi}(T^{hr}_f,T^{hr}_m\circ\phi'_i)
\end{equation}
\begin{equation}
    \begin{aligned}
        \mathcal{L}^R_b(\phi_i) &= \left\|S^{mr}_f-S^{mr}_m\circ\phi_i\right\|_1 + \mathcal{L}_{dsc}(S^{mr}_f,S^{mr}_m\circ\phi_i) \\
        &+ \left\|S^{hr}_f-S^{hr}_m\circ\phi'_i\right\|_1 + \mathcal{L}_{dsc}(S^{hr}_f,S^{hr}_m\circ\phi'_i)
    \end{aligned}
\end{equation}

where $\mathcal{L}^R_i$ refers to intensity-level loss, i.e. SSD, $\mathcal{L}^R_s$ indicates statistic-level loss--mutual information~\citep{guo2019multi} loss $\mathcal{L}_{mi}$, $\mathcal{L}^R_b$ is boundary-level loss and here we employ L1 loss and Dice similarity coefficient (DSC) loss $\mathcal{L}_{dsc}$.
$\phi_i\in\{\phi_{lr}, \phi_{mr}, \phi_{hr}\}$ is the predicted deformation field for different pyramid levels, and $\phi'_i$ refers to the upsampled deformation field of $\phi_i$.
$S^{mr}_m$, $S^{mr}_f$, $S^{hr}_m$, and $S^{hr}_f$ indicate the corresponding breast segmentation masks for $T^{mr}_m$, $T^{mr}_f$, $T^{hr}_m$, and $T^{hr}_f$, respectively. All breast segmentation masks are generated by a 3D U-Net model trained with 115 handcraft annotated samples. And the predicted segmentation masks are then preprocessed with a Gaussian blur with a kernel size of 7 and a standard deviation of 1.

\subsubsection{Structural keypoint loss}
Baseline images and follow-up images sometimes have large deformation and intensity bias due to long acquisition intervals. Registration from the semantic dimension is capable of relieving the intensity and statistical bias. To constrain breast tissue with well-distributed structure representation, we employ a structural keypoint loss based on structural keypoint heatmaps $\Psi_f^S$ and $\Psi_m^S$ extracted by KN-S.
\begin{equation}
	\mathcal{L}^R_{sl}(\phi_i) = \left\|\Psi_f^S-\Psi_m^S\circ\phi_i\right\|_2
\end{equation}
where $\Psi_m^S$, $\Psi_f^S$ are Gaussian-like heatmaps generated based on structural keypoints $\mu_m^S$, $\mu_f^S$ with a standard deviation of 0.01, respectively.

\subsubsection{Volume-preserving loss} One important downstream task for longitudinal breast DCE-MRI registration is to predict response to NAC. The change of tumor before and after treatment is an important indicator for prognosis and may enable image-based treatment changes, particularly when a residual tumor is evident. Achieving a pathological complete response (pCR) would theoretically obviate the need for subsequent surgery.
It is incorrect to alter the tumor size and shape during registration because they are the key indicators for treatment response assessment~\citep{li2009nonrigid}.

To reduce the manual costs of segmenting breast tumors, we automatically localize the tumor area and other unknown treatment-sensitive regions based on the abnormal keypoints detected from KN-A. Furthermore, we assume that keeping the intensity summation of the heatmap $\Psi_m^A$ unchanged contributes to the tumor volume preservation. To suppress deformation on tumor regions, we employ a volume-preserving loss based on abnormal keypoint heatmaps $\Psi^A_m$ to keep the volume of the abnormal area unchanged.
\begin{equation}
	\mathcal{L}^R_{vp}(\phi_i) = \left\|\sum\Psi^A_m-\sum(\Psi^A_m\circ\phi_i)\right\|_1
\end{equation}
where $\Psi_m^A$ is a Gaussian-like heatmap generated based on abnormal keypoints $\mu_m^A$ with a standard deviation of 0.25.

\subsubsection{Total loss}
Besides the losses described above, a gradient-based regularisation term is involved to preserve the smoothness of the deformation field. The total loss is,
\begin{equation}
    \label{eq:total}
    \begin{aligned}
        \mathcal{L}^R = \sum_{\phi_i}&(\mathcal{L}^R_i + \mathcal{L}^R_s + \mathcal{L}^R_b + \mathcal{L}^R_{sl} + \lambda_v\cdot\mathcal{L}^R_{vp} \\
        &+ \lambda_g\cdot\mathbf{Grad}(\phi_i) + \lambda_g\cdot\mathbf{Grad}(\phi'_i))
    \end{aligned}
\end{equation}
where $\lambda_v$ is the weight of volume preserving, $\lambda_g$ balances the similarity term and regularization term. $\lambda_v$ is set to $3\times 10^{-5}$ based on grid research. During the training process, the hyper-parameter $\lambda_g$ is chosen randomly from a uniform distribution with a range of [0, 10] in each step. For the purpose of achieving an optimal balance between similarity and regularization, during testing, grid searching is conducted with a step size of 1 to set the value of $\lambda_g$ to 5.

\section{Experimental Settings}
\subsection{Data}
We utilize a clinical dataset of 314 breast cancer patients treated with NAC from 2017 to 2020 at the Netherlands Cancer Institute in Amsterdam, Netherlands. Two to four visits, before NAC (baseline), during NAC, and after all cycles of NAC, were selected from each patient. The DCE-MRI is acquired with six timepoints obtained within 7.5 min after contrast injection for each visit. And we process T1w pre-contrast images (with fat saturation) and wash-in images (subtraction between the first post-contrast at 1.5 min after contrast injection and pre-contrast images).
The MRIs were acquired with Philips Ingenia 3.0-T scanners. All the data are collected and utilized with the approval of the local ethics committee.
%Besides, to evaluate the generalization ability for the proposed method, we validate the proposed method on an external dataset~\citep{newitt2016single,clark2013cancer}. The external dataset contains longitudinal DCE-MRI studies of 64 patients undergoing NAC for invasive breast cancer and we select 46 patients for validation. Each patient schedules three DCE-MRI scans: before treatment, after one cycle of chemotherapy, after completion of treatment.
%All the breast MRIs in the external dataset are acquired on a 1.5-T scanner (Signa, GE Healthcare, Milwaukee, WI) using a bilateral phased-array breast coil.
Patient characteristics are listed in Table~\ref{tab:data}. The following clinical information was also collected: age, hormone receptor status (estrogen receptor (ER), progesterone receptor (PR), human epidermal growth factor receptor 2 (HER2)), and clinical tumor stage before treatment (T-stage and N-stage). The T-stage refers to the size and extent of the primary tumor, while the N-stage indicates whether cancer cells have spread to nearby lymph nodes. Statistical analyses are performed between patients that achieved pCR and those that did not. The two-sided independent samples T-test is applied for age, and the Chi-square test or Fisher’s exact test is utilized for categorical variables.

\begin{table}
	\centering
	\caption{Characteristics of the patients in the clinical dataset. Age is shown with mean and SD. 
	ER, PR, HER2, T-stage, and N-stage are presented with the number and percentage of patients. \textbf{pCR}: pathological complete response, \textbf{SD}: standard deviation.}
	\label{tab:data}
	\setlength{\tabcolsep}{3pt}
	\begin{tabular}{lccc}
		\hline\hline
		%\multirow{2}*{Characteristics}
		Characteristics
		    %& \multicolumn{3}{c}{Internal dataset} & \multicolumn{3}{c}{External dataset} \\
		    %\cline{2-3}\cline{5-6}
		    & Non-pCR (n=178) & pCR (n=136) & $p$ \\%& Non-pCR (n=36) & pCR (n=10) & $p$ \\
		\hline
		\textbf{Age} (Mean$\pm$SD) & 51.7$\pm$11.4 & 49.2$\pm$12.7 & 0.066 \\%& 48.1$\pm$9.2 & 43.7$\pm$11.1 & 0.207 \\
		\textbf{ER} &  &  & $<$0.001 \\%&  &  & 0.480 \\
		\multicolumn{1}{c}{Positive} & 108 (60.7\%) & 26 (19.1\%) &  \\%& 23 (63.9\%) & 5 (50.0\%) & \\
		\multicolumn{1}{c}{Negative} & 70 (39.3\%) & 110 (80.9\%) &  \\%& 13 (36.1\%) & 5 (50.0\%) & \\
		\textbf{PR} &  &  & $<$0.001 \\%&  &  & $>$0.999 \\
		\multicolumn{1}{c}{Positive} & 134 (75.3\%) & 51 (37.5\%) &  \\%& 18 (50.0\%) & 5 (50.0\%) & \\
		\multicolumn{1}{c}{Negative} & 44 (24.7\%) & 85 (62.5\%) &  \\%& 18 (50.0\%) & 5 (50.0\%) & \\
		\textbf{HER2} &  &  & $<$0.001 \\%&  &  & 0.707 \\
		\multicolumn{1}{c}{Positive} & 49 (27.5\%) & 69 (50.7\%) &  \\%& 11 (30.6\%) & 4 (40.0\%) & \\
		\multicolumn{1}{c}{Negative} & 129 (72.5\%) & 67 (49.3\%) &  \\%& 25 (69.4\%) & 6 (60.0\%) & \\
		\textbf{T-stage} &  &  & 0.057 \\%&  &  & - \\
		\multicolumn{1}{c}{cT1} & 46 (25.8\%) & 50 (36.8\%) &  \\%& - & - & \\
		\multicolumn{1}{c}{cT2} & 97 (54.5\%) & 71 (52.2\%) &  \\%& - & - & \\
		\multicolumn{1}{c}{cT3} & 29 (16.3\%) & 14 (10.3\%) &  \\%& - & - & \\
		\multicolumn{1}{c}{cT4} & 6 (3.4\%) & 1 (0.7\%) &  \\%& - & - & \\
		\textbf{N-stage} &  &  & $<$0.001 \\%&  &  & - \\
		\multicolumn{1}{c}{cN0} & 74 (41.6\%) & 82 (60.3\%) &  \\%& - & - & \\
		\multicolumn{1}{c}{cN1} & 81 (45.5\%) & 33 (24.3\%) \\%&  & - & - & \\
		\multicolumn{1}{c}{cN2} & 1 (0.6\%) & 3 (2.2\%) &  \\%& - & - & \\
		\multicolumn{1}{c}{cN3} & 22 (12.4\%) & 18 (13.2\%) &  \\%& - & - & \\
		\hline\hline
	\end{tabular}
\end{table}

\subsection{Implementation details}
The proposed framework is implemented by PyTorch on NVIDIA Quadro RTX A6000.
In this study, CPRN and KN-S were first trained separately without using $\mathcal{L}^R_{sl}$ and $\mathcal{L}^R_{vp}$, then KN-A was trained based on the pre-trained CPRN, and finally, CPRN was fine-tuned with the loss function in Eq.~\ref{eq:total}.
We optimize these models with Adam under a learning rate of $10^{-4}$, 100 epochs, and a batch size of 1.
From the dataset, we utilize 250 patients for training, 14 patients for validation, and 50 patients for testing.
We align images between every two MRI timepoints from each patient during training and testing.
%And we set all the 46 patients from the external dataset for testing.
All the images are resampled to 1mm$\times$1mm$\times$1mm and then cropped to keep the bilateral breast area with the size of 176$\times$176$\times$352.
Unilateral breast images with the size of 176$\times$176$\times$176 are utilized to train, and bilateral breast images are for testing.
%For the external dataset, the testing image size is set to be 176$\times$176$\times$176.

\subsection{Evaluation metrics}
Four metrics are utilized to evaluate the quality of longitudinal image registration, including average landmark errors, changes in lesion volume between the moving image and the moved image, dice coefficient similarity (DSC), and the norm of the gradient of the Jacobian determinant (NGJD).

\begin{figure*}[!htbp]
    \centering
    \begin{minipage}{0.12\linewidth}
        \centerline{Baseline}
    \end{minipage}
    \begin{minipage}{0.21\linewidth}
        \centerline{\includegraphics[width=\textwidth]{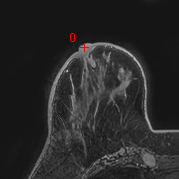}}
    \end{minipage}
    \begin{minipage}{0.21\linewidth}
        \centerline{\includegraphics[width=\textwidth]{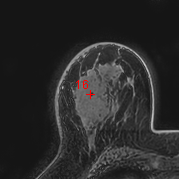}}
    \end{minipage}
    \begin{minipage}{0.21\linewidth}
        \centerline{\includegraphics[width=\textwidth]{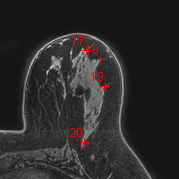}}
    \end{minipage}
    \begin{minipage}{0.21\linewidth}
        \centerline{\includegraphics[width=\textwidth]{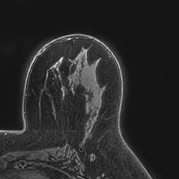}}
    \end{minipage}

    \vspace{2pt}

    \begin{minipage}{0.12\linewidth}
        \centerline{}
    \end{minipage}
    \begin{minipage}{0.21\linewidth}
       \centerline{Slice 54}
    \end{minipage}
    \begin{minipage}{0.21\linewidth}
        \centerline{Slice 79}
    \end{minipage}
    \begin{minipage}{0.21\linewidth}
        \centerline{Slice 82}
    \end{minipage}
    \begin{minipage}{0.21\linewidth}
        \centerline{Slice 85}
    \end{minipage}

    \vspace{5pt}
    
    \begin{minipage}{0.12\linewidth}
        \centerline{Follow-up 1}
    \end{minipage}
    \begin{minipage}{0.21\linewidth}
        \centerline{\includegraphics[width=\textwidth]{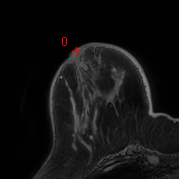}}
    \end{minipage}
    \begin{minipage}{0.21\linewidth}
        \centerline{\includegraphics[width=\textwidth]{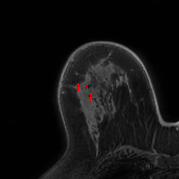}}
    \end{minipage}
    \begin{minipage}{0.21\linewidth}
        \centerline{\includegraphics[width=\textwidth]{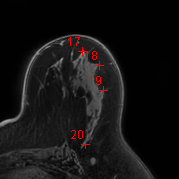}}
    \end{minipage}
    \begin{minipage}{0.21\linewidth}
        \centerline{\includegraphics[width=\textwidth]{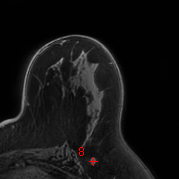}}
    \end{minipage}

    \vspace{2pt}

    \begin{minipage}{0.12\linewidth}
        \centerline{}
    \end{minipage}
    \begin{minipage}{0.21\linewidth}
        \centerline{Slice 59}
    \end{minipage}
    \begin{minipage}{0.21\linewidth}
        \centerline{Slice 80}
    \end{minipage}
    \begin{minipage}{0.21\linewidth}
        \centerline{Slice 83}
    \end{minipage}
    \begin{minipage}{0.21\linewidth}
        \centerline{Slice 84}
    \end{minipage}

    \vspace{5pt}
    
    \begin{minipage}{0.12\linewidth}
        \centerline{Follow-up 2}
    \end{minipage}
    \begin{minipage}{0.21\linewidth}
        \centerline{\includegraphics[width=\textwidth]{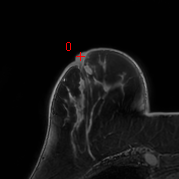}}
    \end{minipage}
    \begin{minipage}{0.21\linewidth}
        \centerline{\includegraphics[width=\textwidth]{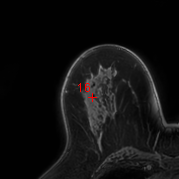}}
    \end{minipage}
    \begin{minipage}{0.21\linewidth}
        \centerline{\includegraphics[width=\textwidth]{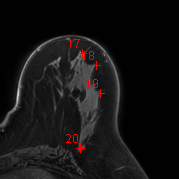}}
    \end{minipage}
    \begin{minipage}{0.21\linewidth}
        \centerline{\includegraphics[width=\textwidth]{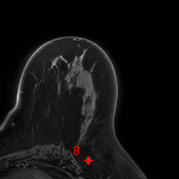}}
    \end{minipage}

    \vspace{2pt}

    \begin{minipage}{0.12\linewidth}
        \centerline{}
    \end{minipage}
    \begin{minipage}{0.21\linewidth}
        \centerline{Slice 52}
    \end{minipage}
    \begin{minipage}{0.21\linewidth}
        \centerline{Slice 80}
    \end{minipage}
    \begin{minipage}{0.21\linewidth}
        \centerline{Slice 80}
    \end{minipage}
    \begin{minipage}{0.21\linewidth}
        \centerline{Slice 87}
    \end{minipage}
    
    \caption{Examples of manually selected landmarks on baseline and follow-up images for one patient. Landmark \#0: right nipple; \#8: left axillary lymph node; \#16: tumor with a marker; \#17-20: outline of fibroglandular tissue.} \label{fig:landmark}
\end{figure*}

Average landmarks error $d_{avg}$ is a metric to measure the registration accuracy for expert-defined landmarks between baseline and follow-up images.
Four radiologists (V.L., S.V., A.D., and R.M.) manually determined the landmarks independently for the testing dataset. V.L. defined 21 anatomical/geometric landmarks in the baseline image for each patient, which other radiologists confirmed. Then, they independently define the same landmarks in the follow-up images corresponding to those in the baseline images. The inter-expert landmark distance for the follow-up images is 3.66$\pm$2.87 mm. The final expert-based landmarks are determined using the average coordinates of four radiologists. As shown in Fig.~\ref{fig:landmark}, these landmarks are usually located at nipples, intramammary/axillary lymph nodes, vessels, glandular tissues, cysts, and tumors. Only the tumor marker for non-pCR patients is annotated as the landmark. The tumors change dramatically during treatment, and consequently, it is difficult for pCR patients to locate their tumors because they disappear completely. In addition, landmarks \#0 and \#1 are fixed to the right and left nipples, respectively.
%Experts annotate $K$ landmarks on baseline images and label the corresponding landmarks in follow-up images.
The average landmarks error is defined as $d_{avg}=\sum_{k=1}^{K}{\Vert y_k - x_k\circ\phi \Vert}_2/K$, where $K=20$ refers to the number of landmarks, $x_k$ and $y_k$ indicate the position of $k$th landmark in baseline images and follow-up images, respectively.
%And $\phi$ is the deformation field calculated by registration algorithm.

The change in lesion volume, denoted as $\Delta V$, is determined by comparing the volume of lesions in the baseline images before and after registration. This helps to evaluate the volume preserving of lesions accurately and is defined as $\Delta V=\| V(I_m\circ\phi) - V(I_m) \|_1/V(I_m)$. $V(\cdot)$ refers to the lesion volume for the input image.
The lesion masks for baseline images in the testing dataset are annotated manually using the wash-in images. R.M. performed the initial annotation, which was then reviewed and confirmed by other radiologists.
The fewer changes in lesion refer to better registration accuracy.
%Because the shape and size of the lesion are key factors for follow-up pCR prediction.

Besides, the DSC of the whole breast segmentation and the NGJD of the deformation field, the widely used image registration metrics, are employed to estimate the similarity for the whole breast and the smoothness of the deformation field, respectively as global registration quality measures.

\section{Results}
\subsection{Alternative comparisons}\label{sec:alternative}
We compare our method with five conventional methods and four learning-based methods. Conventional methods consist of SyN~\citep{avants2008symmetric}, NiftyReg~\citep{modat2010fast}, deedsBCV~\citep{heinrich2013mrf}, DRAMMS~\citep{ou2011dramms,ou2015deformable}, and Elastix~\citep{klein2009elastix,mehrabian2018deformable}. T1w pre-contrast images $T_m$ and $T_f$ are utilized to optimize the conventional methods.
Learning-based methods include VoxelMorph~\citep{balakrishnan2019voxelmorph}, JointRegSeg~\citep{estienne2020deep}, HyperMorph~\citep{hoopes2021hypermorph}, and cLapIRN~\citep{mok2021conditional}. Both T1w pre-contrast images $T_m$, $T_f$ and wash-in images $I_m$, $I_f$ are concatenated as the input of the learning-based methods and only $T_m$ and $T_f$ are utilized to calculate similarity losses.

\begin{table*}
	\centering
	\caption{The quantitative results of longitudinal breast DCE-MRI registration on the clinical dataset. The comparison methods involve state-of-the-art methods and ablation studies with different key components in the proposed method (SK (only) indicates generating the deformation field by interpolating on structural keypoints, CRN indicates conditional registration network with a single scale, CPRN indicates conditional pyramid registration network, +SK indicates training CPRN with structural keypoint loss, +VP (AK) indicates training CPRN with abnormal keypoints-based volume-preserving loss, +VP (Handcraft) indicates training CPRN with handcraft tumor annotation-based volume-preserving loss, and +SK+VP (AK) indicates using both SK and VP (AK). The best result is in bold and the second best one is underlined.}
	\label{tab:results}
	\setlength{\tabcolsep}{3pt}
	\begin{tabular}{lcccc}
		\hline\hline
		%\multirow{2}*{Method}
		Method
		    %& \multicolumn{4}{c}{Internal dataset} && \multicolumn{4}{c}{External dataset} \\
		    %\cline{2-5}\cline{7-10}
		    & DSC$\uparrow$ & $d_{avg}$(mm)$\downarrow$ & $\Delta V(\%)\downarrow$ & NGJD$(\%)\downarrow$ \\%&
		    %& DSC$\uparrow$ & $d_{avg}$(mm)$\downarrow$ & $\Delta V(\%)\downarrow$ & NGJD$(\%)\downarrow$ \\
		\hline
		Rigid & 0.843$\pm$0.064 & 8.47$\pm$3.51 & - & - \\%&& 0.738$\pm$0.106 & 16.91$\pm$8.12 & - & -\\
        SK (only) & 0.878$\pm$0.044 & 7.89$\pm$3.09 & 12.7$\pm$10.5 & 0.294$\pm$0.150 \\
        \hline
		SyN~\citep{avants2008symmetric} & 0.929$\pm$0.030 & 5.63$\pm$4.25 & 36.2$\pm$22.3 & 0.018$\pm$0.047 \\
		NiftyReg~\citep{modat2010fast} & 0.931$\pm$0.036 & 5.53$\pm$3.43 & \underline{9.8$\pm$7.6} & 0.016$\pm$0.035 \\%&& 0.810$\pm$0.110 & 12.92$\pm$8.65 & $\pm$ & 0.013$\pm$0.028 \\
		deedsBCV~\citep{heinrich2013mrf} & 0.904$\pm$0.040 & 5.52$\pm$3.01 & 13.7$\pm$10.0 & 0.065$\pm$0.192 \\%&& 0.910$\pm$0.043 & 6.89$\pm$4.16 & 22.4$\pm$23.8 & 0.008$\pm$0.032 \\
		DRAMMS~\citep{ou2011dramms,ou2015deformable} & 0.932$\pm$0.040 & 5.60$\pm$3.85 & 34.1$\pm$18.9 & 0.040$\pm$0.085 \\
		Elastix~\citep{klein2009elastix,mehrabian2018deformable} & 0.933$\pm$0.028 & \textbf{5.12$\pm$2.93} & 21.2$\pm$14.7 & 0.020$\pm$0.063 \\
		\hline
		VoxelMorph~\citep{balakrishnan2019voxelmorph} & 0.947$\pm$0.022 & 6.11$\pm$4.61 & 19.0$\pm$13.2 & 0.003$\pm$0.007 \\%&& 0.852$\pm$0.076 & 12.71$\pm$7.80 & 19.4$\pm$28.9 & 0.002$\pm$0.006 \\
		JointRegSeg~\citep{estienne2020deep} & 0.915$\pm$0.046 & 7.47$\pm$5.05 & 15.9$\pm$18.5 & 0.014$\pm$0.020 \\%&& 0.797$\pm$0.094 & 16.02$\pm$8.44 & 12.0$\pm$15.0 & 0.024$\pm$0.037 \\
		HyperMorph~\citep{hoopes2021hypermorph} & 0.926$\pm$0.038 & 7.02$\pm$4.55 & 22.9$\pm$15.8 &0.008$\pm$0.013 \\
		cLapIRN~\citep{mok2021conditional} & 0.947$\pm$0.021 & 5.61$\pm$3.72 & 21.0$\pm$15.1 & \textbf{0.002$\pm$0.003} \\
		\hline
		CRN & \underline{0.948$\pm$0.021} & 5.78$\pm$3.50 & 22.0$\pm$16.0 & 0.003$\pm$0.005 \\%&& 0.857$\pm$0.073 & 12.93$\pm$7.97 & 15.9$\pm$21.8 & 0.009$\pm$0.011 \\
		CPRN & \underline{0.948$\pm$0.021} & 5.59$\pm$3.78 & 18.6$\pm$14.6 & \underline{0.002$\pm$0.005} \\%&& 0.857$\pm$0.076 & 12.58$\pm$7.85 & 20.3$\pm$25.2 & 0.002$\pm$0.004 \\
		+SK & \underline{0.948$\pm$0.021} & \underline{5.20$\pm$3.44} & 21.2$\pm$17.0 & 0.004$\pm$0.009 \\%&& 0.864$\pm$0.072 & 12.11$\pm$8.05 & 20.3$\pm$24.8 & 0.007$\pm$0.014 \\
		+VP (AK) & \underline{0.948$\pm$0.021} & 5.45$\pm$3.79 & 10.5$\pm$11.0 & 0.003$\pm$0.005 \\%&& 0.856$\pm$0.075 & 12.68$\pm$8.15 & 15.2$\pm$17.9 & 0.004$\pm$0.007 \\
		+VP (Handcraft) & \textbf{0.948$\pm$0.020} & 5.49$\pm$3.74 & \textbf{9.7$\pm$9.7} & 0.004$\pm$0.009 \\%&& 0.861$\pm$0.072 & 12.47$\pm$7.99 & 18.9$\pm$20.5 & 0.004$\pm$0.008 \\
		+SK+VP (AK) (Proposed) & 0.947$\pm$0.021 & 5.35$\pm$3.46 & 11.0$\pm$10.7 & 0.005$\pm$0.011 \\%&& 0.852$\pm$0.076 & 12.68$\pm$8.20 & 15.4$\pm$14.1 & 0.007$\pm$0.010 \\
		\hline\hline
	\end{tabular}
\end{table*}

\begin{figure*}[!htbp]
    \centering
    \begin{minipage}{0.158\linewidth}
        \begin{minipage}{0.47\textwidth}
            \centerline{\includegraphics[width=\textwidth]{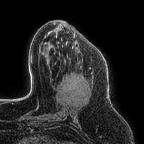}}
            \vspace{2pt}
            \centerline{\includegraphics[width=\textwidth]{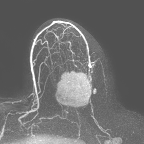}}
            \vspace{2pt}
        \end{minipage}
        \begin{minipage}{0.47\textwidth}
            \centerline{\includegraphics[width=\textwidth]{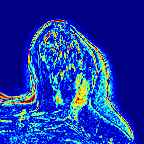}}
            \vspace{2pt}
            \centerline{\includegraphics[width=\textwidth]{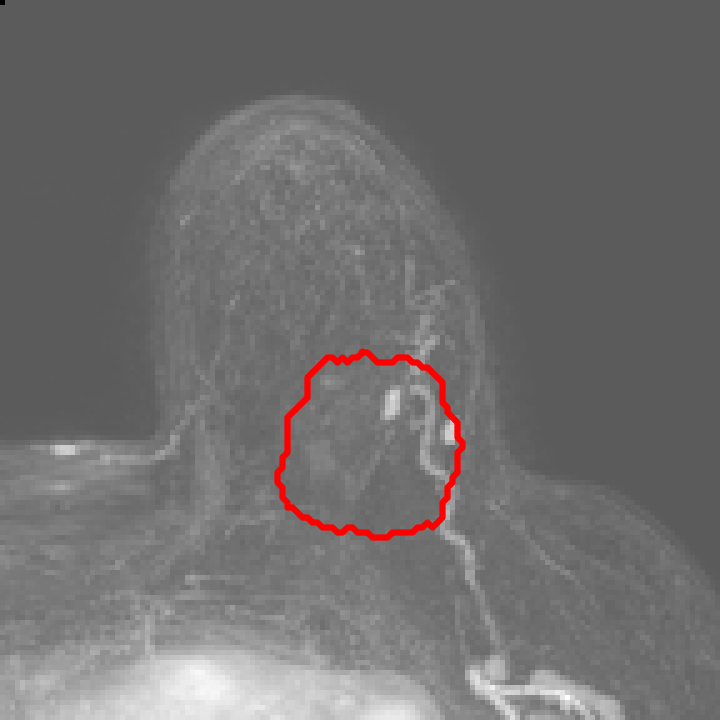}}
            \vspace{2pt}
        \end{minipage}
    \end{minipage}
    \begin{minipage}{0.158\linewidth}
        \begin{minipage}{0.47\textwidth}
            \centerline{\includegraphics[width=\textwidth]{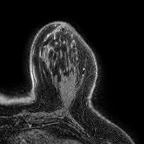}}
            \vspace{2pt}
            \centerline{\includegraphics[width=\textwidth]{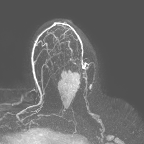}}
            \vspace{2pt}
        \end{minipage}
        \begin{minipage}{0.47\textwidth}
            \centerline{\includegraphics[width=\textwidth]{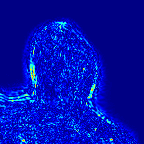}}
            \vspace{2pt}
            \centerline{\includegraphics[width=\textwidth]{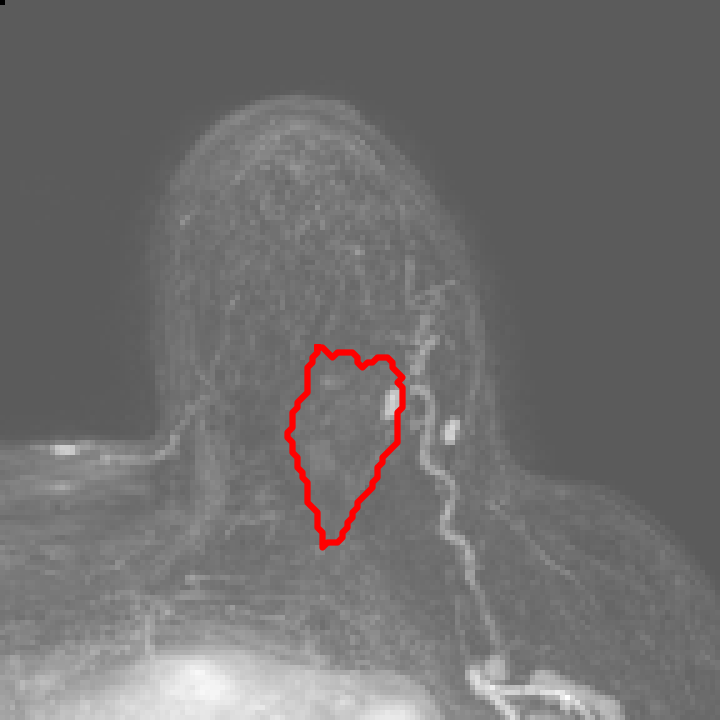}}
            \vspace{2pt}
        \end{minipage}
    \end{minipage}
    \begin{minipage}{0.158\linewidth}
        \begin{minipage}{0.47\textwidth}
            \centerline{\includegraphics[width=\textwidth]{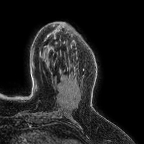}}
            \vspace{2pt}
            \centerline{\includegraphics[width=\textwidth]{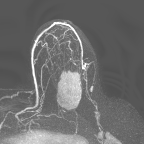}}
            \vspace{2pt}
        \end{minipage}
        \begin{minipage}{0.47\textwidth}
            \centerline{\includegraphics[width=\textwidth]{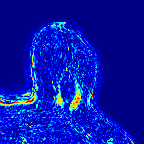}}
            \vspace{2pt}
            \centerline{\includegraphics[width=\textwidth]{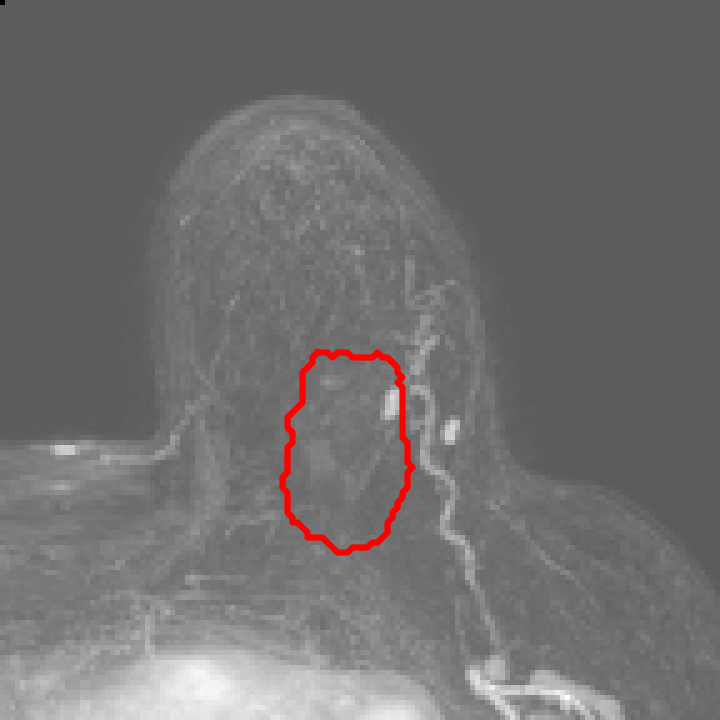}}
            \vspace{2pt}
        \end{minipage}
    \end{minipage}
    \begin{minipage}{0.158\linewidth}
        \begin{minipage}{0.47\textwidth}
            \centerline{\includegraphics[width=\textwidth]{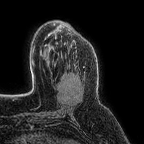}}
            \vspace{2pt}
            \centerline{\includegraphics[width=\textwidth]{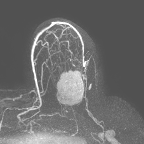}}
            \vspace{2pt}
        \end{minipage}
        \begin{minipage}{0.47\textwidth}
            \centerline{\includegraphics[width=\textwidth]{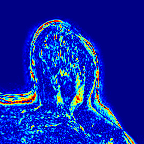}}
            \vspace{2pt}
            \centerline{\includegraphics[width=\textwidth]{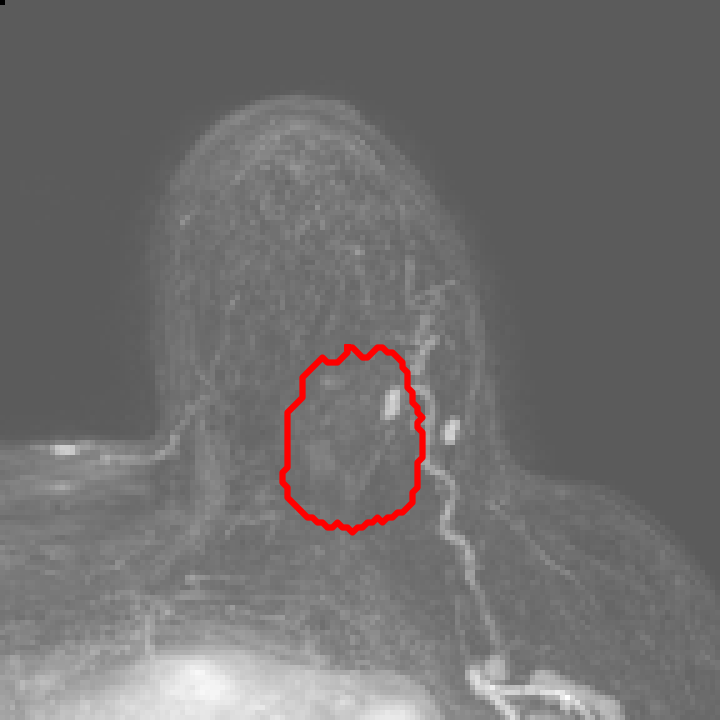}}
            \vspace{2pt}
        \end{minipage}
    \end{minipage}
    \begin{minipage}{0.158\linewidth}
        \begin{minipage}{0.47\textwidth}
            \centerline{\includegraphics[width=\textwidth]{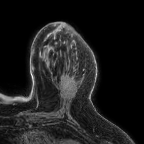}}
            \vspace{2pt}
            \centerline{\includegraphics[width=\textwidth]{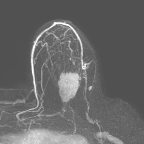}}
            \vspace{2pt}
        \end{minipage}
        \begin{minipage}{0.47\textwidth}
            \centerline{\includegraphics[width=\textwidth]{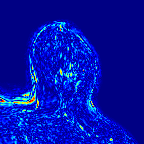}}
            \vspace{2pt}
            \centerline{\includegraphics[width=\textwidth]{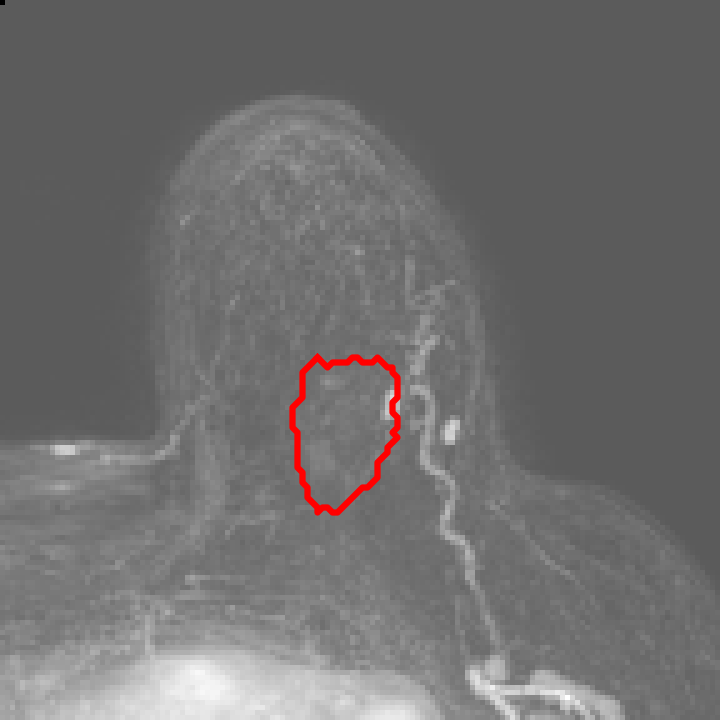}}
            \vspace{2pt}
        \end{minipage}
    \end{minipage}
    \begin{minipage}{0.158\linewidth}
        \begin{minipage}{0.47\textwidth}
            \centerline{\includegraphics[width=\textwidth]{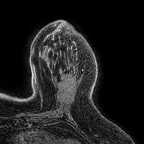}}
            \vspace{2pt}
            \centerline{\includegraphics[width=\textwidth]{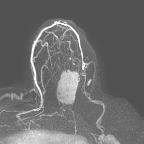}}
            \vspace{2pt}
        \end{minipage}
        \begin{minipage}{0.47\textwidth}
            \centerline{\includegraphics[width=\textwidth]{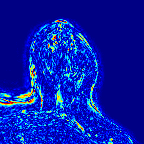}}
            \vspace{2pt}
            \centerline{\includegraphics[width=\textwidth]{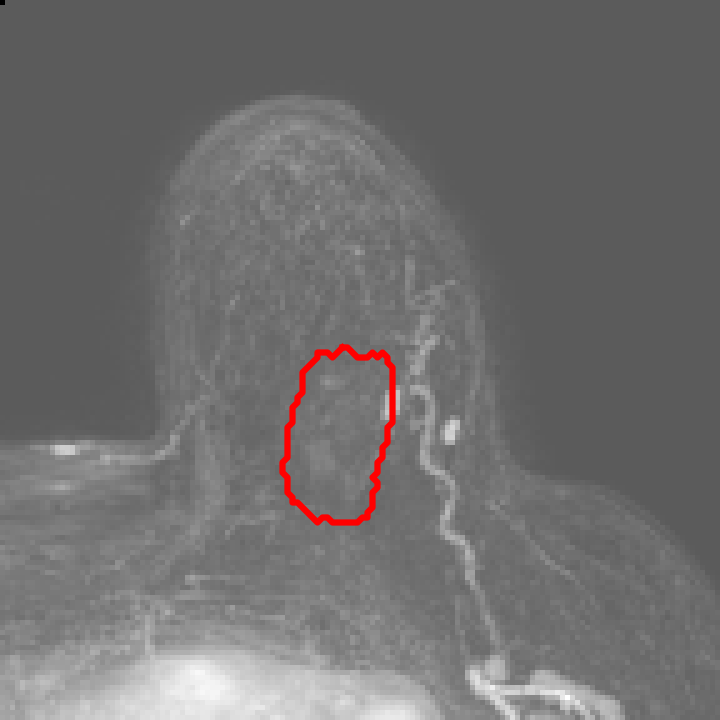}}
            \vspace{2pt}
        \end{minipage}
    \end{minipage}
    \begin{minipage}{0.013\linewidth}
        \centerline{\includegraphics[width=\textwidth]{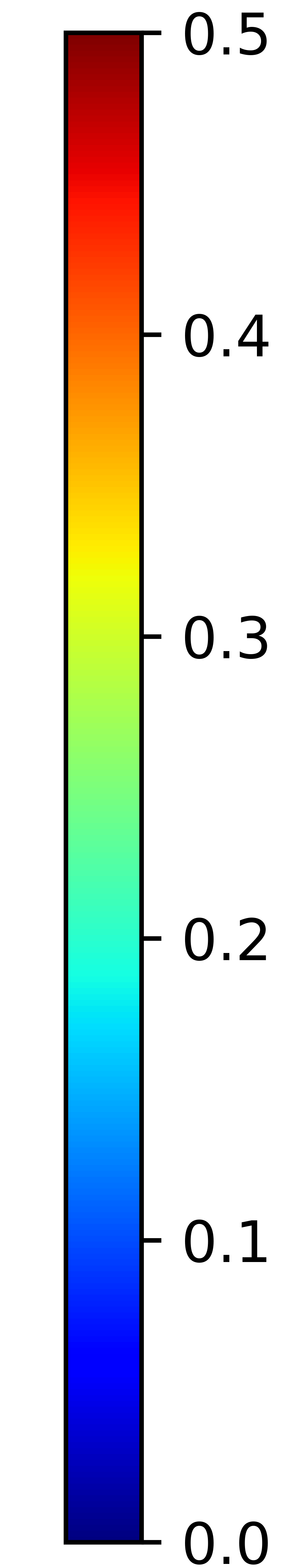}}
        \vspace{40pt}
    \end{minipage}
    
    \begin{minipage}{0.158\linewidth}
        \centerline{Moving}
        \vspace{2pt}
    \end{minipage}
    \begin{minipage}{0.158\linewidth}
        \centerline{SyN}
        \vspace{2pt}
    \end{minipage}
    \begin{minipage}{0.158\linewidth}
        \centerline{NiftyReg}
        \vspace{2pt}
    \end{minipage}
    \begin{minipage}{0.158\linewidth}
        \centerline{deedsBCV}
        \vspace{2pt}
    \end{minipage}
    \begin{minipage}{0.158\linewidth}
        \centerline{DRAMMS}
        \vspace{2pt}
    \end{minipage}
    \begin{minipage}{0.158\linewidth}
        \centerline{Elastix}
        \vspace{2pt}
    \end{minipage}
    \begin{minipage}{0.013\linewidth}
        \centerline{}
    \end{minipage}
    
    \begin{minipage}{0.158\linewidth}
        \begin{minipage}{0.47\textwidth}
            \centerline{\includegraphics[width=\textwidth]{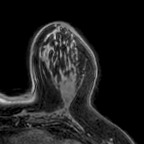}}
            \vspace{2pt}
            \centerline{\includegraphics[width=\textwidth]{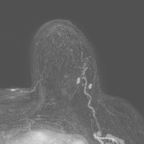}}
            \vspace{2pt}
        \end{minipage}
        \begin{minipage}{0.47\textwidth}
            \centerline{\includegraphics[width=\textwidth]{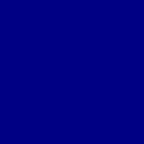}}
            \vspace{2pt}
            \centerline{\includegraphics[width=\textwidth]{figs/compare/B000020339/mov_tumor.png}}
            \vspace{2pt}
        \end{minipage}
    \end{minipage}
    \begin{minipage}{0.158\linewidth}
        \begin{minipage}{0.47\textwidth}
            \centerline{\includegraphics[width=\textwidth]{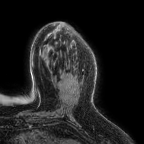}}
            \vspace{2pt}
            \centerline{\includegraphics[width=\textwidth]{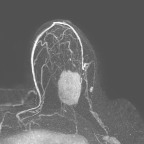}}
            \vspace{2pt}
        \end{minipage}
        \begin{minipage}{0.47\textwidth}
            \centerline{\includegraphics[width=\textwidth]{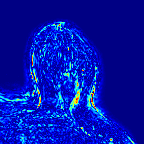}}
            \vspace{2pt}
            \centerline{\includegraphics[width=\textwidth]{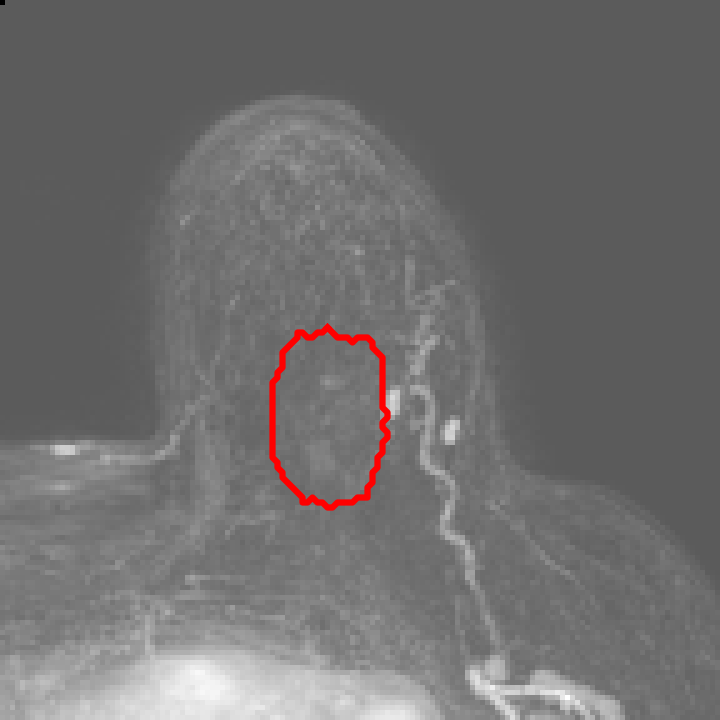}}
            \vspace{2pt}
        \end{minipage}
    \end{minipage}
    \begin{minipage}{0.158\linewidth}
        \begin{minipage}{0.47\textwidth}
            \centerline{\includegraphics[width=\textwidth]{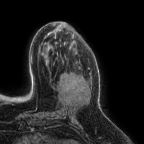}}
            \vspace{2pt}
            \centerline{\includegraphics[width=\textwidth]{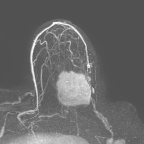}}
            \vspace{2pt}
        \end{minipage}
        \begin{minipage}{0.47\textwidth}
            \centerline{\includegraphics[width=\textwidth]{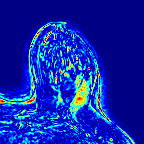}}
            \vspace{2pt}
            \centerline{\includegraphics[width=\textwidth]{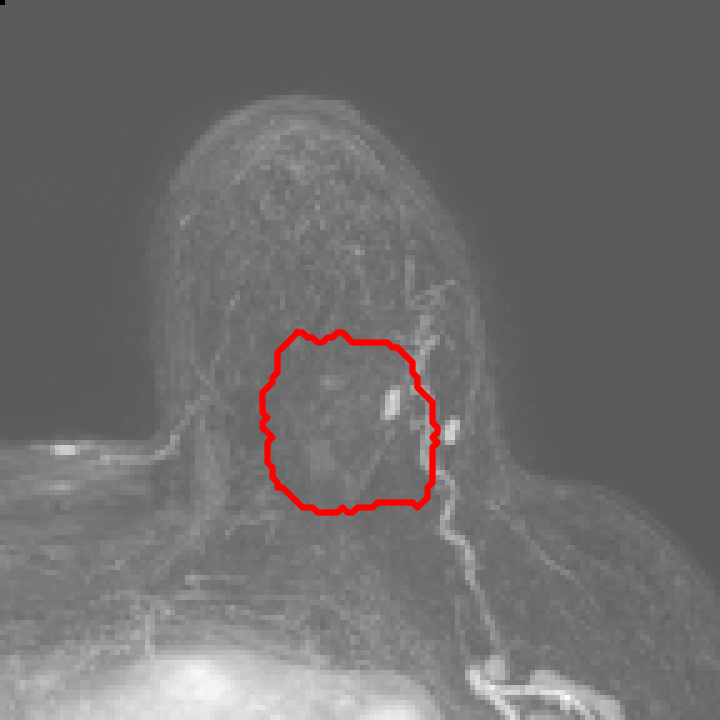}}
            \vspace{2pt}
        \end{minipage}
    \end{minipage}
    \begin{minipage}{0.158\linewidth}
        \begin{minipage}{0.47\textwidth}
            \centerline{\includegraphics[width=\textwidth]{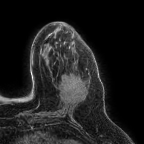}}
            \vspace{2pt}
            \centerline{\includegraphics[width=\textwidth]{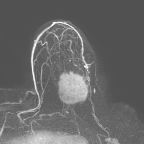}}
            \vspace{2pt}
        \end{minipage}
        \begin{minipage}{0.47\textwidth}
            \centerline{\includegraphics[width=\textwidth]{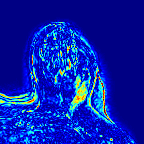}}
            \vspace{2pt}
            \centerline{\includegraphics[width=\textwidth]{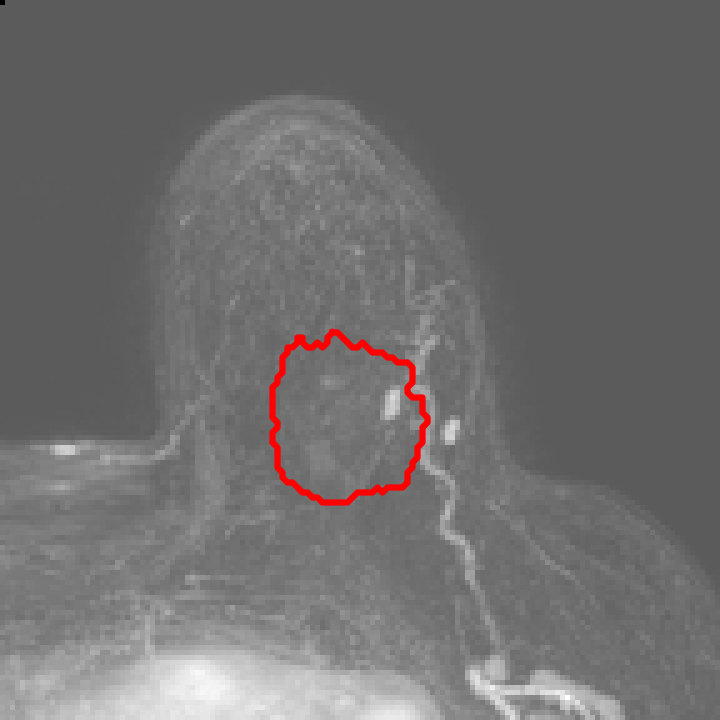}}
            \vspace{2pt}
        \end{minipage}
    \end{minipage}
    \begin{minipage}{0.158\linewidth}
        \begin{minipage}{0.47\textwidth}
            \centerline{\includegraphics[width=\textwidth]{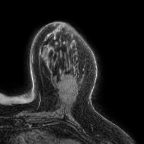}}
            \vspace{2pt}
            \centerline{\includegraphics[width=\textwidth]{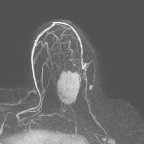}}
            \vspace{2pt}
        \end{minipage}
        \begin{minipage}{0.47\textwidth}
            \centerline{\includegraphics[width=\textwidth]{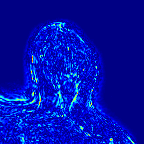}}
            \vspace{2pt}
            \centerline{\includegraphics[width=\textwidth]{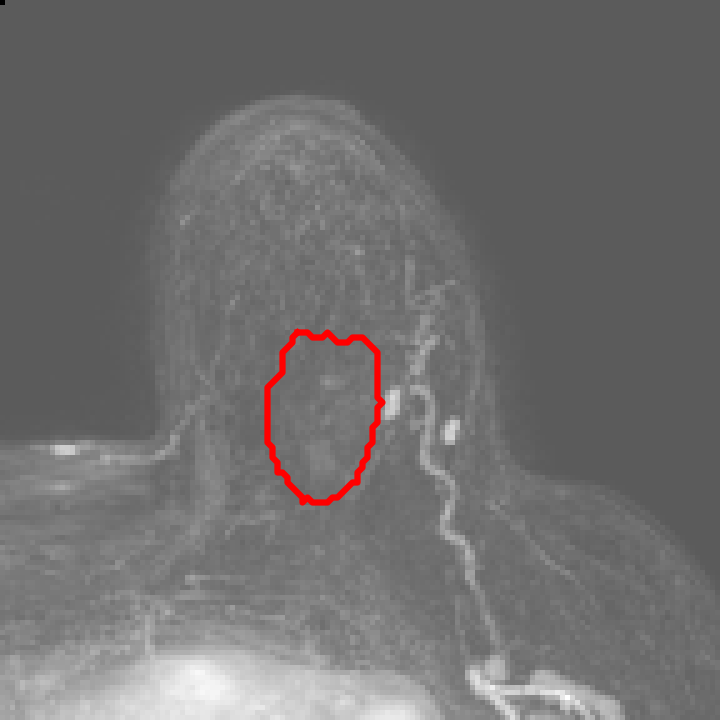}}
            \vspace{2pt}
        \end{minipage}
    \end{minipage}
    \begin{minipage}{0.158\linewidth}
        \begin{minipage}{0.47\textwidth}
            \centerline{\includegraphics[width=\textwidth]{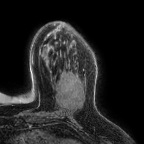}}
            \vspace{2pt}
            \centerline{\includegraphics[width=\textwidth]{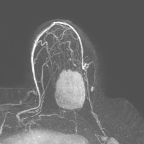}}
            \vspace{2pt}
        \end{minipage}
        \begin{minipage}{0.47\textwidth}
            \centerline{\includegraphics[width=\textwidth]{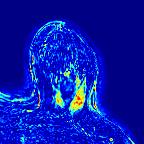}}
            \vspace{2pt}
            \centerline{\includegraphics[width=\textwidth]{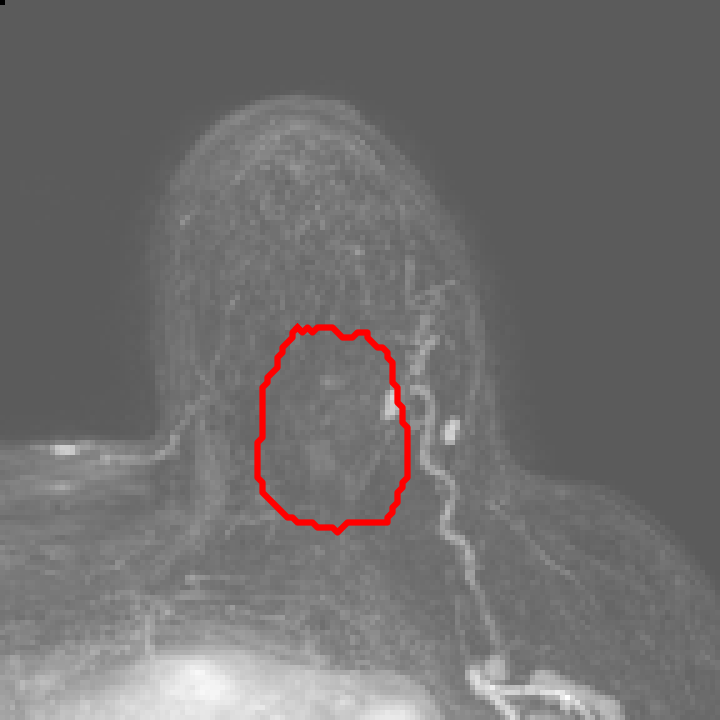}}
            \vspace{2pt}
        \end{minipage}
    \end{minipage}
    \begin{minipage}{0.013\linewidth}
        \centerline{\includegraphics[width=\textwidth]{figs/compare/colormap.png}}
        \vspace{40pt}
    \end{minipage}
    
    \begin{minipage}{0.158\linewidth}
        \centerline{Fixed}
        \vspace{2pt}
    \end{minipage}
    \begin{minipage}{0.158\linewidth}
        \centerline{VoxelMorph}
        \vspace{2pt}
    \end{minipage}
    \begin{minipage}{0.158\linewidth}
        \centerline{JointRegSeg}
        \vspace{2pt}
    \end{minipage}
    \begin{minipage}{0.158\linewidth}
        \centerline{HyperMorph}
        \vspace{2pt}
    \end{minipage}
    \begin{minipage}{0.158\linewidth}
        \centerline{cLapIRN}
        \vspace{2pt}
    \end{minipage}
    \begin{minipage}{0.158\linewidth}
        \centerline{Proposed}
        \vspace{2pt}
    \end{minipage}
    \begin{minipage}{0.013\linewidth}
        \centerline{}
    \end{minipage}
    
    \centerline{\textbf{Case 1}}
    
    % case 2
    \begin{minipage}{0.158\linewidth}
        \vspace{4pt}
        \begin{minipage}{0.47\textwidth}
            \centerline{\includegraphics[width=\textwidth]{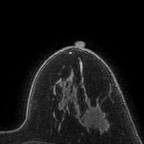}}
            \vspace{2pt}
            \centerline{\includegraphics[width=\textwidth]{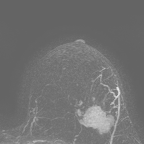}}
            \vspace{2pt}
        \end{minipage}
        \begin{minipage}{0.47\textwidth}
            \centerline{\includegraphics[width=\textwidth]{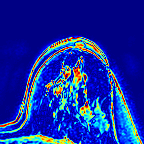}}
            \vspace{2pt}
            \centerline{\includegraphics[width=\textwidth]{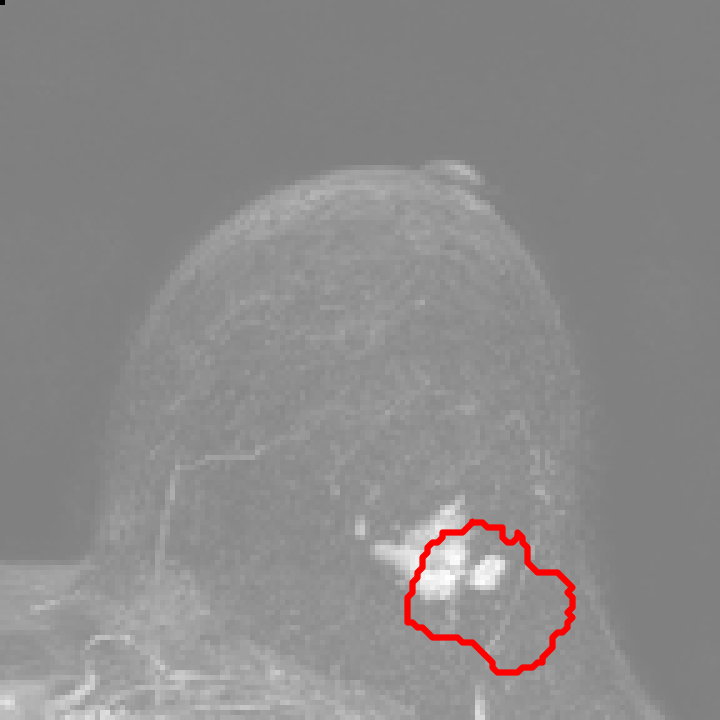}}
            \vspace{2pt}
        \end{minipage}
    \end{minipage}
    \begin{minipage}{0.158\linewidth}
        \vspace{4pt}
        \begin{minipage}{0.47\textwidth}
            \centerline{\includegraphics[width=\textwidth]{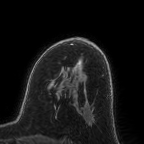}}
            \vspace{2pt}
            \centerline{\includegraphics[width=\textwidth]{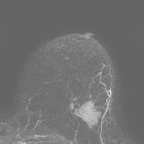}}
            \vspace{2pt}
        \end{minipage}
        \begin{minipage}{0.47\textwidth}
            \centerline{\includegraphics[width=\textwidth]{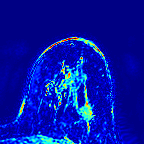}}
            \vspace{2pt}
            \centerline{\includegraphics[width=\textwidth]{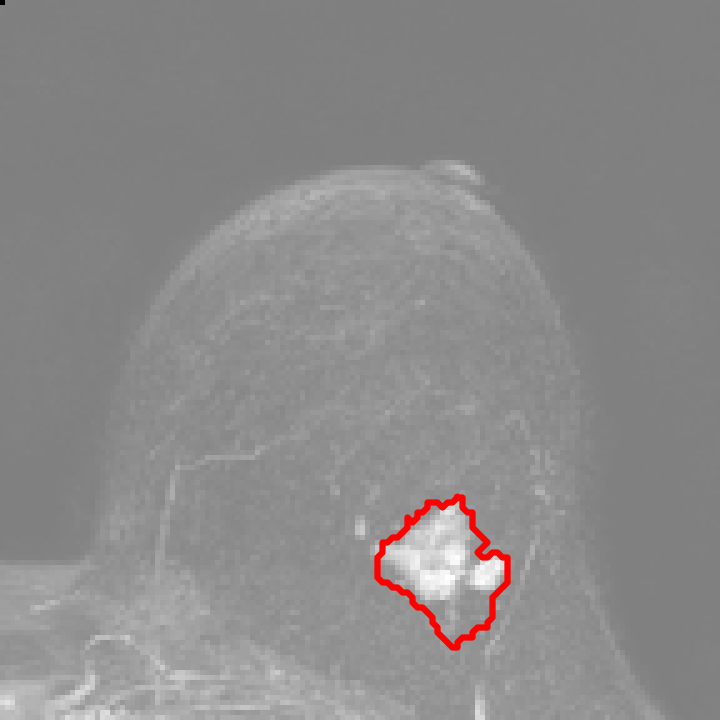}}
            \vspace{2pt}
        \end{minipage}
    \end{minipage}
    \begin{minipage}{0.158\linewidth}
        \vspace{4pt}
        \begin{minipage}{0.47\textwidth}
            \centerline{\includegraphics[width=\textwidth]{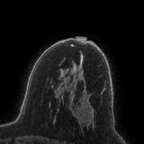}}
            \vspace{2pt}
            \centerline{\includegraphics[width=\textwidth]{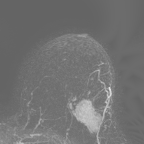}}
            \vspace{2pt}
        \end{minipage}
        \begin{minipage}{0.47\textwidth}
            \centerline{\includegraphics[width=\textwidth]{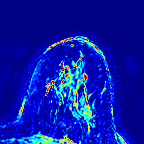}}
            \vspace{2pt}
            \centerline{\includegraphics[width=\textwidth]{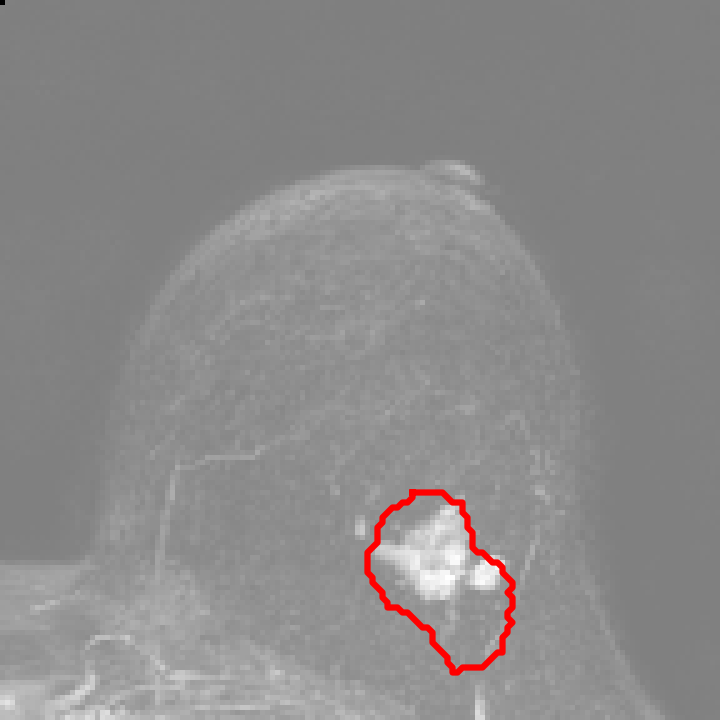}}
            \vspace{2pt}
        \end{minipage}
    \end{minipage}
    \begin{minipage}{0.158\linewidth}
        \vspace{4pt}
        \begin{minipage}{0.47\textwidth}
            \centerline{\includegraphics[width=\textwidth]{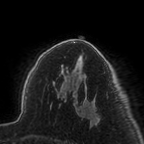}}
            \vspace{2pt}
            \centerline{\includegraphics[width=\textwidth]{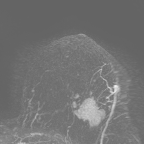}}
            \vspace{2pt}
        \end{minipage}
        \begin{minipage}{0.47\textwidth}
            \centerline{\includegraphics[width=\textwidth]{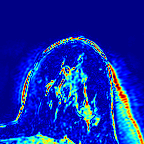}}
            \vspace{2pt}
            \centerline{\includegraphics[width=\textwidth]{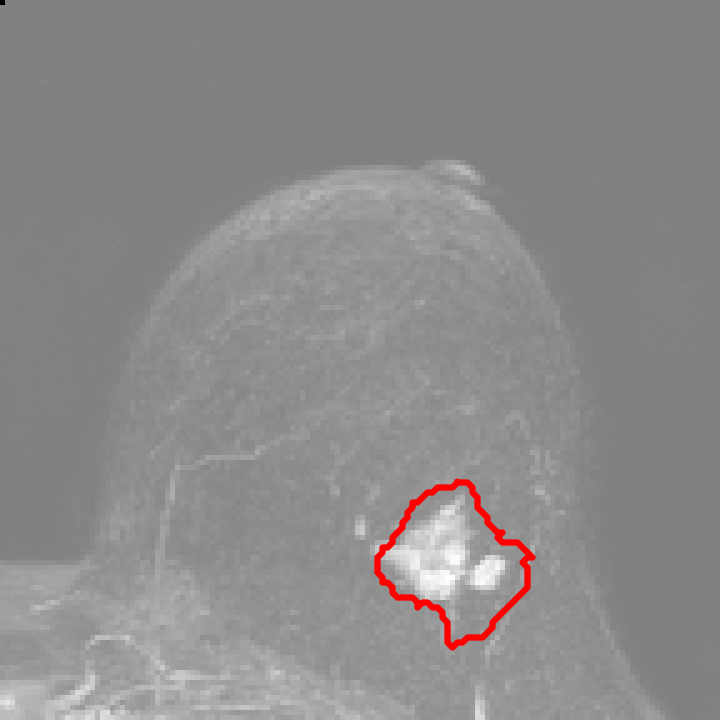}}
            \vspace{2pt}
        \end{minipage}
    \end{minipage}
    \begin{minipage}{0.158\linewidth}
        \vspace{4pt}
        \begin{minipage}{0.47\textwidth}
            \centerline{\includegraphics[width=\textwidth]{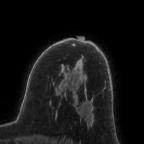}}
            \vspace{2pt}
            \centerline{\includegraphics[width=\textwidth]{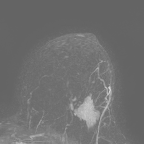}}
            \vspace{2pt}
        \end{minipage}
        \begin{minipage}{0.47\textwidth}
            \centerline{\includegraphics[width=\textwidth]{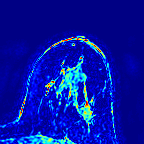}}
            \vspace{2pt}
            \centerline{\includegraphics[width=\textwidth]{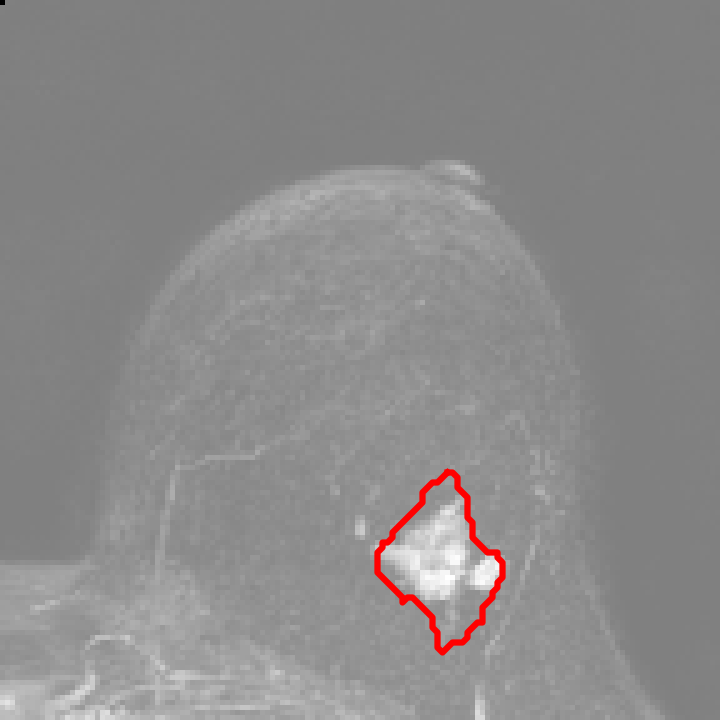}}
            \vspace{2pt}
        \end{minipage}
    \end{minipage}
    \begin{minipage}{0.158\linewidth}
        \vspace{4pt}
        \begin{minipage}{0.47\textwidth}
            \centerline{\includegraphics[width=\textwidth]{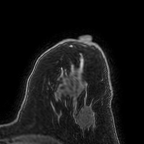}}
            \vspace{2pt}
            \centerline{\includegraphics[width=\textwidth]{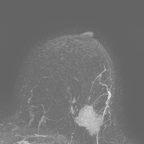}}
            \vspace{2pt}
        \end{minipage}
        \begin{minipage}{0.47\textwidth}
            \centerline{\includegraphics[width=\textwidth]{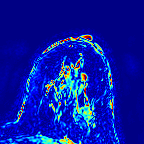}}
            \vspace{2pt}
            \centerline{\includegraphics[width=\textwidth]{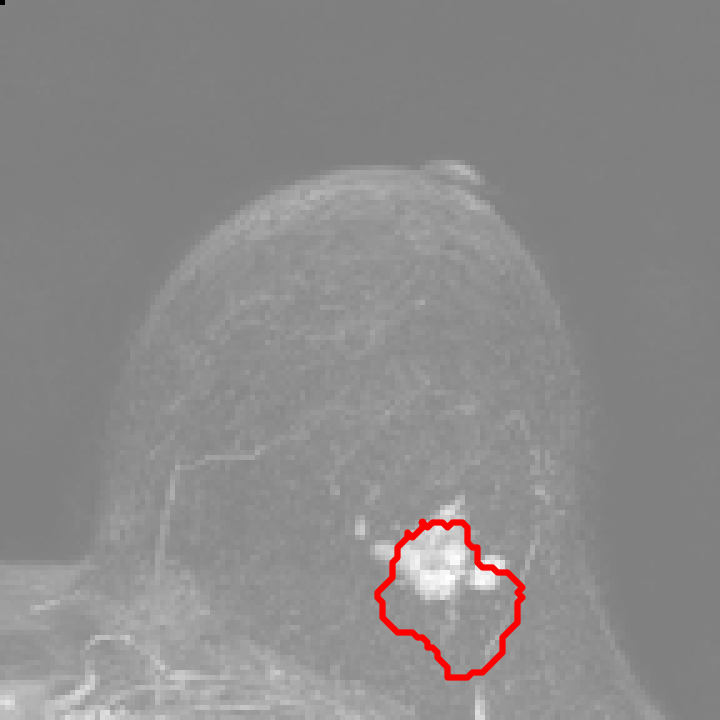}}
            \vspace{2pt}
        \end{minipage}
    \end{minipage}
    \begin{minipage}{0.013\linewidth}
        \centerline{\includegraphics[width=\textwidth]{figs/compare/colormap.png}}
        \vspace{40pt}
    \end{minipage}
    
    \begin{minipage}{0.158\linewidth}
        \centerline{Moving}
        \vspace{2pt}
    \end{minipage}
    \begin{minipage}{0.158\linewidth}
        \centerline{SyN}
        \vspace{2pt}
    \end{minipage}
    \begin{minipage}{0.158\linewidth}
        \centerline{NiftyReg}
        \vspace{2pt}
    \end{minipage}
    \begin{minipage}{0.158\linewidth}
        \centerline{deedsBCV}
        \vspace{2pt}
    \end{minipage}
    \begin{minipage}{0.158\linewidth}
        \centerline{DRAMMS}
        \vspace{2pt}
    \end{minipage}
    \begin{minipage}{0.158\linewidth}
        \centerline{Elastix}
        \vspace{2pt}
    \end{minipage}
    \begin{minipage}{0.013\linewidth}
        \centerline{}
    \end{minipage}
    
    \begin{minipage}{0.158\linewidth}
        \begin{minipage}{0.47\textwidth}
            \centerline{\includegraphics[width=\textwidth]{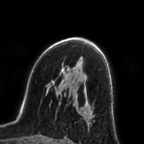}}
            \vspace{2pt}
            \centerline{\includegraphics[width=\textwidth]{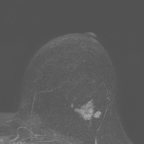}}
            \vspace{2pt}
        \end{minipage}
        \begin{minipage}{0.47\textwidth}
            \centerline{\includegraphics[width=\textwidth]{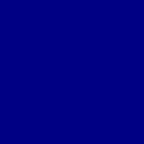}}
            \vspace{2pt}
            \centerline{\includegraphics[width=\textwidth]{figs/compare/B000020083/mov_tumor.png}}
            \vspace{2pt}
        \end{minipage}
    \end{minipage}
    \begin{minipage}{0.158\linewidth}
        \begin{minipage}{0.47\textwidth}
            \centerline{\includegraphics[width=\textwidth]{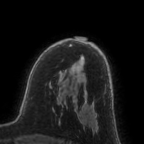}}
            \vspace{2pt}
            \centerline{\includegraphics[width=\textwidth]{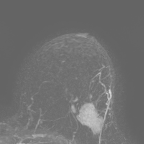}}
            \vspace{2pt}
        \end{minipage}
        \begin{minipage}{0.47\textwidth}
            \centerline{\includegraphics[width=\textwidth]{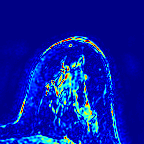}}
            \vspace{2pt}
            \centerline{\includegraphics[width=\textwidth]{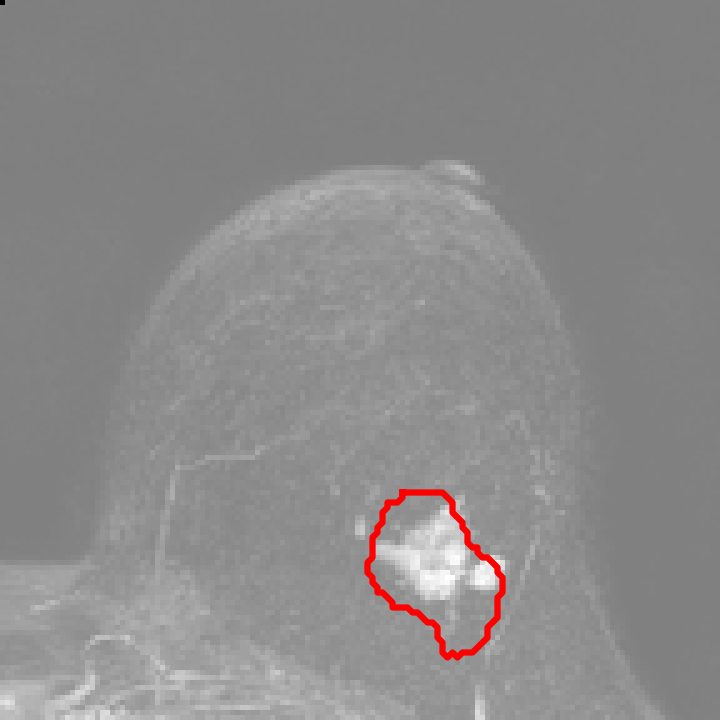}}
            \vspace{2pt}
        \end{minipage}
    \end{minipage}
    \begin{minipage}{0.158\linewidth}
        \begin{minipage}{0.47\textwidth}
            \centerline{\includegraphics[width=\textwidth]{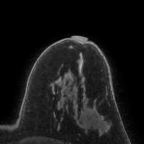}}
            \vspace{2pt}
            \centerline{\includegraphics[width=\textwidth]{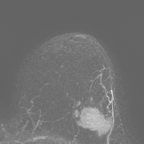}}
            \vspace{2pt}
        \end{minipage}
        \begin{minipage}{0.47\textwidth}
            \centerline{\includegraphics[width=\textwidth]{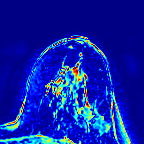}}
            \vspace{2pt}
            \centerline{\includegraphics[width=\textwidth]{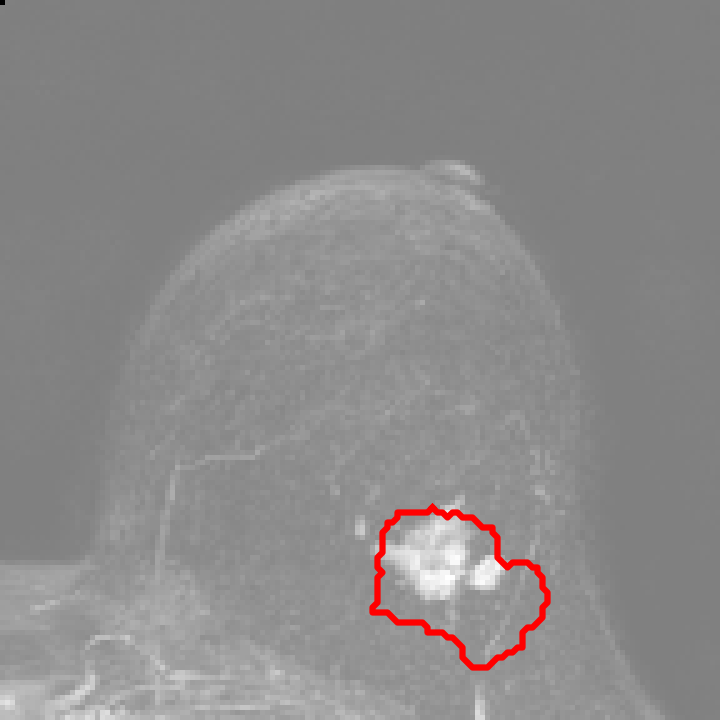}}
            \vspace{2pt}
        \end{minipage}
    \end{minipage}
    \begin{minipage}{0.158\linewidth}
        \begin{minipage}{0.47\textwidth}
            \centerline{\includegraphics[width=\textwidth]{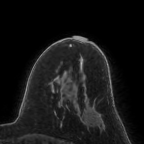}}
            \vspace{2pt}
            \centerline{\includegraphics[width=\textwidth]{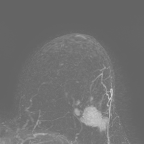}}
            \vspace{2pt}
        \end{minipage}
        \begin{minipage}{0.47\textwidth}
            \centerline{\includegraphics[width=\textwidth]{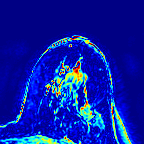}}
            \vspace{2pt}
            \centerline{\includegraphics[width=\textwidth]{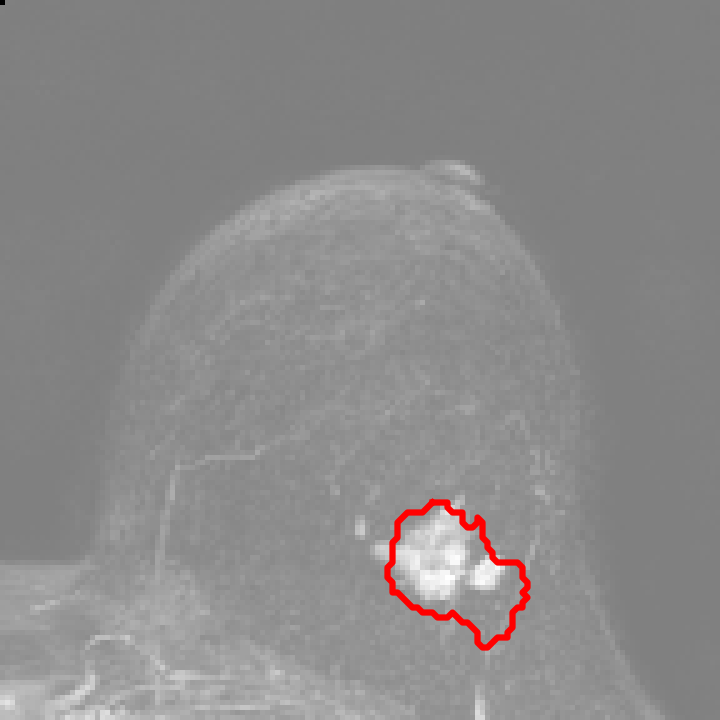}}
            \vspace{2pt}
        \end{minipage}
    \end{minipage}
    \begin{minipage}{0.158\linewidth}
        \begin{minipage}{0.47\textwidth}
            \centerline{\includegraphics[width=\textwidth]{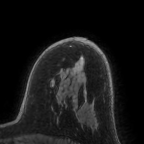}}
            \vspace{2pt}
            \centerline{\includegraphics[width=\textwidth]{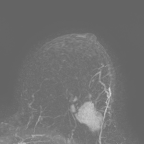}}
            \vspace{2pt}
        \end{minipage}
        \begin{minipage}{0.47\textwidth}
            \centerline{\includegraphics[width=\textwidth]{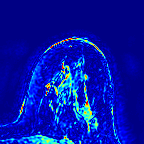}}
            \vspace{2pt}
            \centerline{\includegraphics[width=\textwidth]{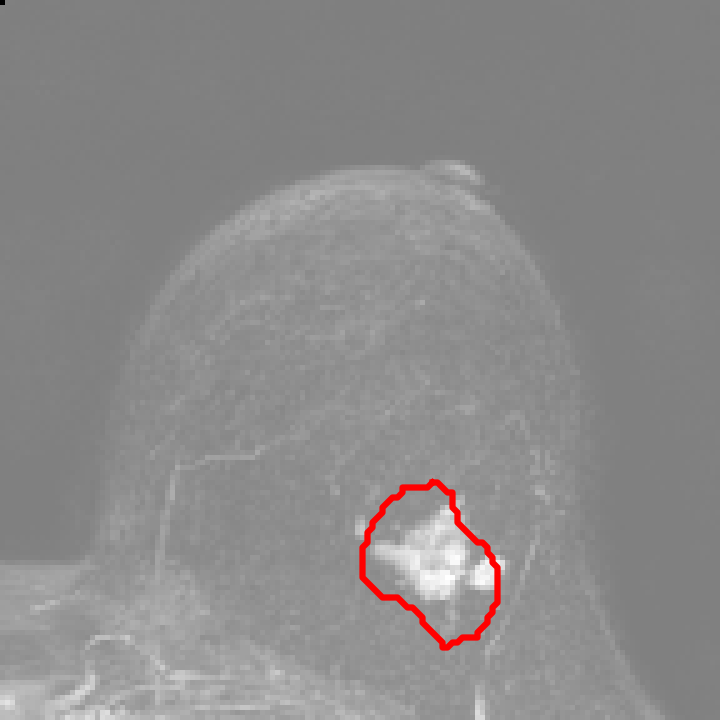}}
            \vspace{2pt}
        \end{minipage}
    \end{minipage}
    \begin{minipage}{0.158\linewidth}
        \begin{minipage}{0.47\textwidth}
            \centerline{\includegraphics[width=\textwidth]{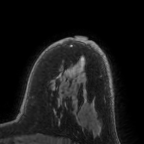}}
            \vspace{2pt}
            \centerline{\includegraphics[width=\textwidth]{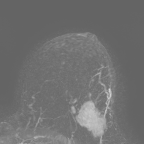}}
            \vspace{2pt}
        \end{minipage}
        \begin{minipage}{0.47\textwidth}
            \centerline{\includegraphics[width=\textwidth]{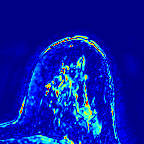}}
            \vspace{2pt}
            \centerline{\includegraphics[width=\textwidth]{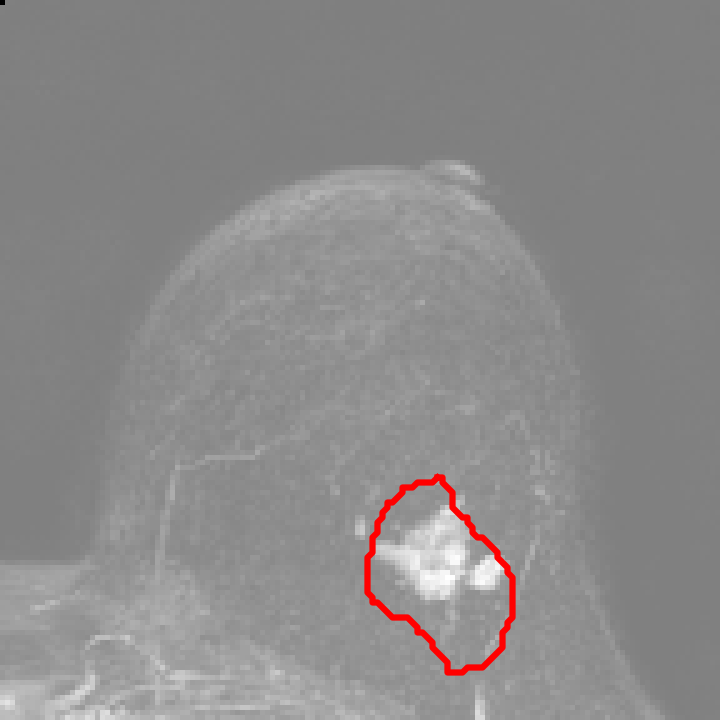}}
            \vspace{2pt}
        \end{minipage}
    \end{minipage}
    \begin{minipage}{0.013\linewidth}
        \centerline{\includegraphics[width=\textwidth]{figs/compare/colormap.png}}
        \vspace{40pt}
    \end{minipage}
    
    \begin{minipage}{0.158\linewidth}
        \centerline{Fixed}
        \vspace{2pt}
    \end{minipage}
    \begin{minipage}{0.158\linewidth}
        \centerline{VoxelMorph}
        \vspace{2pt}
    \end{minipage}
    \begin{minipage}{0.158\linewidth}
        \centerline{JointRegSeg}
        \vspace{2pt}
    \end{minipage}
    \begin{minipage}{0.158\linewidth}
        \centerline{HyperMorph}
        \vspace{2pt}
    \end{minipage}
    \begin{minipage}{0.158\linewidth}
        \centerline{cLapIRN}
        \vspace{2pt}
    \end{minipage}
    \begin{minipage}{0.158\linewidth}
        \centerline{Proposed}
        \vspace{2pt}
    \end{minipage}
    \begin{minipage}{0.013\linewidth}
        \centerline{}
    \end{minipage}
    
    \centerline{\textbf{Case 2}}
    \caption{Breast DCE-MRI image registration results for alternative methods. Case 1 is for a patient who achieves pCR, and case 2 is for non-pCR. For each method, the image in the top left, top right, bottom left, and bottom right refer to the warped T1w pre-contrast image, the error map between the warped image and fixed image, warped MIP, and fixed MIP drew with warped tumor contour (in red), respectively. Note that, the intensity errors in the tumor region are expected because the volume preservation of the tumor would lead to the difference between warped and fixed images if the tumor changes during treatment.} \label{fig:compare}
\end{figure*}
%\begin{figure*}[!htbp]
%    \centering
%    \includegraphics[width=0.9\textwidth]{figs/compare.png}
%    \caption{Breast DCE-MRI image registration results for alternative methods. Case 1 is for a patient who achieves pCR and case 2 is for non-pCR. For each method, the image in the top left, top right, bottom left, and bottom right refer to warped T1w pre-contrast image, the colormap of the residual image between warped image and fixed image, warped MIP, and fixed MIP drawn with warped tumor contour (in red), respectively.} \label{fig:compare}
%\end{figure*}

Table~\ref{tab:results} and Fig.~\ref{fig:compare} illustrate the quantitative and visual results of comparison methods. The proposed method achieves state-of-the-art performance with DSC of 0.947, $d_{avg}$ of 5.35 mm, $\Delta V$ of 11.0$\%$, and NGJD of 0.005$\%$.
Although the $d_{avg}$ of Elastix is 5.12 mm, lightly lower (not significant) than of the proposed method, the DSC, $\Delta V$, and NGJD results of the proposed method are substantially better than those of Elastix. It means that Elastix is capable of aligning breast tissue correctly but blindly makes large deformations on the tumor in the meantime.
The proposed method achieves slightly higher $\Delta V$ than NiftyReg but obtains a 0.016 increase of DSC, a 0.18 mm drop of $d_{avg}$, and a 0.011$\%$ decrease of NGJD. It is indicated that the proposed method introduces better breast shape and tissue alignment and predicts a smoother deformation field.

As shown in Fig.~\ref{fig:compare}, we randomly choose two studies from the testing dataset, including a patient who achieved pCR and another non-pCR. The registration results of the T1w pre-contrast image for the proposed method have better visualization in overview shape and local tissue than other methods and have a lower difference in the error map (top right) except for tumor region.
We display the contour of the tumor (in red) from the moving image into the fixed MIP (bottom right). By showing the pre-NAC tumor region on the post-NAC MIP, radiologists can easily observe the response area for NAC treatment.
As illustrated in case 2, apart from the proposed method, the contours of the tumors (in red) warped by other methods are forced close to the tumor in the fixed image, which can change the pre-NAC region and mislead the follow-up pCR prediction.

\begin{figure*}[!htbp]
\centering
    \begin{minipage}{0.158\linewidth}
        \begin{minipage}{0.47\textwidth}
            \centerline{\includegraphics[width=\textwidth]{figs/compare/B000020339/mov_T1.png}}
            \vspace{2pt}
            \centerline{\includegraphics[width=\textwidth]{figs/compare/B000020339/mov_mip.png}}
            \vspace{2pt}
        \end{minipage}
        \begin{minipage}{0.47\textwidth}
            \centerline{\includegraphics[width=\textwidth]{figs/compare/B000020339/mov_res.png}}
            \vspace{2pt}
            \centerline{\includegraphics[width=\textwidth]{figs/compare/B000020339/mov_tumor.png}}
            \vspace{2pt}
        \end{minipage}
    \end{minipage}
    \begin{minipage}{0.158\linewidth}
        \begin{minipage}{0.47\textwidth}
            \centerline{\includegraphics[width=\textwidth]{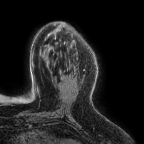}}
            \vspace{2pt}
            \centerline{\includegraphics[width=\textwidth]{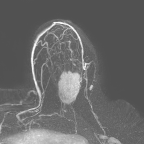}}
            \vspace{2pt}
        \end{minipage}
        \begin{minipage}{0.47\textwidth}
            \centerline{\includegraphics[width=\textwidth]{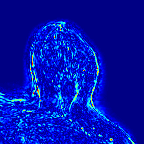}}
            \vspace{2pt}
            \centerline{\includegraphics[width=\textwidth]{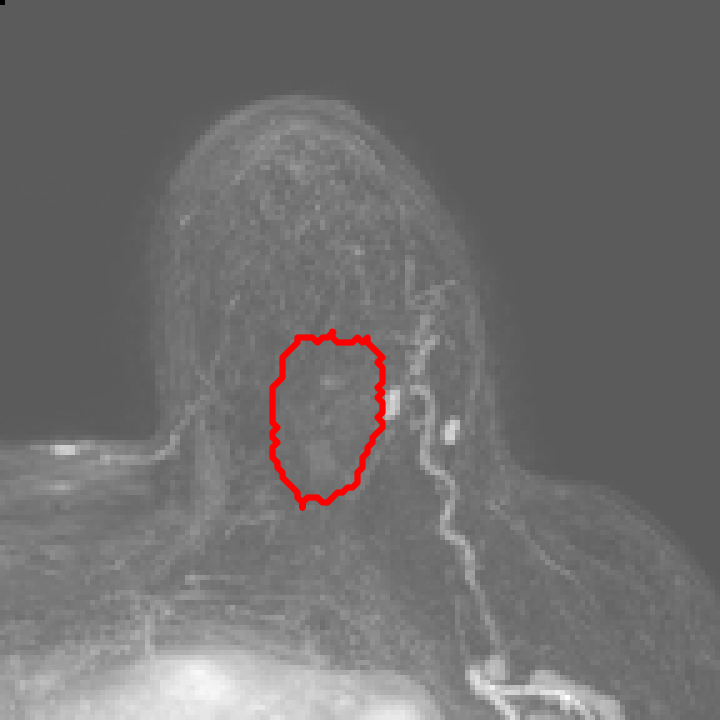}}
            \vspace{2pt}
        \end{minipage}
    \end{minipage}
    \begin{minipage}{0.158\linewidth}
        \begin{minipage}{0.47\textwidth}
            \centerline{\includegraphics[width=\textwidth]{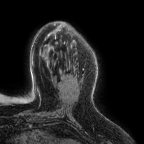}}
            \vspace{2pt}
            \centerline{\includegraphics[width=\textwidth]{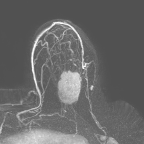}}
            \vspace{2pt}
        \end{minipage}
        \begin{minipage}{0.47\textwidth}
            \centerline{\includegraphics[width=\textwidth]{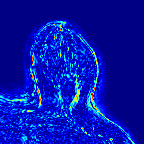}}
            \vspace{2pt}
            \centerline{\includegraphics[width=\textwidth]{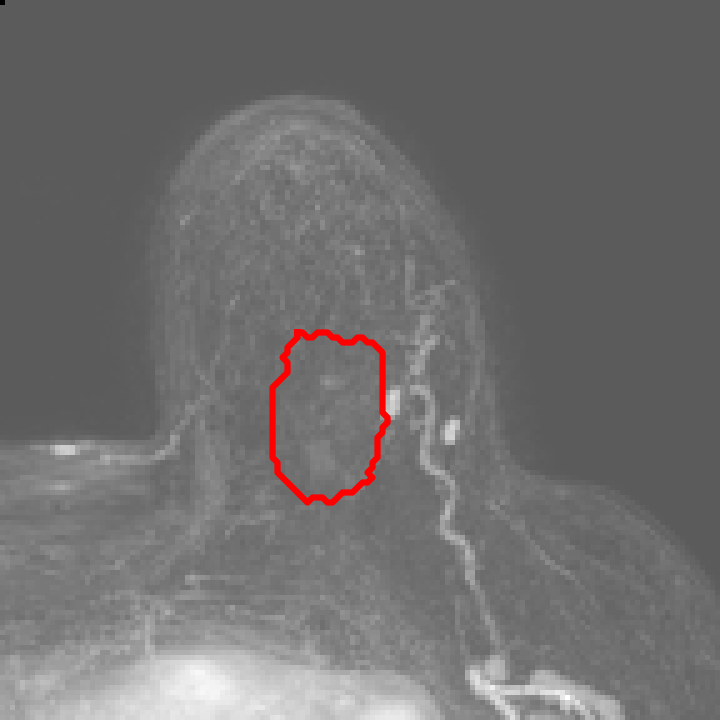}}
            \vspace{2pt}
        \end{minipage}
    \end{minipage}
    \begin{minipage}{0.158\linewidth}
        \begin{minipage}{0.47\textwidth}
            \centerline{\includegraphics[width=\textwidth]{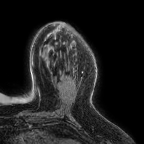}}
            \vspace{2pt}
            \centerline{\includegraphics[width=\textwidth]{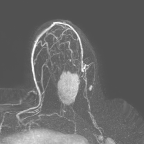}}
            \vspace{2pt}
        \end{minipage}
        \begin{minipage}{0.47\textwidth}
            \centerline{\includegraphics[width=\textwidth]{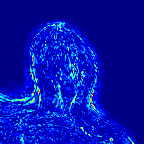}}
            \vspace{2pt}
            \centerline{\includegraphics[width=\textwidth]{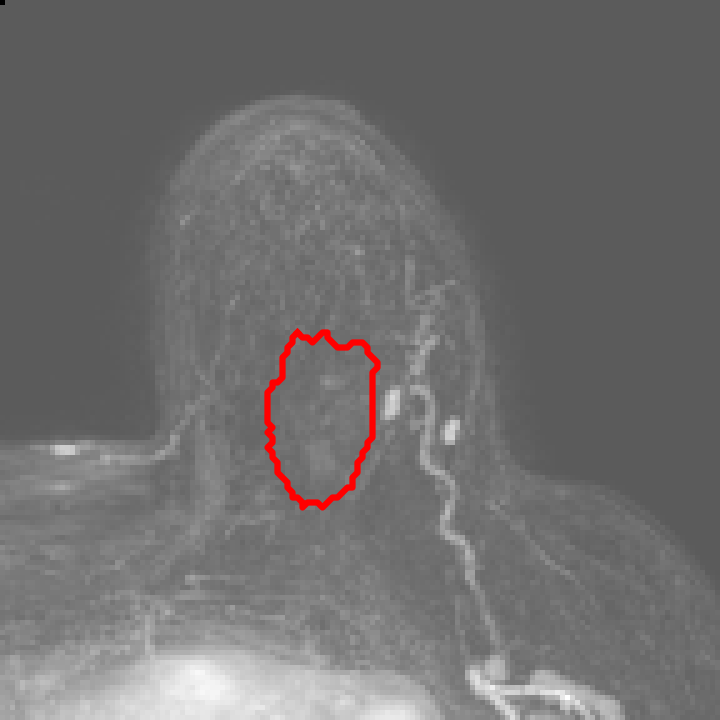}}
            \vspace{2pt}
        \end{minipage}
    \end{minipage}
    \begin{minipage}{0.013\linewidth}
        \centerline{\includegraphics[width=\textwidth]{figs/compare/colormap.png}}
        \vspace{40pt}
    \end{minipage}
    
    \begin{minipage}{0.158\linewidth}
        \centerline{Moving}
        \vspace{2pt}
    \end{minipage}
    \begin{minipage}{0.158\linewidth}
        \centerline{CRN}
        \vspace{2pt}
    \end{minipage}
    \begin{minipage}{0.158\linewidth}
        \centerline{CPRN}
        \vspace{2pt}
    \end{minipage}
    \begin{minipage}{0.158\linewidth}
        \centerline{+SL}
        \vspace{2pt}
    \end{minipage}
    \begin{minipage}{0.013\linewidth}
        \centerline{}
    \end{minipage}
    
    \begin{minipage}{0.158\linewidth}
        \begin{minipage}{0.47\textwidth}
            \centerline{\includegraphics[width=\textwidth]{figs/compare/B000020339/fix_T1.png}}
            \vspace{2pt}
            \centerline{\includegraphics[width=\textwidth]{figs/compare/B000020339/fix_mip.png}}
            \vspace{2pt}
        \end{minipage}
        \begin{minipage}{0.47\textwidth}
            \centerline{\includegraphics[width=\textwidth]{figs/compare/B000020339/fix_res.png}}
            \vspace{2pt}
            \centerline{\includegraphics[width=\textwidth]{figs/compare/B000020339/mov_tumor.png}}
            \vspace{2pt}
        \end{minipage}
    \end{minipage}
    \begin{minipage}{0.158\linewidth}
        \begin{minipage}{0.47\textwidth}
            \centerline{\includegraphics[width=\textwidth]{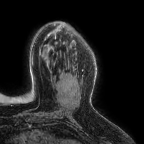}}
            \vspace{2pt}
            \centerline{\includegraphics[width=\textwidth]{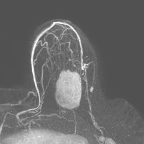}}
            \vspace{2pt}
        \end{minipage}
        \begin{minipage}{0.47\textwidth}
            \centerline{\includegraphics[width=\textwidth]{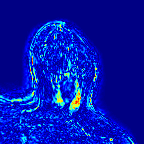}}
            \vspace{2pt}
            \centerline{\includegraphics[width=\textwidth]{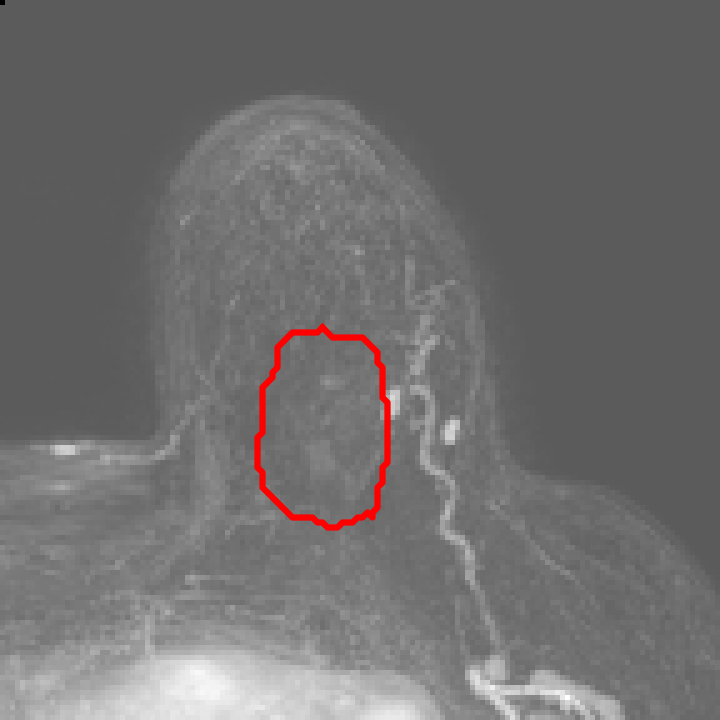}}
            \vspace{2pt}
        \end{minipage}
    \end{minipage}
    \begin{minipage}{0.158\linewidth}
        \begin{minipage}{0.47\textwidth}
            \centerline{\includegraphics[width=\textwidth]{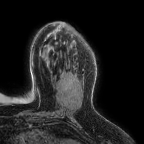}}
            \vspace{2pt}
            \centerline{\includegraphics[width=\textwidth]{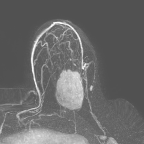}}
            \vspace{2pt}
        \end{minipage}
        \begin{minipage}{0.47\textwidth}
            \centerline{\includegraphics[width=\textwidth]{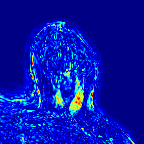}}
            \vspace{2pt}
            \centerline{\includegraphics[width=\textwidth]{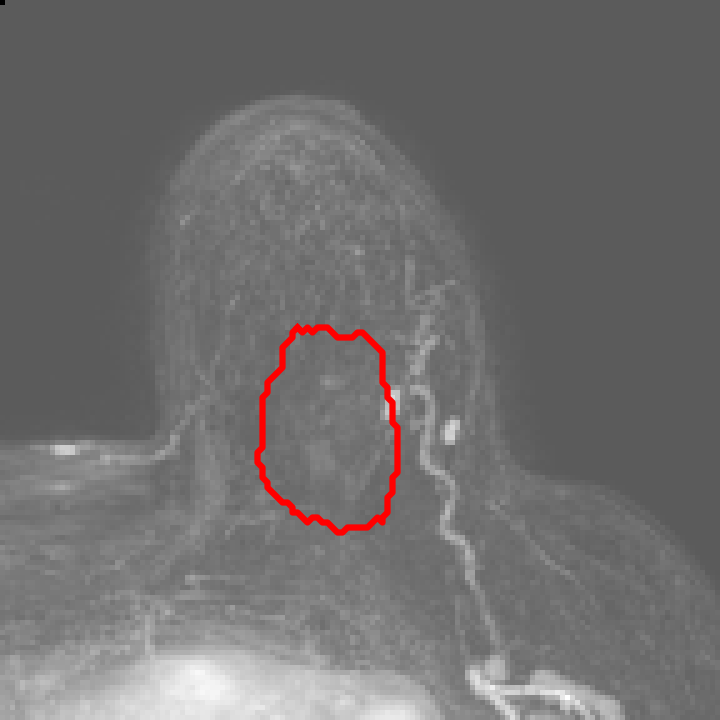}}
            \vspace{2pt}
        \end{minipage}
    \end{minipage}
    \begin{minipage}{0.158\linewidth}
        \begin{minipage}{0.47\textwidth}
            \centerline{\includegraphics[width=\textwidth]{figs/compare/B000020339/pro_T1.png}}
            \vspace{2pt}
            \centerline{\includegraphics[width=\textwidth]{figs/compare/B000020339/pro_mip.png}}
            \vspace{2pt}
        \end{minipage}
        \begin{minipage}{0.47\textwidth}
            \centerline{\includegraphics[width=\textwidth]{figs/compare/B000020339/pro_res.png}}
            \vspace{2pt}
            \centerline{\includegraphics[width=\textwidth]{figs/compare/B000020339/pro_tumor.png}}
            \vspace{2pt}
        \end{minipage}
    \end{minipage}
    \begin{minipage}{0.013\linewidth}
        \centerline{\includegraphics[width=\textwidth]{figs/compare/colormap.png}}
        \vspace{40pt}
    \end{minipage}
    
    \begin{minipage}{0.158\linewidth}
        \centerline{Fixed}
        \vspace{2pt}
    \end{minipage}
    \begin{minipage}{0.158\linewidth}
        \centerline{+VP (AL)}
        \vspace{2pt}
    \end{minipage}
    \begin{minipage}{0.158\linewidth}
        \centerline{+VP (Handcraft)}
        \vspace{2pt}
    \end{minipage}
    \begin{minipage}{0.158\linewidth}
        \centerline{+SL+VP (AL)}
        \vspace{2pt}
    \end{minipage}
    \begin{minipage}{0.013\linewidth}
        \centerline{}
    \end{minipage}
    
    \centerline{\textbf{Case 1}}
    
    % case 2
    \begin{minipage}{0.158\linewidth}
        \vspace{4pt}
        \begin{minipage}{0.47\textwidth}
            \centerline{\includegraphics[width=\textwidth]{figs/compare/B000020083/mov_T1.png}}
            \vspace{2pt}
            \centerline{\includegraphics[width=\textwidth]{figs/compare/B000020083/mov_mip.png}}
            \vspace{2pt}
        \end{minipage}
        \begin{minipage}{0.47\textwidth}
            \centerline{\includegraphics[width=\textwidth]{figs/compare/B000020083/mov_res.png}}
            \vspace{2pt}
            \centerline{\includegraphics[width=\textwidth]{figs/compare/B000020083/mov_tumor.png}}
            \vspace{2pt}
        \end{minipage}
    \end{minipage}
    \begin{minipage}{0.158\linewidth}
        \vspace{4pt}
        \begin{minipage}{0.47\textwidth}
            \centerline{\includegraphics[width=\textwidth]{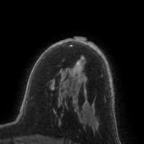}}
            \vspace{2pt}
            \centerline{\includegraphics[width=\textwidth]{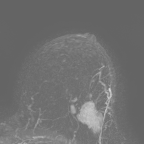}}
            \vspace{2pt}
        \end{minipage}
        \begin{minipage}{0.47\textwidth}
            \centerline{\includegraphics[width=\textwidth]{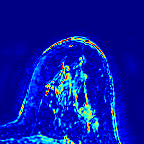}}
            \vspace{2pt}
            \centerline{\includegraphics[width=\textwidth]{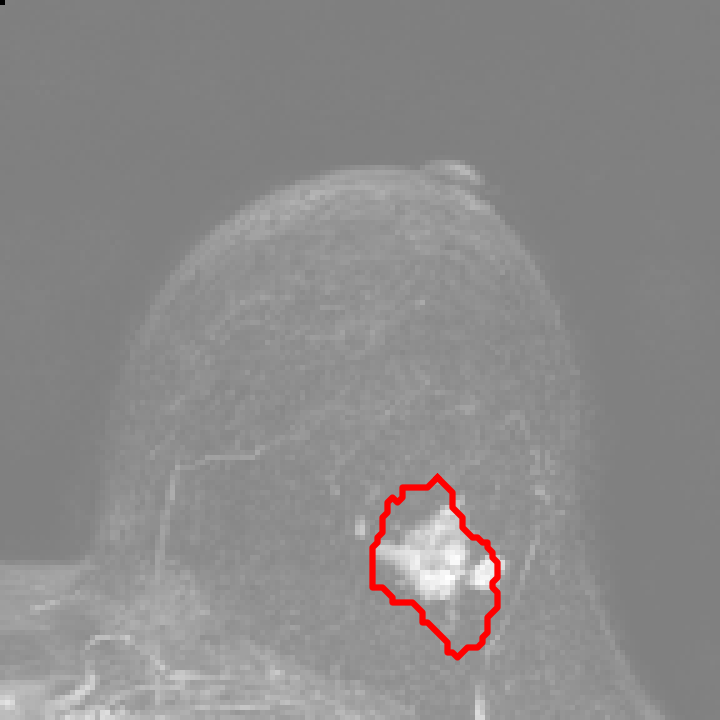}}
            \vspace{2pt}
        \end{minipage}
    \end{minipage}
    \begin{minipage}{0.158\linewidth}
        \vspace{4pt}
        \begin{minipage}{0.47\textwidth}
            \centerline{\includegraphics[width=\textwidth]{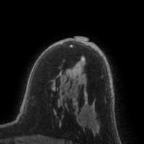}}
            \vspace{2pt}
            \centerline{\includegraphics[width=\textwidth]{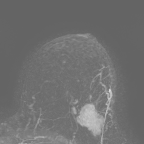}}
            \vspace{2pt}
        \end{minipage}
        \begin{minipage}{0.47\textwidth}
            \centerline{\includegraphics[width=\textwidth]{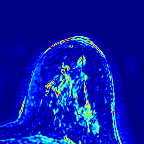}}
            \vspace{2pt}
            \centerline{\includegraphics[width=\textwidth]{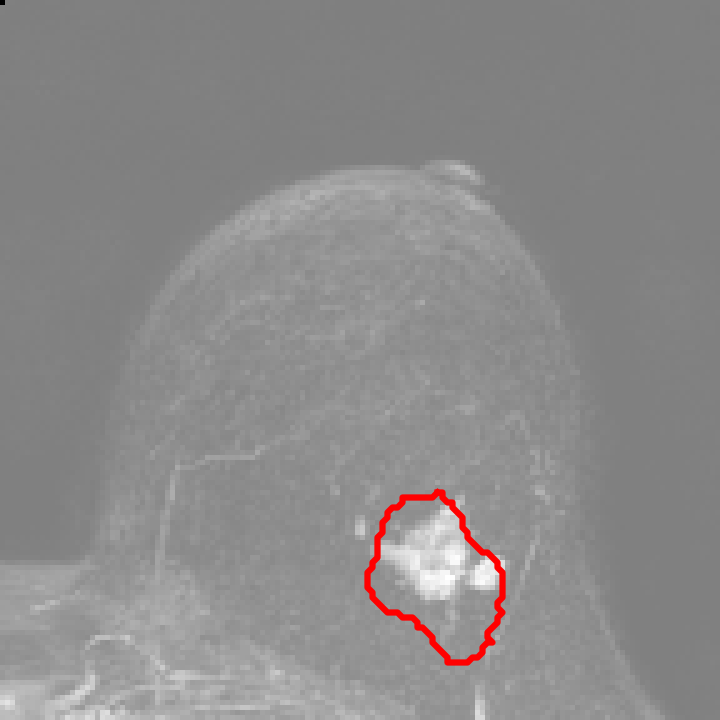}}
            \vspace{2pt}
        \end{minipage}
    \end{minipage}
    \begin{minipage}{0.158\linewidth}
        \vspace{4pt}
        \begin{minipage}{0.47\textwidth}
            \centerline{\includegraphics[width=\textwidth]{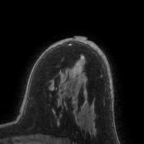}}
            \vspace{2pt}
            \centerline{\includegraphics[width=\textwidth]{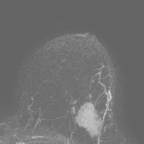}}
            \vspace{2pt}
        \end{minipage}
        \begin{minipage}{0.47\textwidth}
            \centerline{\includegraphics[width=\textwidth]{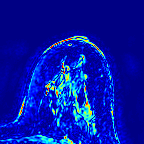}}
            \vspace{2pt}
            \centerline{\includegraphics[width=\textwidth]{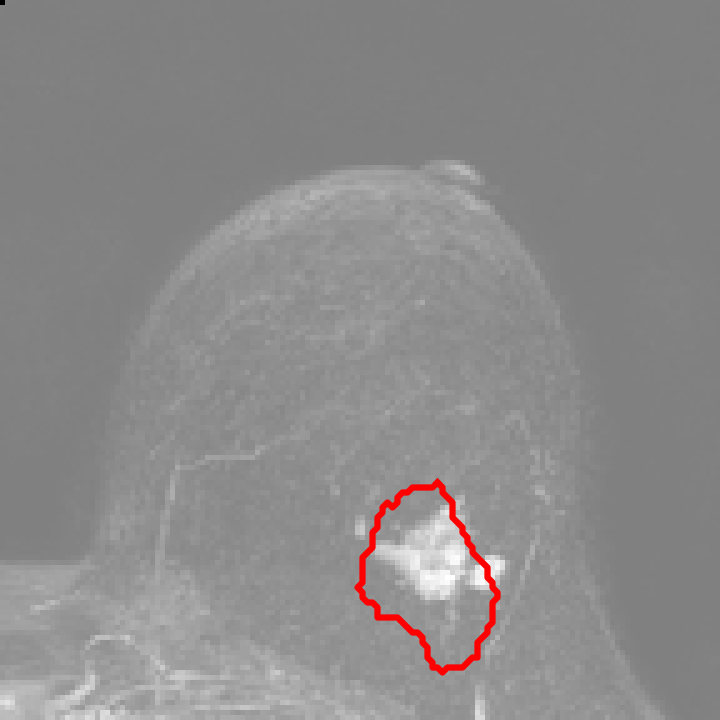}}
            \vspace{2pt}
        \end{minipage}
    \end{minipage}
    \begin{minipage}{0.013\linewidth}
        \centerline{\includegraphics[width=\textwidth]{figs/compare/colormap.png}}
        \vspace{40pt}
    \end{minipage}
    
    \begin{minipage}{0.158\linewidth}
        \centerline{Moving}
        \vspace{2pt}
    \end{minipage}
    \begin{minipage}{0.158\linewidth}
        \centerline{CRN}
        \vspace{2pt}
    \end{minipage}
    \begin{minipage}{0.158\linewidth}
        \centerline{CPRN}
        \vspace{2pt}
    \end{minipage}
    \begin{minipage}{0.158\linewidth}
        \centerline{+SL}
        \vspace{2pt}
    \end{minipage}
    \begin{minipage}{0.013\linewidth}
        \centerline{}
    \end{minipage}
    
    \begin{minipage}{0.158\linewidth}
        \begin{minipage}{0.47\textwidth}
            \centerline{\includegraphics[width=\textwidth]{figs/compare/B000020083/fix_T1.png}}
            \vspace{2pt}
            \centerline{\includegraphics[width=\textwidth]{figs/compare/B000020083/fix_mip.png}}
            \vspace{2pt}
        \end{minipage}
        \begin{minipage}{0.47\textwidth}
            \centerline{\includegraphics[width=\textwidth]{figs/compare/B000020083/fix_res.png}}
            \vspace{2pt}
            \centerline{\includegraphics[width=\textwidth]{figs/compare/B000020083/mov_tumor.png}}
            \vspace{2pt}
        \end{minipage}
    \end{minipage}
    \begin{minipage}{0.158\linewidth}
        \begin{minipage}{0.47\textwidth}
            \centerline{\includegraphics[width=\textwidth]{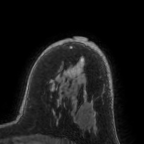}}
            \vspace{2pt}
            \centerline{\includegraphics[width=\textwidth]{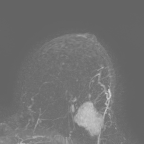}}
            \vspace{2pt}
        \end{minipage}
        \begin{minipage}{0.47\textwidth}
            \centerline{\includegraphics[width=\textwidth]{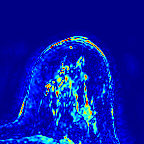}}
            \vspace{2pt}
            \centerline{\includegraphics[width=\textwidth]{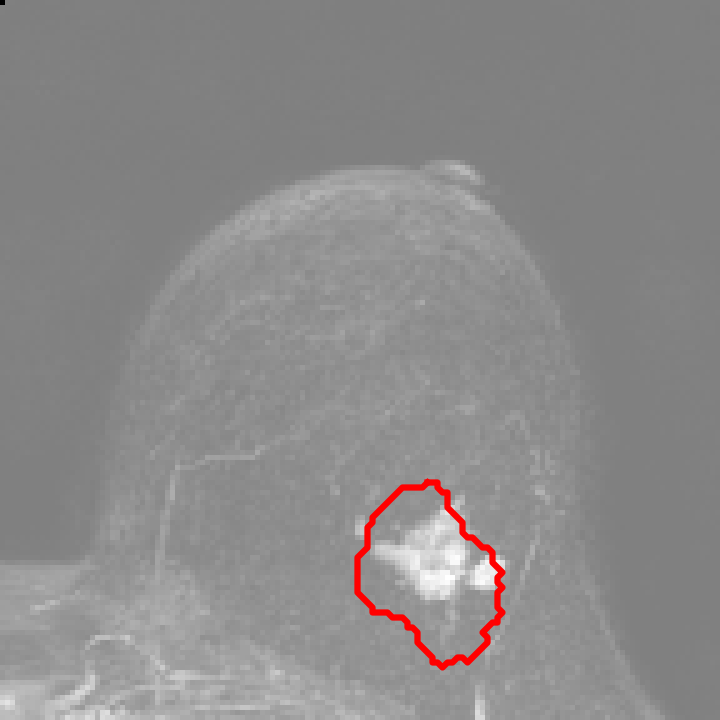}}
            \vspace{2pt}
        \end{minipage}
    \end{minipage}
    \begin{minipage}{0.158\linewidth}
        \begin{minipage}{0.47\textwidth}
            \centerline{\includegraphics[width=\textwidth]{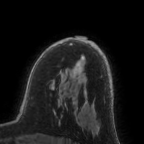}}
            \vspace{2pt}
            \centerline{\includegraphics[width=\textwidth]{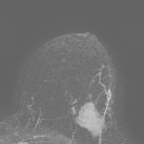}}
            \vspace{2pt}
        \end{minipage}
        \begin{minipage}{0.47\textwidth}
            \centerline{\includegraphics[width=\textwidth]{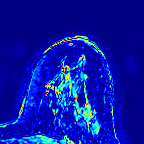}}
            \vspace{2pt}
            \centerline{\includegraphics[width=\textwidth]{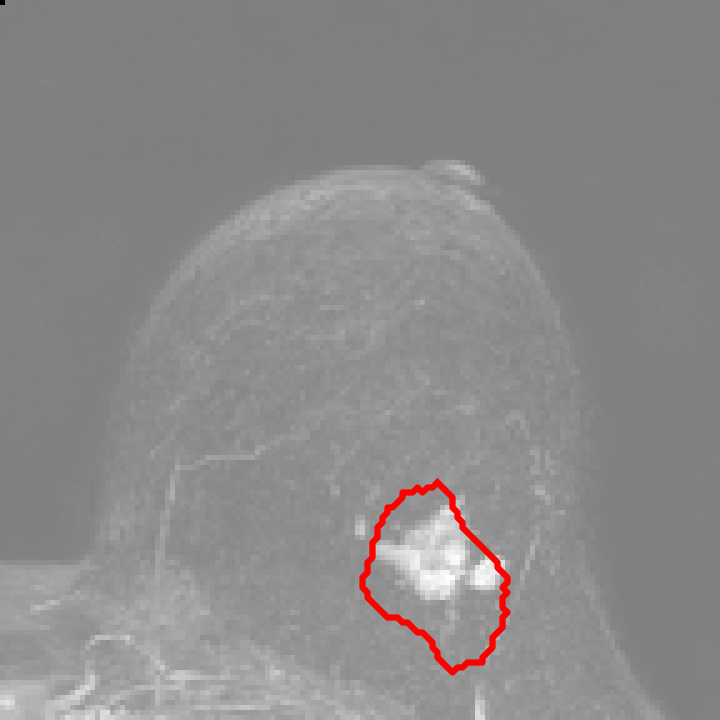}}
            \vspace{2pt}
        \end{minipage}
    \end{minipage}
    \begin{minipage}{0.158\linewidth}
        \begin{minipage}{0.47\textwidth}
            \centerline{\includegraphics[width=\textwidth]{figs/compare/B000020083/pro_T1.png}}
            \vspace{2pt}
            \centerline{\includegraphics[width=\textwidth]{figs/compare/B000020083/pro_mip.png}}
            \vspace{2pt}
        \end{minipage}
        \begin{minipage}{0.47\textwidth}
            \centerline{\includegraphics[width=\textwidth]{figs/compare/B000020083/pro_res.png}}
            \vspace{2pt}
            \centerline{\includegraphics[width=\textwidth]{figs/compare/B000020083/pro_tumor.png}}
            \vspace{2pt}
        \end{minipage}
    \end{minipage}
    \begin{minipage}{0.013\linewidth}
        \centerline{\includegraphics[width=\textwidth]{figs/compare/colormap.png}}
        \vspace{40pt}
    \end{minipage}

    \begin{minipage}{0.158\linewidth}
        \centerline{Fixed}
        \vspace{2pt}
    \end{minipage}
    \begin{minipage}{0.158\linewidth}
        \centerline{+VP (AL)}
        \vspace{2pt}
    \end{minipage}
    \begin{minipage}{0.158\linewidth}
        \centerline{+VP (Handcraft)}
        \vspace{2pt}
    \end{minipage}
    \begin{minipage}{0.158\linewidth}
        \centerline{+SL+VP (AL)}
        \vspace{2pt}
    \end{minipage}
    \begin{minipage}{0.013\linewidth}
        \centerline{}
    \end{minipage}
    
    \centerline{\textbf{Case 2}}
	\caption{Breast DCE-MRI image registration results of the ablation study. Case 1 is for a patient who achieves pCR, and case 2 is for non-pCR. For each method, the image in the top left, top right, bottom left, and bottom right refer to the warped T1w pre-contrast image, the error map between the warped image and fixed image, warped MIP, and fixed MIP drew with warped tumor contour (in red), respectively. Note that, the intensity errors in the tumor region are expected because the volume preservation of the tumor would lead to the difference between warped and fixed images if the tumor changes during treatment.} \label{fig:ablation}
\end{figure*}

\subsection{Ablation study}
We investigate the effectiveness of three key components in the proposed method: (1) conditional pyramid architecture; (2) structural keypoint loss; and (3) volume-preserving loss based on abnormal keypoints.
The experiments with and without pyramid architecture are performed to demonstrate the effectiveness of conditional pyramid architecture. And the effect of structural keypoint loss and volume-preserving loss is illustrated by utilizing or disabling these losses.
The detailed experimental settings are CRN, CPRN, CPRN+SK, CPRN+VP (Handcraft), CPRN+VP (AK), and the proposed method CPRN+SK+VP (AK).

In the experiment of CRN, we select the subnetwork with the highest pyramid level from CPRN as an independent registration model and cut off any skip connection from other pyramid levels. And the model is trained with the following loss function,
\begin{equation}
    \begin{aligned}
        \mathcal{L}^R_{CRN}(\phi_{hr}) &= \mathcal{L}^R_i(\phi_{hr}) + \mathcal{L}^R_s(\phi_{hr}) + \mathcal{L}^R_b(\phi_{hr}) \\
        &+ \lambda_g\cdot\mathbf{Grad}(\phi_{hr}) + \lambda_g\cdot\mathbf{Grad}(\phi'_{hr})
    \end{aligned}
    \label{eq:single}
\end{equation}
where $\phi'_{hr}$ refers to the upsampled deformation field of $\phi_{hr}$.
In the experiment of CPRN, we take the CPRN mentioned in Section~\ref{sec:regnet} as the baseline model and apply it for the follow-up experiments. Based on Eq.~\ref{eq:single}, the pyramid baseline model can be optimized by loss of $\sum_{\phi_i}\mathcal{L}^R_{CRN}$, where $\phi_i\in\{\phi_{lr}, \phi_{mr}, \phi_{hr}\}$ refers to the predicted deformation field for different pyramid levels.
In the experiment of CPRN+SK, we train the baseline model with additional structural keypoint loss, which can be formulated as $\sum_{\phi_i}(\mathcal{L}^R_{CRN}+\mathcal{L}^R_{sl})$.
In the experiment of CPRN+VP (AK), volume-preserving loss based on the abnormal keypoints is employed to train the baseline model, whose loss function is defined as $\sum_{\phi_i}(\mathcal{L}^R_{CRN}+\lambda_v\cdot\mathcal{L}^R_{vp})$.
In the experiment of CPRN+VP (Handcraft), we replace the abnormal keypoints in the volume-preserving loss with handcraft breast tumor annotation to train the baseline model. An updated loss function can be written as $\sum_{\phi_i}(\mathcal{L}^R_{CRN}+\lambda_v\cdot\left\|\sum M_m-\sum(M_m\circ\phi_i)\right\|_1)$, where $M_m$ refers to the annotated tumor masks for moving images.
We compare the quantitative results by adding the above components in turn and evaluate with the metrics of DSC, $d_{avg}$, $\Delta V$, and NGJD.

Table~\ref{tab:results} and Fig.~\ref{fig:ablation} show the quantitative and visual results of the ablation study of three key components. All the methods in the ablation study have similar DSC and NGJD due to boundary-level similarity loss $\mathcal{L}^R_b$ and hyper-parameter-based CRM.
Compared with CRN and CPRN, the $d_{avg}$ is reduced by nearly 0.18 mm, the $\Delta V$ decrease of 3.4\%, which shows that introducing pyramid architecture can improve registration performance.
Adding $\mathcal{L}^R_{sl}$ can reduce the $d_{avg}$ by 0.39 mm compared to baseline CPRN, illustrating that the tissue difference decreases by restricting the structural keypoint loss.
Volume-preserving loss based on the abnormal keypoint and handcraft tumor annotation can obviously decrease the $\Delta V$ to 10.5\% and 9.7\%, respectively. But +VP (AK) results in the increase of $d_{avg}$, indicating the loss in the similarity of breast tissue.
Combining with the structural keypoint loss and the volume-preserving loss, our proposed method can balance $d_{avg}$ and $\Delta V$, and achieve state-of-the-art results on both metrics.

As shown in Fig.~\ref{fig:ablation}, two studies are chosen from the testing dataset to compare the registration performance for the ablation study visually.
When adding with $\mathcal{L}^R_{sl}$, we obtain registration results with less residue in the colormap but shrinkage for the tumor.
Adding with the volume-preserving loss alone, the tumor shrinks less, but the position of the tumor may be offset due to the large deformation of breast tissue. As shown in case 1, the tumor in the moving image is attached to the vessel, but the tumor after registration (excluding the proposed method) is detached from the vessel.
Note that, for +SK, it is because the tumor shrinks, and for +VP, it is due to the large deformation.
The proposed method can effectively avoid both shortages and achieve better registration performance.

\subsection{Statistical tests}
%Fig.~\ref{fig:ttest} illustrates the box plots for DSC, $d_{avg}$, $\Delta V$, and NGJD of the proposed method and comparisons.
Paired samples T-test shows that $d_{avg}$ and $\Delta V$ improvement given by the proposed method are statistically significant ($p<0.05$) against SyN, DRAMMS, JointRegSeg, cLapIRN, and HyperMorph.
Elastix achieves the best $d_{avg}$ but has no statistically significant difference against the proposed method ($p=0.55$), and also statistically significantly ($p<0.05$) worse DSC and $\Delta V$ than our method.
Although the proposed method does not give statistically significant improvements on $d_{avg}$ and $\Delta V$ against NiftyReg ($p=0.25$ and $p=0.33$) and deedsBCV ($p=0.39$ and $p=0.05$), the proposed method achieves statistically significant improvements on DSC ($p<0.05$).

%\begin{figure*}[!htbp]
%\centering
%\begin{minipage}{0.03\linewidth}
%        \centerline{}
%    \end{minipage}
%    \begin{minipage}{0.45\linewidth}
%        \centerline{\includegraphics[width=\textwidth]{figs/static/DSC.eps}}
%        \vspace{-10pt}
%        \centerline{(a)}
%    \end{minipage}
%    \begin{minipage}{0.45\linewidth}
%        \centerline{\includegraphics[width=\textwidth]{figs/static/davg.eps}}
%        \vspace{-10pt}
%        \centerline{(b)}
%    \end{minipage}
%    
%    \begin{minipage}{0.03\linewidth}
%        \centerline{}
%    \end{minipage}
%    \begin{minipage}{0.45\linewidth}
%        \centerline{\includegraphics[width=\textwidth]{figs/static/dV.eps}}
%        \vspace{-10pt}
%        \centerline{(c)}
%    \end{minipage}
%    \begin{minipage}{0.45\linewidth}
%        \centerline{\includegraphics[width=\textwidth]{figs/static/NGJD.eps}}
%        \vspace{-10pt}
%        \centerline{(d)}
%    \end{minipage}
%\caption{Boxplots of the metrics (a) DSC, (b) $d_{avg}$, (c) $\Delta V$, and (d) NGJD achieved by the comparison methods over the testing dataset. The boxes with backgrounds in orange colors from light to dark refer to conventional methods, learning-based methods, and ablation studies of our proposed method, respectively. The two‐sided paired samples T-test compares the results between the proposed method and alternative methods; *, **, and *** marked on the top of the boxes indicate statistically significant differences of $0.01<p<0.05$, $0.001<p<0.01$, and $p<0.001$, respectively.} \label{fig:ttest}
%\end{figure*}

\subsection{Unsupervised keypoint detection}
\subsubsection{Structural keypoints}
To quantitatively evaluate the structural keypoints extracted by KN-S, we perform two experiments: (1) only use the structural keypoints for registration; (2) use the coordinates of the structural keypoints to predict the coordinate of the nipple in the corresponding image.
To realize structural keypoints registration, we calculate the displacement between structural keypoints in both the moving and fixed images. Then we interpolate the coordinates displacement into a 3D mesh grid to create the deformation field. SK (only) in Table~\ref{tab:results} illustrates the results of generating the deformation field by interpolating on structural keypoints, which improves DSC and $d_{avg}$ compared to rigid registration.
On the other hand, we propose a landmark distance metric $d_{nipple}$ presents the distance between the handcraft annotated nipple and the predicted coordinate. We apply a fully connected layer to predict the nipple coordinate with the shape of $[1,3]$ by inputting structural keypoints with the shape of $[1,K_S\times 3]$. We randomly choose 25 patients from the testing set to train the fully connected layer and another 25 patients for testing. Table~\ref{tab:sk-pred} illustrates the results on the different numbers of structural keypoints $K_S$.
With the increase of the number of structural keypoints, it becomes more accurate to predict the coordinate of nipples, which means these keypoints can present the breast structure better.
%Predicted structural keypoints are difficult to evaluate due to a lack of ground truth. We assume that the structural keypoints perform better when they have more similar distributions with the handcraft keypoints. Hence, we employ a statistical metric NSD--the percentage of studies with no significant difference for inter-image keypoint distance between structural and handcraft keypoints.
%For each moving and fixed image pair, we first calculate inter-image keypoint distance for structural keypoints and handcraft keypoints, respectively. Then independent samples t-test is applied for predicted keypoint-based distance and handcraft keypoint-based distance and output the percentage of studies with no significant difference.

\begin{table}
    \centering
    \caption{The results of nipple coordinate prediction based on the structural keypoints.}
    \begin{tabular}{ccc}
        \hline\hline
        Method & $K_S$ & $d_{nipple}$(mm)$\downarrow$ \\
        \hline
        %Inter-expert & - & 2.61$\pm$2.32 \\
        KN-S & 8 & 13.35$\pm$6.53 \\
        KN-S & 16 & 10.08$\pm$6.94 \\
        KN-S & 32 & 7.93$\pm$5.14 \\
        KN-S & 64 & 7.19$\pm$5.99 \\
        KN-S & 128 & 7.17$\pm$6.91 \\
        \hline\hline
    \end{tabular}
    \label{tab:sk-pred}
\end{table}

\begin{figure*}[!htbp]
\centering
    \begin{minipage}{0.49\linewidth}
        \centerline{\includegraphics[width=\textwidth]{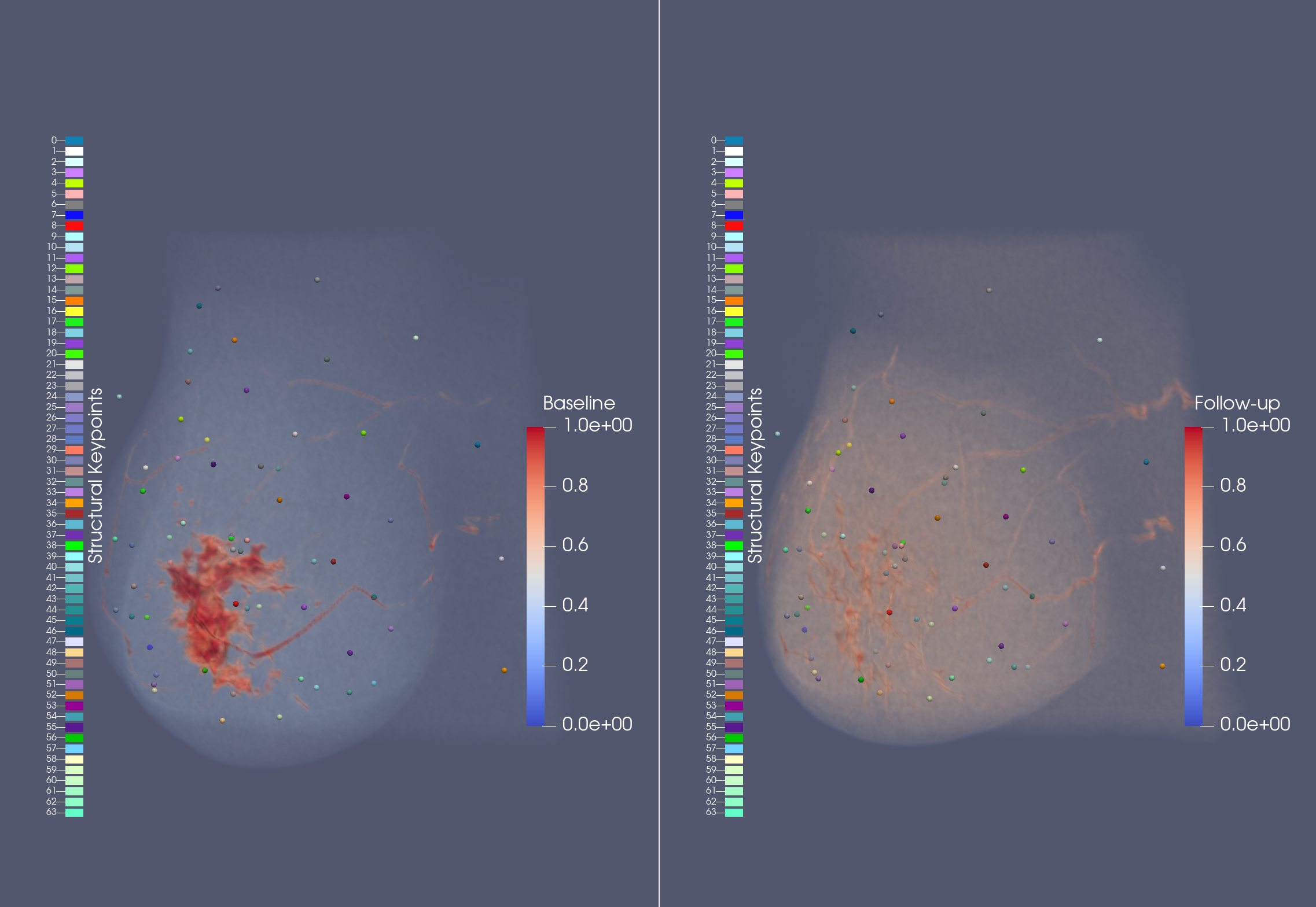}}
        \centerline{Case 1}
    \end{minipage}
    \begin{minipage}{0.49\linewidth}
        \centerline{\includegraphics[width=\textwidth]{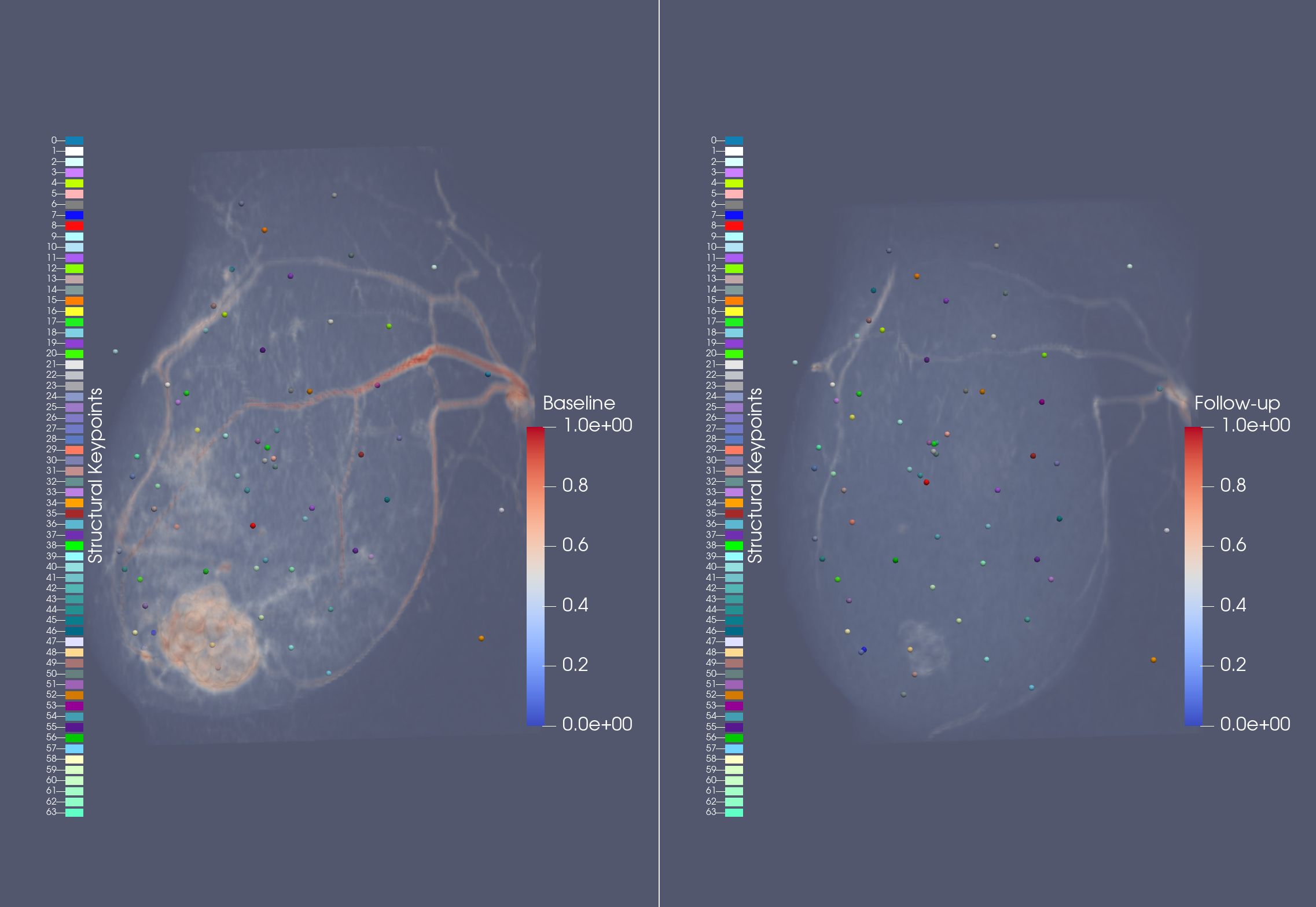}}
        \centerline{Case 2}
    \end{minipage}

    \vspace{5pt}
    
    \begin{minipage}{0.49\linewidth}
        \centerline{\includegraphics[width=\textwidth]{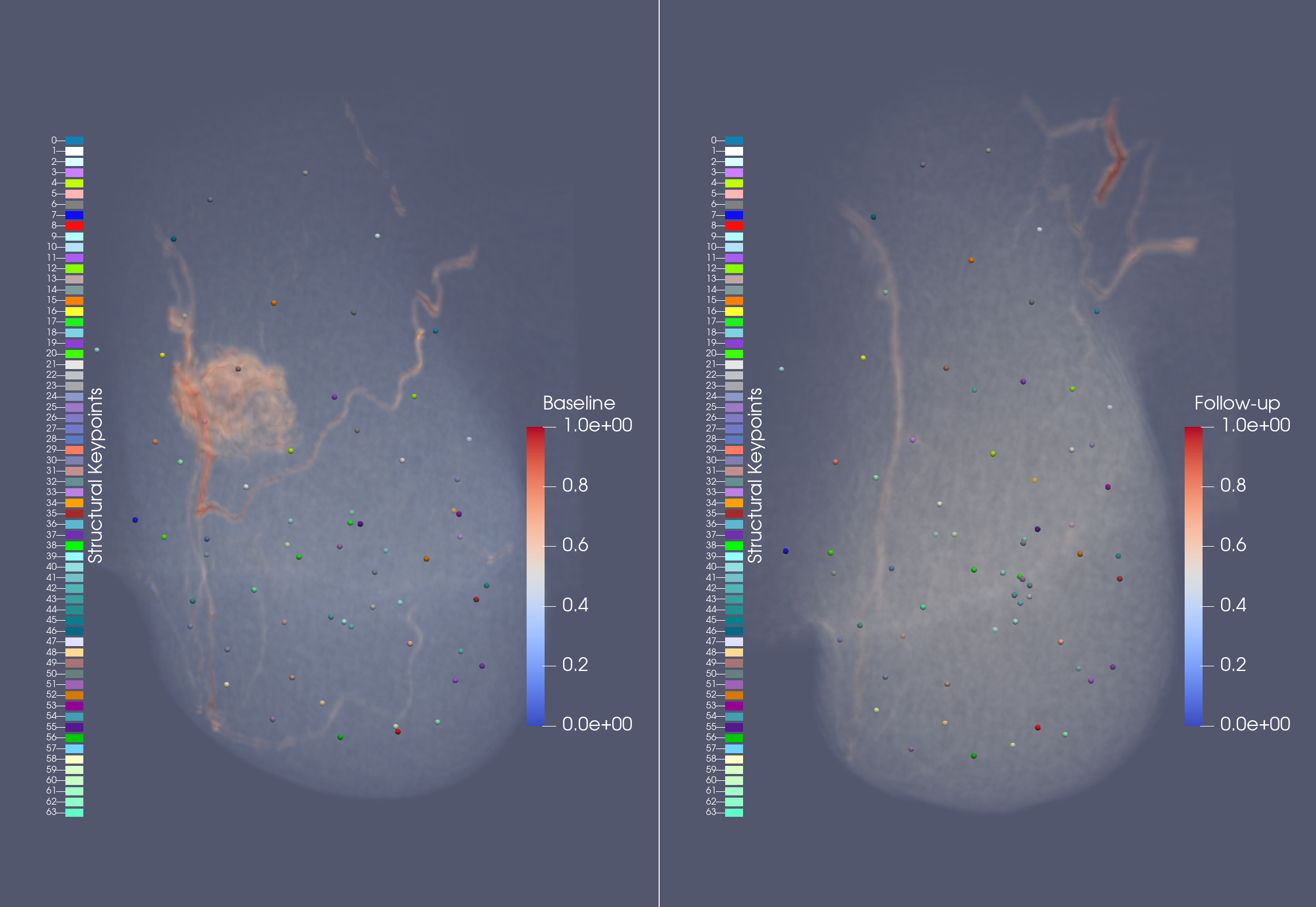}}
        \centerline{Case 3}
    \end{minipage}
    \begin{minipage}{0.49\linewidth}
        \centerline{\includegraphics[width=\textwidth]{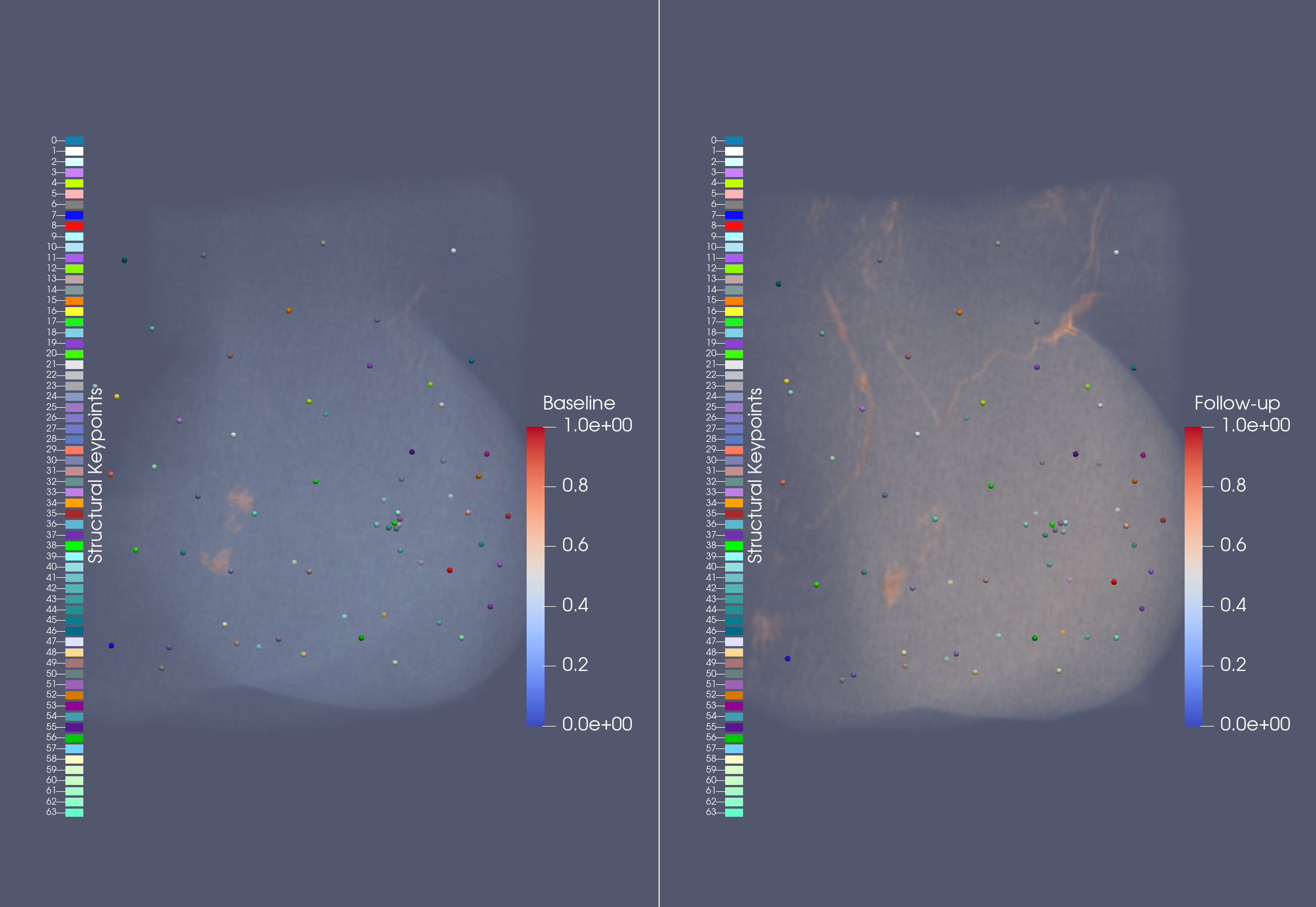}}
        \centerline{Case 4}
    \end{minipage}
    
	\caption{Visualization of structural keypoints on the volume rendering for breast wash-in images. For each case, the baseline image is on the left, and the follow-up image is on the right. Colorful points indicate structural keypoints detected by KN-S. Points with the same color refer to the corresponding points between baseline and follow-up images.} \label{fig:landmarks-S}
\end{figure*}

%The comparison results between Jakab \emph{et al.}~\citep{jakab2018unsupervised} and KN-S are illustrated in Table~\ref{tab:results-sl}. KN-S achieve the best NSD of 82.3$\%$ in the testing dataset.
Fig.~\ref{fig:landmarks-S} shows the visual results of structural keypoints detected by KN-S on volume rendering for breast wash-in images. Colorful points indicate structural keypoints, and points with the same color refer to the corresponding points between moving and fixed images.
And corresponding structural keypoints are pointed at similar positions for the cases shown.

%\begin{table}
%	\centering
%	\caption{The percentage of no significant difference for keypoint distances between handcraft keypoints and learning-based keypoints.}
%	\label{tab:results-sl}
%	\setlength{\tabcolsep}{3pt}
%	\begin{tabular}{lcc}
%		\hline\hline
%		\multirow{2}*{Method}
%		    & \multicolumn{2}{c}{NSD(\%)$\uparrow$} \\
%		    \cline{2-3}
%		    & Internal dataset & External dataset \\
%		\hline
%		Jakab \emph{et al.}~\citep{jakab2018unsupervised} \\
%		KN-S & 82.3 &  \\
%		\hline\hline
%	\end{tabular}
%\end{table}

%\begin{figure*}[!htbp]
%\centering
%\includegraphics[width=0.9\textwidth]{figs/structural.png}
%\caption{Visualization of structural keypoints on the volume rendering for breast T1-weighted images. Colorful points indicate structural keypoints detected by KN-S. Points with the same color refer to the corresponding points between baseline and follow-up images.} \label{fig:landmarks-S}
%\end{figure*}

\subsubsection{Abnormal keypoints}
The abnormal keypoints are used to locate the tumor regions and preserve their volume. Therefore, it is essential to evaluate the performance of KN-A in finding the tumor. We employ the abnormal keypoints in tumor segmentation task comparing with supervised learning-based method JointRegSeg~\citep{estienne2020deep}, and conventional thresholding segmentation methods Otsu~\citep{otsu1979threshold} and Yen~\citep{yen1995new}.
For comparison, we utilize abnormal keypoints to guide the thresholding segmentation methods to refine the tumor area further. Specifically, we extract connected components from thresholding segmentation results and preserve some of which overlap with abnormal keypoints. Two metrics are utilized to evaluate the segmentation results, including DSC of tumor segmentation and corresponding hitting percentage--the percentage of samples with DSC higher than 0.4.

\begin{table}
	\centering
	\caption{The quantitative results of tumor segmentation comparing with the supervised learning-based methods, conventional thresholding-based methods, and composition of KN-A and thresholding-based methods. The best result is in bold and the second best one is underlined.}
	\label{tab:results-al}
	\setlength{\tabcolsep}{3pt}
	\begin{tabular}{lcc}
		\hline\hline
		%\multirow{2}*{Method}
		Method
		    %& \multicolumn{2}{c}{Internal dataset} && \multicolumn{2}{c}{External dataset} \\
		    %\cline{2-3}\cline{5-6}
		    & DSC$\uparrow$ & Hit(\%)$\uparrow$ \\%&
		    %& DSC$\uparrow$ & Hit(\%)$\uparrow$ \\
		\hline
		JointRegSeg~\citep{estienne2020deep} & \textbf{0.624$\pm$0.284} & \textbf{82.0} \\
		thresholding-Otsu~\citep{otsu1979threshold} & 0.436$\pm$0.230 & 56.0 \\
		thresholding-Yen~\citep{yen1995new} & 0.550$\pm$0.227 & 74.0 \\
		KN-A + Otsu & 0.510$\pm$0.286 & 68.0 \\
		KN-A + Yen & \underline{0.589$\pm$0.267} & \underline{80.0} \\
		\hline\hline
	\end{tabular}
\end{table}

\begin{figure*}[!htbp]
\centering
    \begin{minipage}{0.49\linewidth}
        \centerline{\includegraphics[width=\textwidth]{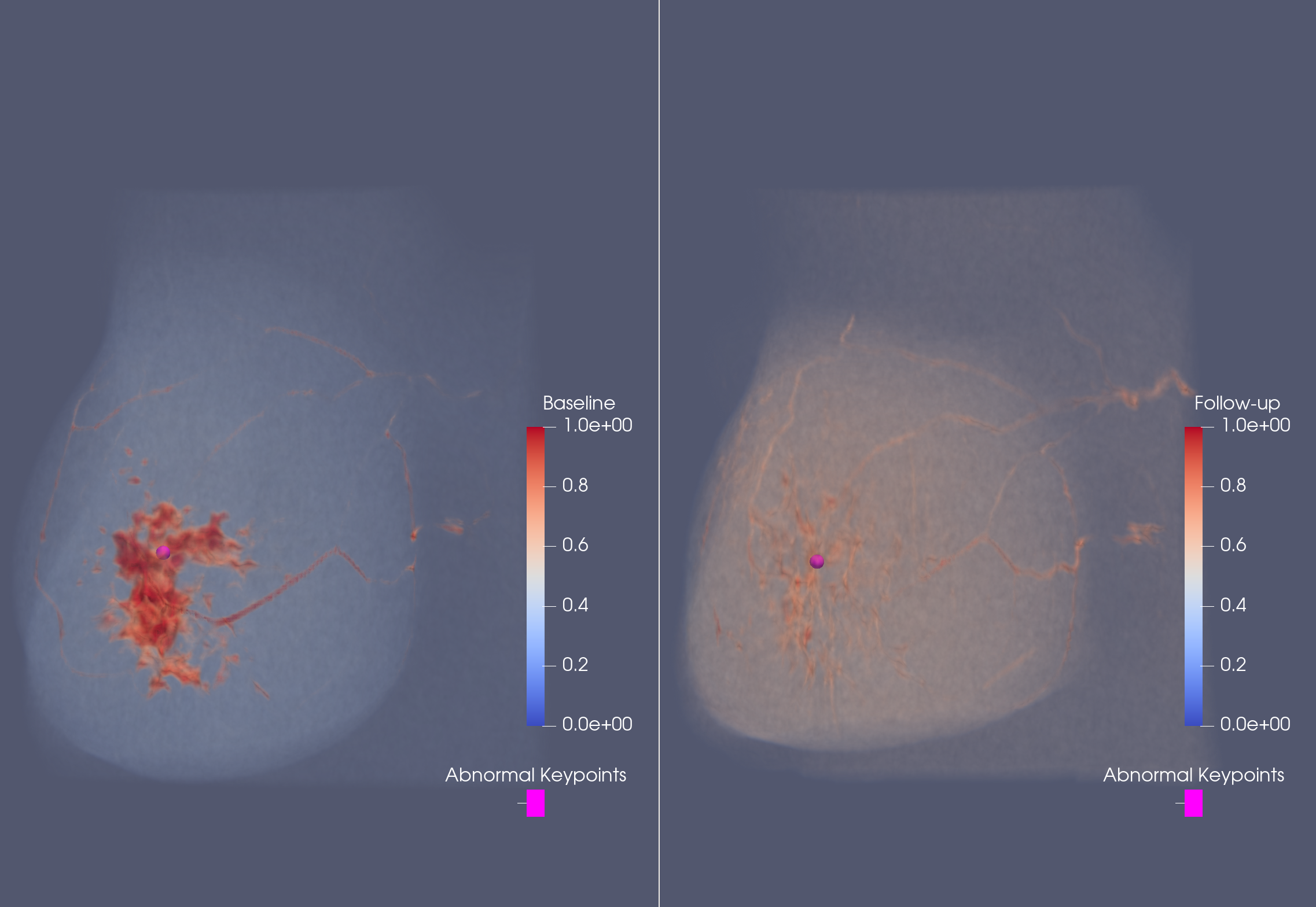}}
        \centerline{Case 1}
    \end{minipage}
    \begin{minipage}{0.49\linewidth}
        \centerline{\includegraphics[width=\textwidth]{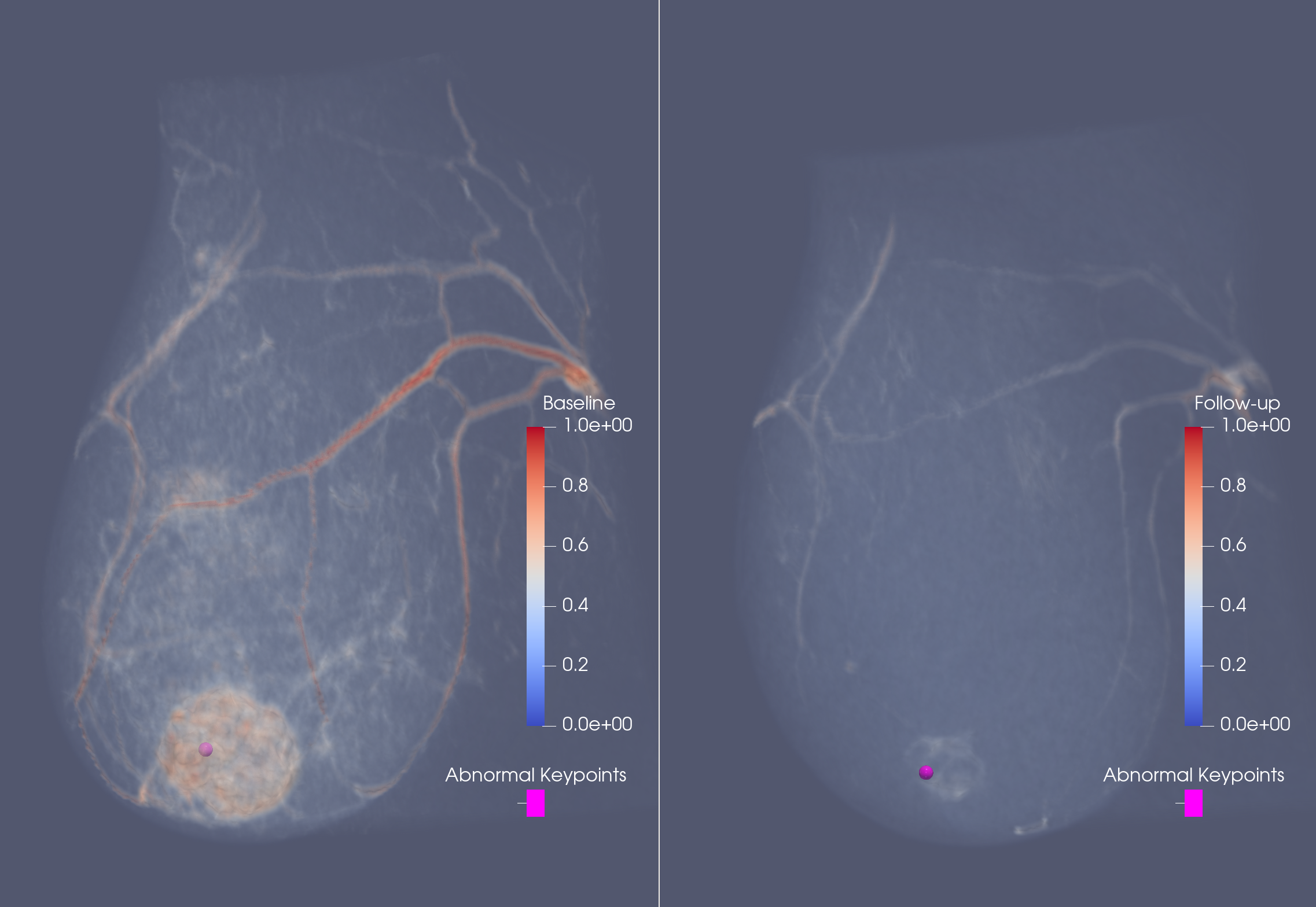}}
        \centerline{Case 2}
    \end{minipage}

    \vspace{5pt}
    
    \begin{minipage}{0.49\linewidth}
        \centerline{\includegraphics[width=\textwidth]{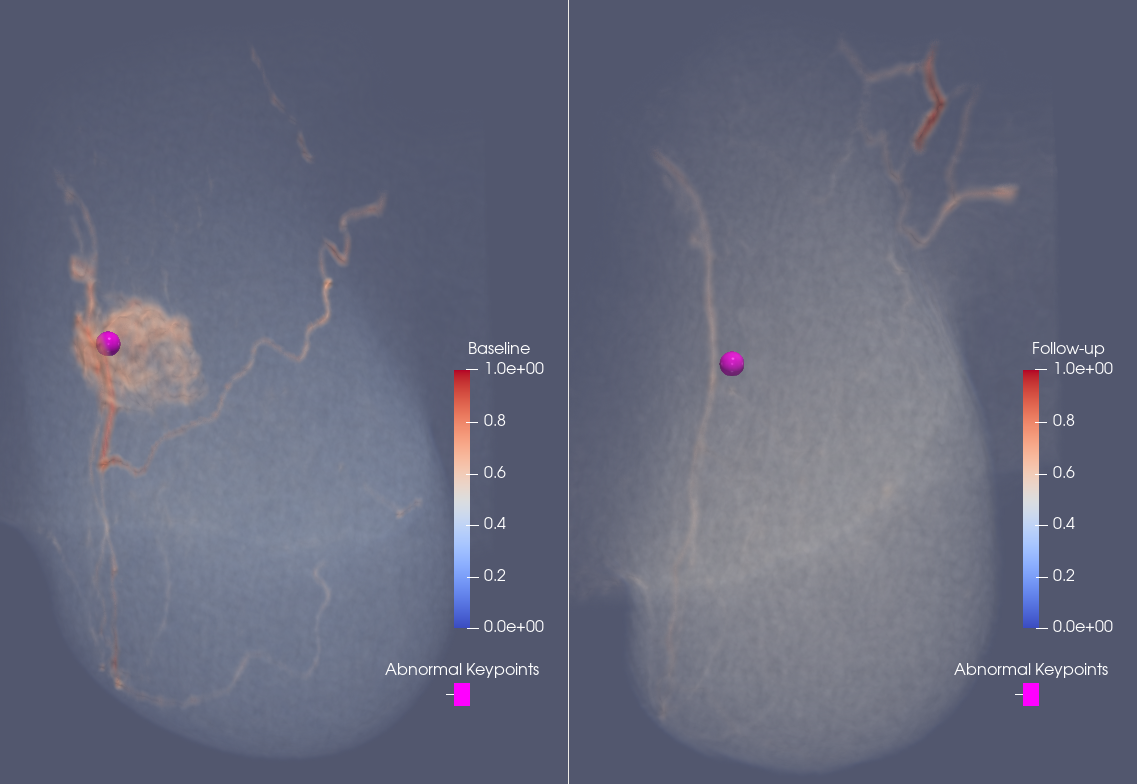}}
        \centerline{Case 3}
    \end{minipage}
    \begin{minipage}{0.49\linewidth}
        \centerline{\includegraphics[width=\textwidth]{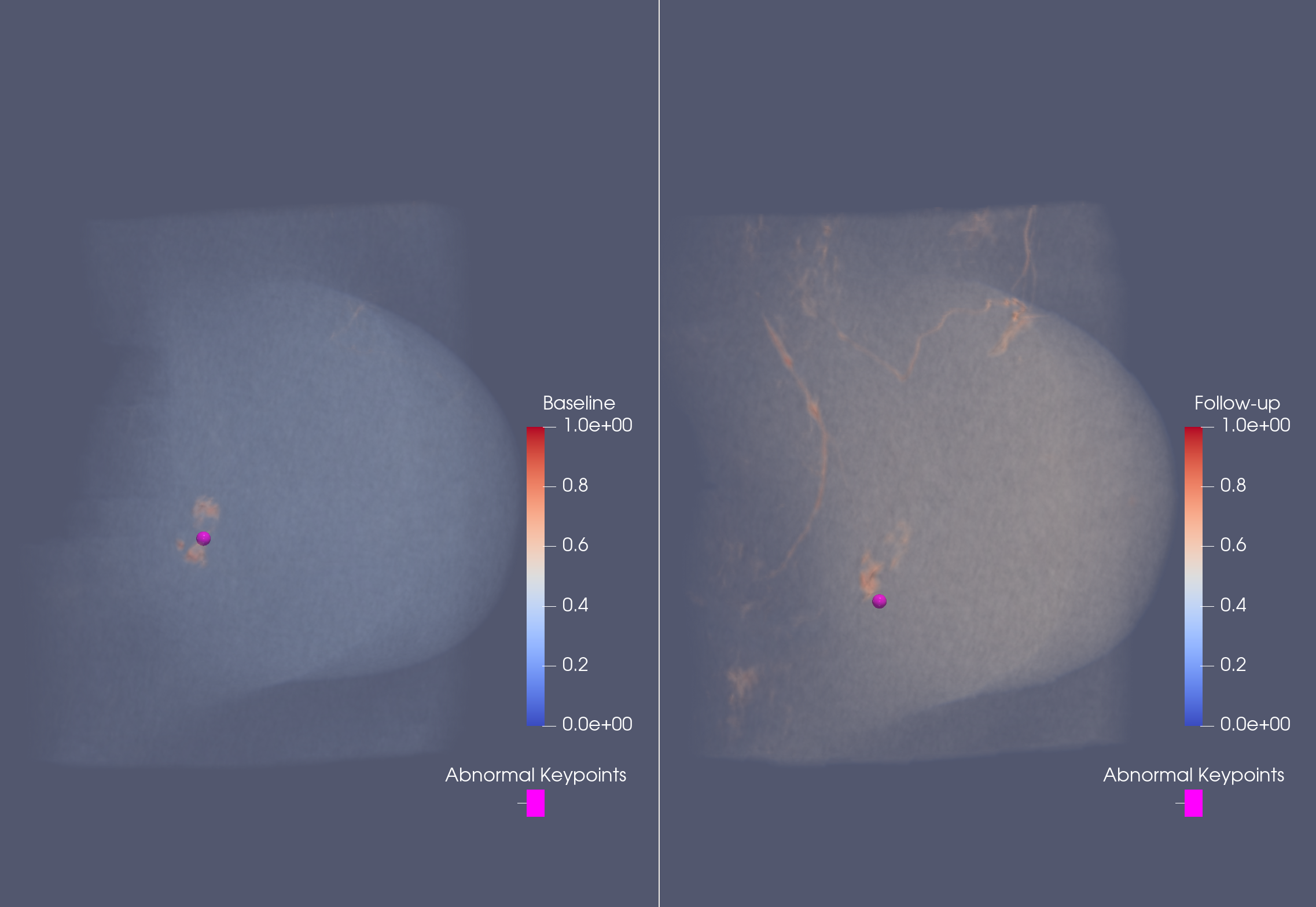}}
        \centerline{Case 4}
    \end{minipage}
    
	\caption{Visualization of abnormal keypoints on the volume rendering for breast wash-in images. For each case, the baseline image is on the left, and the follow-up image is on the right. Purple points indicate the top 1 abnormal keypoints detected by KN-A. Abnormal keypoints focus on the different areas between baseline and follow-up images, especially the tumor areas (the region in red).} \label{fig:landmarks-A}
\end{figure*}
%\begin{figure*}[!htbp]
%\centering
%\includegraphics[width=0.9\textwidth]{figs/abnormal.png}
%\caption{Visualization of abnormal keypoints on the volume rendering for breast wash-in images. Green points indicate abnormal keypoints detected by KN-A. Abnormal keypoints focus on the different areas between baseline and follow-up images especially the tumor areas (region in red).} \label{fig:landmarks-A}
%\end{figure*}

Table~\ref{tab:results-al} illustrates the quantitative segmentation results of different methods and Fig.~\ref{fig:landmarks-A} shows the visual results for abnormal keypoints.
Based on conventional thresholding methods, utilizing the abnormal keypoints can improve the DSC over 0.04 and increase the hitting percentage by more than 6$\%$.
Combining abnormal keypoints with Yen's method can achieve results slightly lower than the supervised learning-based method with a DSC of 0.589 and a hitting percentage of 80.0$\%$.
As shown in Fig.~\ref{fig:landmarks-A}, we visualize the abnormal keypoints on the volume rendering for breast wash-in images.
Purple points indicate the abnormal keypoints detected by KN-A, and orange/red areas refer to high tissue blood flow, including mass-like tumors and long thin vessels. Abnormal keypoints are overlapping with the tumor area for the cases shown.
It is supported that abnormal keypoints can locate tumor areas and improve segmentation accuracy for breast DCE-MRI images.

\begin{table*}
	\centering
	\caption{The quantitative results of predicting pCR vs. non-pCR with comparison biomarkers. The best result is in bold and the second best one is underlined.}
	\label{tab:radiomics}
	\setlength{\tabcolsep}{3pt}
	\begin{tabular}{lcccccc}
		\hline\hline
		Method & AUC$\uparrow$ & Accuracy$\uparrow$ & Sensitivity$\uparrow$ & Specificity$\uparrow$ & PPV$\uparrow$ & NPV$\uparrow$ \\
		\hline
		Patient characteristics & 0.723$\pm$0.029 & 0.681$\pm$0.045 & 0.632$\pm$0.113 & 0.715$\pm$0.049 & 0.611$\pm$0.090 & 0.733$\pm$0.073 \\
		+PRM (Local) & \underline{0.791$\pm$0.024} & \underline{0.715$\pm$0.023} & \underline{0.648$\pm$0.059} & \underline{0.771$\pm$0.060} & \underline{0.669$\pm$0.109} & \underline{0.750$\pm$0.054} \\
		+PRM (Global) & 0.729$\pm$0.044 & 0.673$\pm$0.046 & 0.606$\pm$0.088 & 0.731$\pm$0.023 & 0.616$\pm$0.044 & 0.714$\pm$0.096\\
		+PRM (Local\&Global) & \textbf{0.809$\pm$0.039} & \textbf{0.721$\pm$0.050} & \textbf{0.653$\pm$0.068} & \textbf{0.775$\pm$0.045} & \textbf{0.672$\pm$0.094} & \textbf{0.753$\pm$0.075} \\
		\hline\hline
	\end{tabular}
\end{table*}

\begin{figure*}[!htbp]
\centering
\begin{minipage}{0.33\linewidth}
        \centerline{\includegraphics[width=\textwidth]{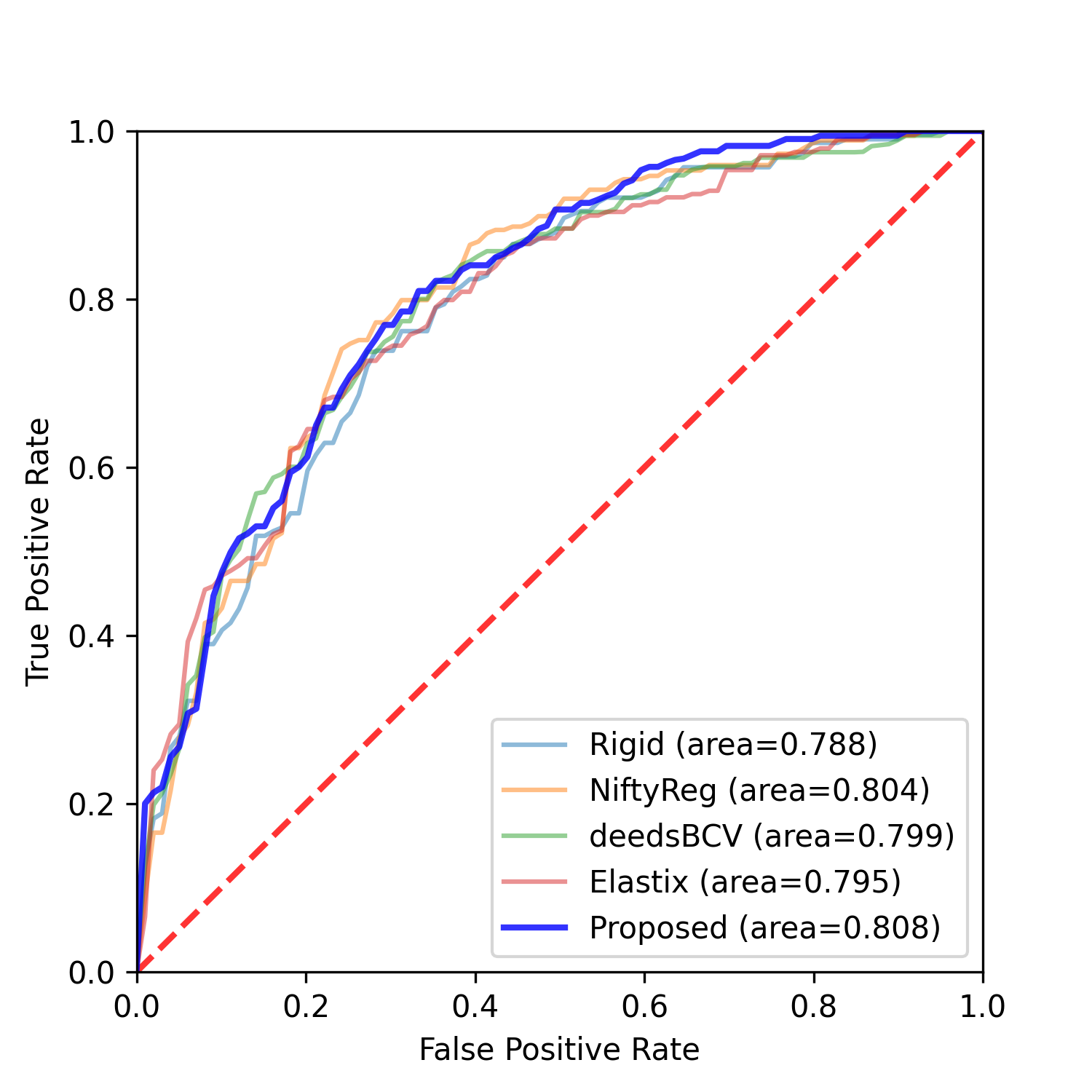}}
        \centerline{(a)}
    \end{minipage}
    \begin{minipage}{0.33\linewidth}
        \centerline{\includegraphics[width=\textwidth]{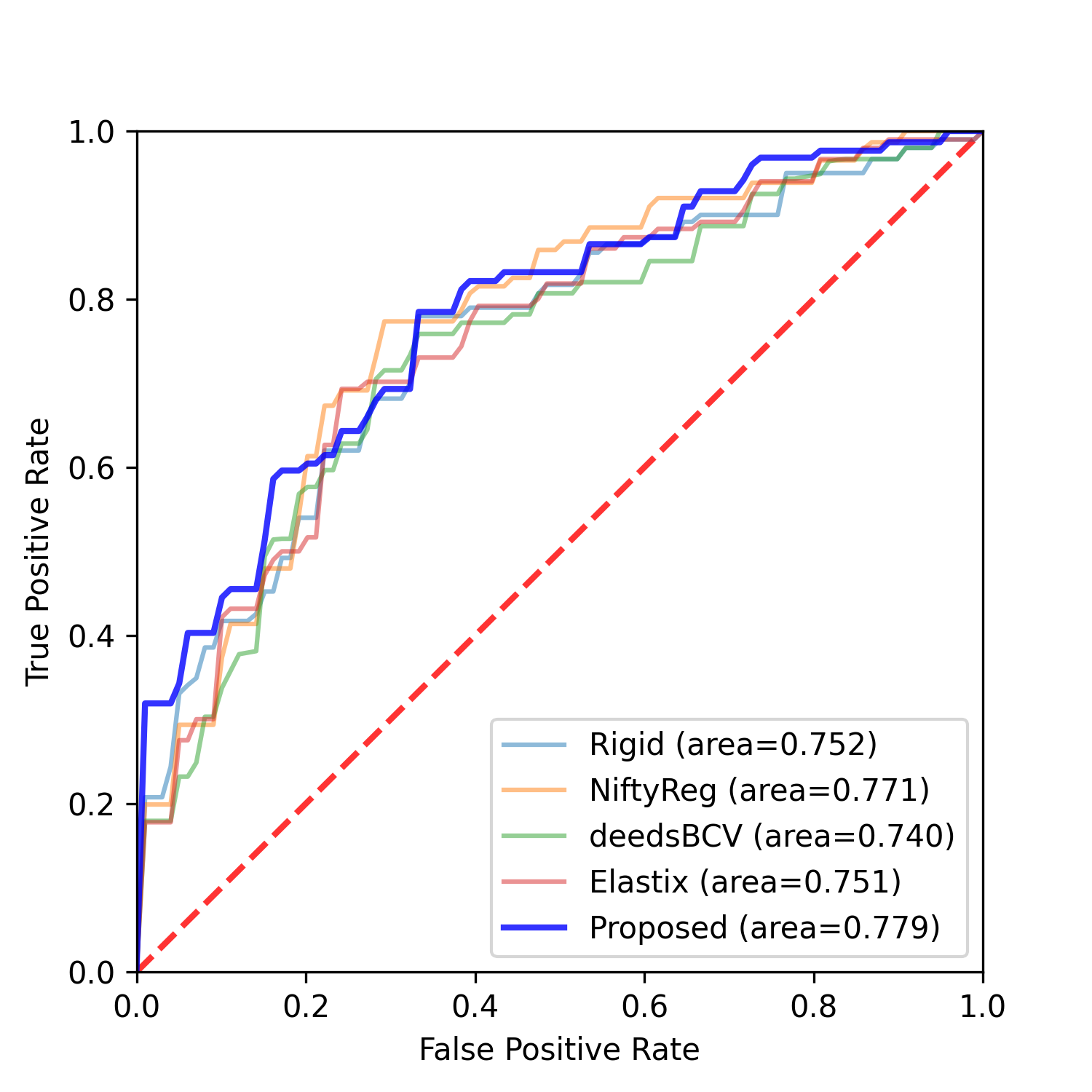}}
        \centerline{(b)}
    \end{minipage}
    \begin{minipage}{0.33\linewidth}
        \centerline{\includegraphics[width=\textwidth]{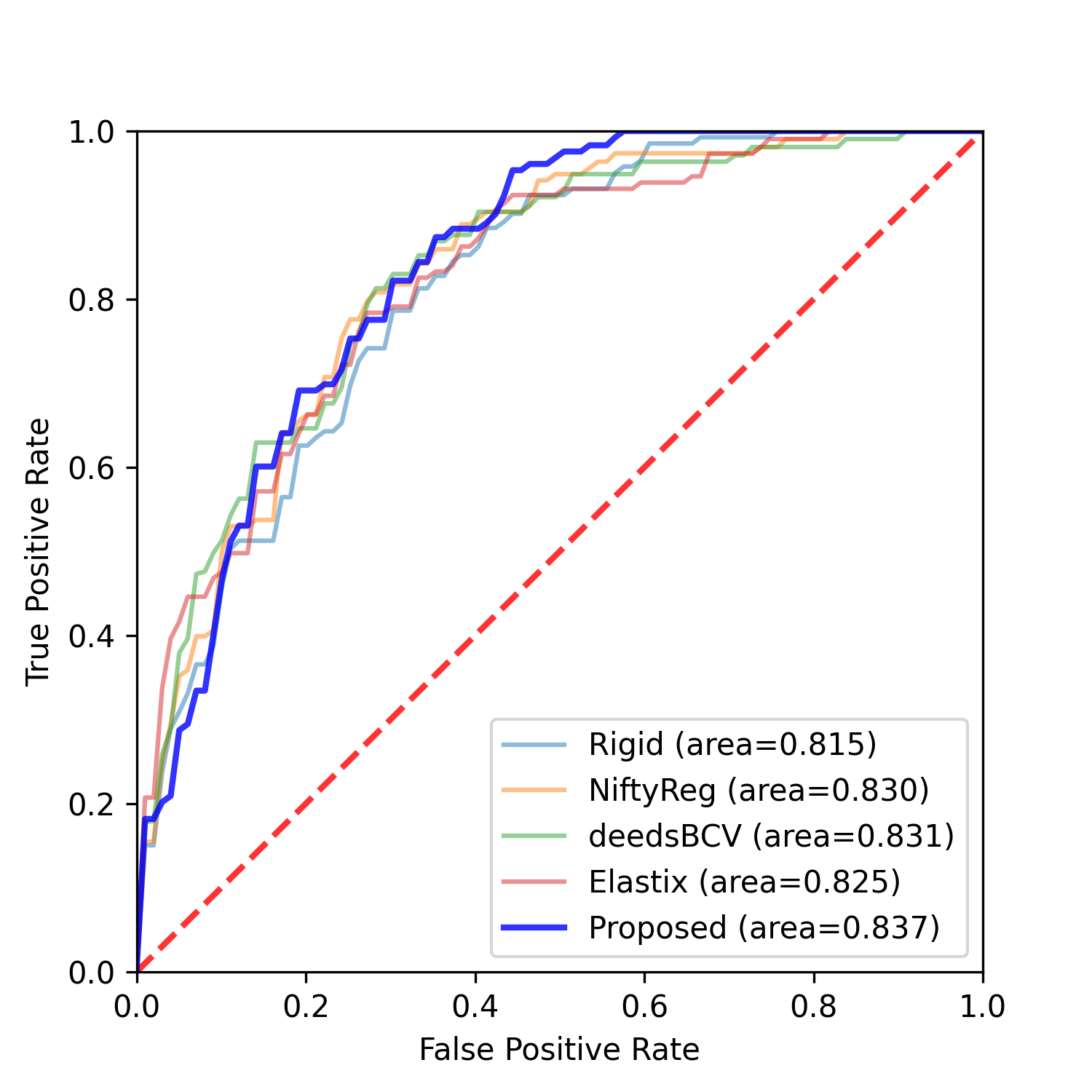}}
        \centerline{(c)}
    \end{minipage}
\caption{ROC curves for comparison registration methods on the testing set with different interval days. (a) All scans; (b) Scans with interval days less than 90; (c) Scans with interval days more than 90.} \label{fig:roc}
\end{figure*}

\subsection{Downstream applications}
The parametric response map (PRM)~\citep{el2018prm} is a biomarker calculated by subtraction between pre-NAC and follow-up DCE-MRIs, which is sensitive to the preprocessing registration methods. 
PRM is defined as,

\begin{equation}
    \begin{aligned}
        PRM(M,S) = \frac{1}{\sum{S}} \cdot
        \left[
            \begin{matrix}
                \sum{(M>\delta)\cdot S} \\
                \sum{(|M|\leq\delta)\cdot S} \\
                \sum{(M<-\delta)\cdot S}
            \end{matrix}
        \right] 
    \end{aligned}
    \label{eq:prm}
\end{equation}

where $M=(I_m\circ\phi-I_f)/(I_m\circ\phi)$ with a given deformation field $\phi$, $S$ refers to the segmentation mask of the region to calculate PRM, $\delta$ indicates the threshold to divide the response area and was set to 0.1 in the experiment.
We employ a downstream application of pCR prediction based on the improved PRM to evaluate the quality of the registration methods.
In this section, pre-NAC and follow-up scans for each patient are selected in the experiment. This yields a dataset with 314 patients and 501 paired scans. A two-level nested cross-validation is employed based on patients to select an optimal classifier from multiple classification models.
For each 'outer' five-fold cross-validation in the 5$\times$4 two-level nested cross-validation approach, an 'inner' four-fold cross-validation is employed on the training samples (80$\%$) to find the best model, and the other 20$\%$ samples remain for testing with the selected best model.
An XGBoost model is applied for pCR and non-pCR prediction. And we compare the classification results using the metrics of the area under curve (AUC), accuracy, sensitivity, specificity, positive predictive value (PPV), and negative predictive value (NPV).

We first evaluated the performance of the PRM by calculating it on the tumor area (Local) and breast area (Global). To discriminate follow-up scans at different times for the patient, we combine the interval days for the scan with the PRM when training the model. We propose four combinations including: (1) patient characteristics: age, ER, PR, HER2, T-stage, and N-stage; (2) +PRM (Local); (3) +PRM (Global); (4) +PRM (Local\&Global).
All the biomarkers are calculated based on the pre-NAC and follow-up wash-in images aligned with the proposed method.
The results are illustrated in Table~\ref{tab:radiomics}. Combining local PRM and global PRM with patient characteristics achieves state-of-the-art results in pCR classification.
Utilizing local PRM achieves a 6.8$\%$ increase in AUC and 3.4$\%$ in accuracy than using only patient characteristics. Concatenating information from the tumor and breast area obtains a performance boost of 0.018 in average AUC and 0.006 in average accuracy than local PRM. This indicates that global information from breast tissue contributes to predicting response to NAC.

In order to demonstrate the effectiveness of comparison registration methods for pCR prediction, we utilize patient characteristics and PRM with local and global information as the input features to predict response to NAC.
We compare our method with NiftyReg, deedsBCV, and Elastix, which achieve state-of-the-art registration performance in Section~\ref{sec:alternative}, and further we analyze the performance of the classification model under different interval days. The receiver operating characteristic (ROC) curves are illustrated in Fig.~\ref{fig:roc}. Utilizing follow-up scans with longer interval days from pre-NAC scans helps the model achieve higher AUC and accuracy in predicting pCR from non-pCR. Utilizing the images aligned by the proposed method can obtain the best pCR prediction AUC and accuracy than other types of input preprocessing.

\section{Discussion}
In this study, we develop a conditional pyramid framework based on unsupervised-detected keypoints for breast DCE-MRI registration during NAC. The registration model is sensible to decide where to deform and where not to deform. 
Our proposed method is specially designed to handle the challenges -- large global deformation and tumor changes in different scans -- in breast registration.
We employ a coarse-to-fine conditional pyramid architecture to fit the complex non-linear deformation field in different scales. And conditional registration module inside the network can control the smoothness of the resulting deformation field during inference without sacrificing runtime benefits or registration accuracy, greatly easing the difficulty of hyperparameter optimization.
To further cover large deformation in registration, we introduce structural keypoints which are detected by KN-S with a self-supervised framework.
Furthermore, a volume-preserving loss based on abnormal keypoints is introduced to ensure the tumor size remains unchanged.
We demonstrate the effectiveness of our proposed method by comparing it with state-of-the-art methods. Sufficient ablation studies are conducted to validate the improvement of each key component in the proposed method.

For the learning-based methods, large deformation is difficult to capture due to the limitation of the field of view (FOV) for convolutional operations.
Stacking convolutional layers in depth can enlarge the FOV but is capable of losing details and causing accumulative errors.
A coarse-to-fine conditional pyramid architecture provides different pyramid-level FOVs to fit the complex non-linear deformation field on different scales. The coarser level focuses on larger deformation and the finer level contributes to local deformation.
Quantitative results listed in Table~\ref{tab:results} and predictions shown in Fig.~\ref{fig:compare} prove that the conditional pyramid architecture is efficient to improve the accuracy of registration.
The pyramid framework like CPRN and cLapIRN achieve superior performance from both statistical and visual results than those of methods with a single resolution level, e.g., CRN, VoxelMorph, HyperMorph, and JointRegSeg.

Structural keypoint loss and volume-preserving loss are key factors for longitudinal registration between breast DCE-MRIs.
Breast tissue is often sparsely distributed in breast MRI images. Large deformations may lead toward local minimum convergence for the model, resulting in incorrect alignment.
Structural keypoints loss is capable of relieving the intensity and statistical bias from the semantic dimension by constraining breast tissue with well-distributed structure representation.
As shown in Table~\ref{tab:results}, methods employed with structural keypoint loss, e.g., +SK and +SK+VP (AK), achieve much lower $d_{avg}$ than that of CPRN and +VP (AK) in ablation study.
Besides, the volume of breast tumors may or may not change (become bigger, smaller, or stable) due to different responses to NAC during treatment. 
%Keeping tumor size unchanged helps follow-up pCR prediction task.
As shown in Table~\ref{tab:results}, optimizing the model with volume-preserving loss results in a drop on $\Delta V$.
%In meantime, Table~\ref{tab:downstream} illustrates that better pCR prediction is related to lower $\Delta V$ for the registration.

Besides the performance of image registration, we compare our proposed registration method with other methods by a downstream application -- pCR prediction.
We also propose a new biomarker PRM (Local\&Global) for pCR prediction. As shown in Table~\ref{tab:radiomics}, combining both local and global information achieves better AUC and accuracy than using local information on tumor area only. It supports that image-based registration contains more information than tumor-based alignment and contributes to predicting the response to NAC.
%Table~\ref{tab:downstream} illustrates that the better registration performance the method achieves, the higher AUC and accuracy for pCR prediction is obtained. 
It is because the pixel-level biomarker PRM (Local\&Global) is related to breast MRI images with finer pixel alignment.
From Fig.~\ref{fig:roc}, within the treatment phase, our AI model can effectively determine that 10$\%$ patients will not achieve pCR and a different treatment plan might be needed. At the end of the treatment, our AI model would potentially be able to exclude 18$\%$ patients for further surgical treatments.

\section{Conclusion}
In this paper, we propose a treatment-sensible registration network to align breast DCE-MRIs for a patient undergoing sequential evaluations during treatment. The proposed approach is capable of detecting structural keypoints and abnormal keypoints, restricting large deformation, and preserving the tumor volume. Furthermore, our downstream model on the top of registration can effectively select patients who need a different treatment method because they are not responding to the current therapy and also can help to avoid unnecessary surgical treatment for selected patients who achieved pCR.

\section*{Acknowledgments}
Luyi Han was funded by the Chinese Scholarship Council (CSC) scholarship.

%\section*{References}
%%Harvard
\bibliographystyle{model2-names.bst}\biboptions{authoryear}
\bibliography{refs}

\begin{thebibliography}{43}
\expandafter\ifx\csname natexlab\endcsname\relax\def\natexlab#1{#1}\fi
\providecommand{\url}[1]{\texttt{#1}}
\providecommand{\href}[2]{#2}
\providecommand{\path}[1]{#1}
\providecommand{\DOIprefix}{doi:}
\providecommand{\ArXivprefix}{arXiv:}
\providecommand{\URLprefix}{URL: }
\providecommand{\Pubmedprefix}{pmid:}
\providecommand{\doi}[1]{\href{http://dx.doi.org/#1}{\path{#1}}}
\providecommand{\Pubmed}[1]{\href{pmid:#1}{\path{#1}}}
\providecommand{\bibinfo}[2]{#2}
\ifx\xfnm\relax \def\xfnm[#1]{\unskip,\space#1}\fi
%Type = Article
\bibitem[{Andersen et~al.(2010)Andersen, Rapcsak and Beeson}]{andersen2010cost}
\bibinfo{author}{Andersen, S.M.}, \bibinfo{author}{Rapcsak, S.Z.},
  \bibinfo{author}{Beeson, P.M.}, \bibinfo{year}{2010}.
\newblock \bibinfo{title}{Cost function masking during normalization of brains
  with focal lesions: still a necessity?}
\newblock \bibinfo{journal}{Neuroimage} \bibinfo{volume}{53},
  \bibinfo{pages}{78--84}.
%Type = Article
\bibitem[{Avants et~al.(2008)Avants, Epstein, Grossman and
  Gee}]{avants2008symmetric}
\bibinfo{author}{Avants, B.B.}, \bibinfo{author}{Epstein, C.L.},
  \bibinfo{author}{Grossman, M.}, \bibinfo{author}{Gee, J.C.},
  \bibinfo{year}{2008}.
\newblock \bibinfo{title}{Symmetric diffeomorphic image registration with
  cross-correlation: evaluating automated labeling of elderly and
  neurodegenerative brain}.
\newblock \bibinfo{journal}{Medical image analysis} \bibinfo{volume}{12},
  \bibinfo{pages}{26--41}.
%Type = Article
\bibitem[{Balakrishnan et~al.(2019)Balakrishnan, Zhao, Sabuncu, Guttag and
  Dalca}]{balakrishnan2019voxelmorph}
\bibinfo{author}{Balakrishnan, G.}, \bibinfo{author}{Zhao, A.},
  \bibinfo{author}{Sabuncu, M.R.}, \bibinfo{author}{Guttag, J.},
  \bibinfo{author}{Dalca, A.V.}, \bibinfo{year}{2019}.
\newblock \bibinfo{title}{Voxelmorph: a learning framework for deformable
  medical image registration}.
\newblock \bibinfo{journal}{IEEE transactions on medical imaging}
  \bibinfo{volume}{38}, \bibinfo{pages}{1788--1800}.
%Type = Article
\bibitem[{Bonadonna et~al.(1998)Bonadonna, Valagussa, Brambilla, Ferrari,
  Moliterni, Terenziani and Zambetti}]{bonadonna1998primary}
\bibinfo{author}{Bonadonna, G.}, \bibinfo{author}{Valagussa, P.},
  \bibinfo{author}{Brambilla, C.}, \bibinfo{author}{Ferrari, L.},
  \bibinfo{author}{Moliterni, A.}, \bibinfo{author}{Terenziani, M.},
  \bibinfo{author}{Zambetti, M.}, \bibinfo{year}{1998}.
\newblock \bibinfo{title}{Primary chemotherapy in operable breast cancer:
  eight-year experience at the milan cancer institute}.
\newblock \bibinfo{journal}{Journal of Clinical Oncology} \bibinfo{volume}{16},
  \bibinfo{pages}{93--100}.
%Type = Article
\bibitem[{Brett et~al.(2001)Brett, Leff, Rorden and
  Ashburner}]{brett2001spatial}
\bibinfo{author}{Brett, M.}, \bibinfo{author}{Leff, A.P.},
  \bibinfo{author}{Rorden, C.}, \bibinfo{author}{Ashburner, J.},
  \bibinfo{year}{2001}.
\newblock \bibinfo{title}{Spatial normalization of brain images with focal
  lesions using cost function masking}.
\newblock \bibinfo{journal}{Neuroimage} \bibinfo{volume}{14},
  \bibinfo{pages}{486--500}.
%Type = Article
\bibitem[{Cao et~al.(2018)Cao, Yang, Zhang, Wang, Yap and
  Shen}]{cao2018deformable}
\bibinfo{author}{Cao, X.}, \bibinfo{author}{Yang, J.}, \bibinfo{author}{Zhang,
  J.}, \bibinfo{author}{Wang, Q.}, \bibinfo{author}{Yap, P.T.},
  \bibinfo{author}{Shen, D.}, \bibinfo{year}{2018}.
\newblock \bibinfo{title}{Deformable image registration using a cue-aware deep
  regression network}.
\newblock \bibinfo{journal}{IEEE Transactions on Biomedical Engineering}
  \bibinfo{volume}{65}, \bibinfo{pages}{1900--1911}.
%Type = Article
\bibitem[{Cho et~al.(2013)Cho, Park, Park, Park, Kim and
  Park}]{cho2013oncologic}
\bibinfo{author}{Cho, J.H.}, \bibinfo{author}{Park, J.M.},
  \bibinfo{author}{Park, H.S.}, \bibinfo{author}{Park, S.},
  \bibinfo{author}{Kim, S.I.}, \bibinfo{author}{Park, B.W.},
  \bibinfo{year}{2013}.
\newblock \bibinfo{title}{Oncologic safety of breast-conserving surgery
  compared to mastectomy in patients receiving neoadjuvant chemotherapy for
  locally advanced breast cancer}.
\newblock \bibinfo{journal}{Journal of surgical oncology}
  \bibinfo{volume}{108}, \bibinfo{pages}{531--536}.
%Type = Article
\bibitem[{Curigliano et~al.(2017)Curigliano, Burstein, Winer, Gnant, Dubsky,
  Loibl, Colleoni, Regan, Piccart-Gebhart, Senn
  et~al.}]{curigliano2017escalating}
\bibinfo{author}{Curigliano, G.}, \bibinfo{author}{Burstein, H.J.},
  \bibinfo{author}{Winer, E.P.}, \bibinfo{author}{Gnant, M.},
  \bibinfo{author}{Dubsky, P.}, \bibinfo{author}{Loibl, S.},
  \bibinfo{author}{Colleoni, M.}, \bibinfo{author}{Regan, M.M.},
  \bibinfo{author}{Piccart-Gebhart, M.}, \bibinfo{author}{Senn, H.J.}, et~al.,
  \bibinfo{year}{2017}.
\newblock \bibinfo{title}{De-escalating and escalating treatments for
  early-stage breast cancer: the st. gallen international expert consensus
  conference on the primary therapy of early breast cancer 2017}.
\newblock \bibinfo{journal}{Annals of Oncology} \bibinfo{volume}{28},
  \bibinfo{pages}{1700--1712}.
%Type = Article
\bibitem[{De~Vos et~al.(2019)De~Vos, Berendsen, Viergever, Sokooti, Staring and
  I{\v{s}}gum}]{de2019deep}
\bibinfo{author}{De~Vos, B.D.}, \bibinfo{author}{Berendsen, F.F.},
  \bibinfo{author}{Viergever, M.A.}, \bibinfo{author}{Sokooti, H.},
  \bibinfo{author}{Staring, M.}, \bibinfo{author}{I{\v{s}}gum, I.},
  \bibinfo{year}{2019}.
\newblock \bibinfo{title}{A deep learning framework for unsupervised affine and
  deformable image registration}.
\newblock \bibinfo{journal}{Medical image analysis} \bibinfo{volume}{52},
  \bibinfo{pages}{128--143}.
%Type = Article
\bibitem[{Earl et~al.(2015)Earl, Provenzano, Abraham, Dunn, Vallier, Gounaris
  and Hiller}]{earl2015neoadjuvant}
\bibinfo{author}{Earl, H.}, \bibinfo{author}{Provenzano, E.},
  \bibinfo{author}{Abraham, J.}, \bibinfo{author}{Dunn, J.},
  \bibinfo{author}{Vallier, A.L.}, \bibinfo{author}{Gounaris, I.},
  \bibinfo{author}{Hiller, L.}, \bibinfo{year}{2015}.
\newblock \bibinfo{title}{Neoadjuvant trials in early breast cancer:
  pathological response at surgery and correlation to longer term
  outcomes--what does it all mean?}
\newblock \bibinfo{journal}{BMC medicine} \bibinfo{volume}{13},
  \bibinfo{pages}{1--11}.
%Type = Article
\bibitem[{El~Adoui et~al.(2018)El~Adoui, Drisis and Benjelloun}]{el2018prm}
\bibinfo{author}{El~Adoui, M.}, \bibinfo{author}{Drisis, S.},
  \bibinfo{author}{Benjelloun, M.}, \bibinfo{year}{2018}.
\newblock \bibinfo{title}{A prm approach for early prediction of breast cancer
  response to chemotherapy based on registered mr images}.
\newblock \bibinfo{journal}{International journal of computer assisted
  radiology and surgery} \bibinfo{volume}{13}, \bibinfo{pages}{1233--1243}.
%Type = Article
\bibitem[{Eppenhof and Pluim(2018)}]{eppenhof2018pulmonary}
\bibinfo{author}{Eppenhof, K.A.}, \bibinfo{author}{Pluim, J.P.},
  \bibinfo{year}{2018}.
\newblock \bibinfo{title}{Pulmonary ct registration through supervised learning
  with convolutional neural networks}.
\newblock \bibinfo{journal}{IEEE transactions on medical imaging}
  \bibinfo{volume}{38}, \bibinfo{pages}{1097--1105}.
%Type = Article
\bibitem[{Estienne et~al.(2020)Estienne, Lerousseau, Vakalopoulou,
  Alvarez~Andres, Battistella, Carr{\'e}, Chandra, Christodoulidis,
  Sahasrabudhe, Sun et~al.}]{estienne2020deep}
\bibinfo{author}{Estienne, T.}, \bibinfo{author}{Lerousseau, M.},
  \bibinfo{author}{Vakalopoulou, M.}, \bibinfo{author}{Alvarez~Andres, E.},
  \bibinfo{author}{Battistella, E.}, \bibinfo{author}{Carr{\'e}, A.},
  \bibinfo{author}{Chandra, S.}, \bibinfo{author}{Christodoulidis, S.},
  \bibinfo{author}{Sahasrabudhe, M.}, \bibinfo{author}{Sun, R.}, et~al.,
  \bibinfo{year}{2020}.
\newblock \bibinfo{title}{Deep learning-based concurrent brain registration and
  tumor segmentation}.
\newblock \bibinfo{journal}{Frontiers in computational neuroscience}
  \bibinfo{volume}{14}, \bibinfo{pages}{17}.
%Type = Article
\bibitem[{Gigengack et~al.(2011)Gigengack, Ruthotto, Burger, Wolters, Jiang and
  Schafers}]{gigengack2011motion}
\bibinfo{author}{Gigengack, F.}, \bibinfo{author}{Ruthotto, L.},
  \bibinfo{author}{Burger, M.}, \bibinfo{author}{Wolters, C.H.},
  \bibinfo{author}{Jiang, X.}, \bibinfo{author}{Schafers, K.P.},
  \bibinfo{year}{2011}.
\newblock \bibinfo{title}{Motion correction in dual gated cardiac pet using
  mass-preserving image registration}.
\newblock \bibinfo{journal}{IEEE transactions on medical imaging}
  \bibinfo{volume}{31}, \bibinfo{pages}{698--712}.
%Type = Article
\bibitem[{Gooya et~al.(2010)Gooya, Biros and Davatzikos}]{gooya2010deformable}
\bibinfo{author}{Gooya, A.}, \bibinfo{author}{Biros, G.},
  \bibinfo{author}{Davatzikos, C.}, \bibinfo{year}{2010}.
\newblock \bibinfo{title}{Deformable registration of glioma images using em
  algorithm and diffusion reaction modeling}.
\newblock \bibinfo{journal}{IEEE transactions on medical imaging}
  \bibinfo{volume}{30}, \bibinfo{pages}{375--390}.
%Type = Article
\bibitem[{Gooya et~al.(2012)Gooya, Pohl, Bilello, Cirillo, Biros, Melhem and
  Davatzikos}]{gooya2012glistr}
\bibinfo{author}{Gooya, A.}, \bibinfo{author}{Pohl, K.M.},
  \bibinfo{author}{Bilello, M.}, \bibinfo{author}{Cirillo, L.},
  \bibinfo{author}{Biros, G.}, \bibinfo{author}{Melhem, E.R.},
  \bibinfo{author}{Davatzikos, C.}, \bibinfo{year}{2012}.
\newblock \bibinfo{title}{Glistr: glioma image segmentation and registration}.
\newblock \bibinfo{journal}{IEEE transactions on medical imaging}
  \bibinfo{volume}{31}, \bibinfo{pages}{1941--1954}.
%Type = Phdthesis
\bibitem[{Guo(2019)}]{guo2019multi}
\bibinfo{author}{Guo, C.K.}, \bibinfo{year}{2019}.
\newblock \bibinfo{title}{Multi-modal image registration with unsupervised deep
  learning}.
\newblock Ph.D. thesis. Massachusetts Institute of Technology.
%Type = Inproceedings
\bibitem[{Han et~al.(2021)Han, Dou, Huang and Yap}]{han2021deformable}
\bibinfo{author}{Han, L.}, \bibinfo{author}{Dou, H.}, \bibinfo{author}{Huang,
  Y.}, \bibinfo{author}{Yap, P.T.}, \bibinfo{year}{2021}.
\newblock \bibinfo{title}{Deformable registration of brain mr images via a
  hybrid loss}, in: \bibinfo{booktitle}{International Conference on Medical
  Image Computing and Computer-Assisted Intervention},
  \bibinfo{organization}{Springer}. pp. \bibinfo{pages}{141--146}.
%Type = Article
\bibitem[{Heinrich et~al.(2013)Heinrich, Jenkinson, Brady and
  Schnabel}]{heinrich2013mrf}
\bibinfo{author}{Heinrich, M.P.}, \bibinfo{author}{Jenkinson, M.},
  \bibinfo{author}{Brady, M.}, \bibinfo{author}{Schnabel, J.A.},
  \bibinfo{year}{2013}.
\newblock \bibinfo{title}{Mrf-based deformable registration and ventilation
  estimation of lung ct}.
\newblock \bibinfo{journal}{IEEE transactions on medical imaging}
  \bibinfo{volume}{32}, \bibinfo{pages}{1239--1248}.
%Type = Inproceedings
\bibitem[{Hoopes et~al.(2021)Hoopes, Hoffmann, Fischl, Guttag and
  Dalca}]{hoopes2021hypermorph}
\bibinfo{author}{Hoopes, A.}, \bibinfo{author}{Hoffmann, M.},
  \bibinfo{author}{Fischl, B.}, \bibinfo{author}{Guttag, J.},
  \bibinfo{author}{Dalca, A.V.}, \bibinfo{year}{2021}.
\newblock \bibinfo{title}{Hypermorph: amortized hyperparameter learning for
  image registration}, in: \bibinfo{booktitle}{International Conference on
  Information Processing in Medical Imaging}, \bibinfo{organization}{Springer}.
  pp. \bibinfo{pages}{3--17}.
%Type = Article
\bibitem[{Hu et~al.(2018)Hu, Modat, Gibson, Li, Ghavami, Bonmati, Wang,
  Bandula, Moore, Emberton et~al.}]{hu2018weakly}
\bibinfo{author}{Hu, Y.}, \bibinfo{author}{Modat, M.}, \bibinfo{author}{Gibson,
  E.}, \bibinfo{author}{Li, W.}, \bibinfo{author}{Ghavami, N.},
  \bibinfo{author}{Bonmati, E.}, \bibinfo{author}{Wang, G.},
  \bibinfo{author}{Bandula, S.}, \bibinfo{author}{Moore, C.M.},
  \bibinfo{author}{Emberton, M.}, et~al., \bibinfo{year}{2018}.
\newblock \bibinfo{title}{Weakly-supervised convolutional neural networks for
  multimodal image registration}.
\newblock \bibinfo{journal}{Medical image analysis} \bibinfo{volume}{49},
  \bibinfo{pages}{1--13}.
%Type = Inproceedings
\bibitem[{Jakab et~al.(2018)Jakab, Gupta, Bilen and
  Vedaldi}]{jakab2018unsupervised}
\bibinfo{author}{Jakab, T.}, \bibinfo{author}{Gupta, A.},
  \bibinfo{author}{Bilen, H.}, \bibinfo{author}{Vedaldi, A.},
  \bibinfo{year}{2018}.
\newblock \bibinfo{title}{Unsupervised learning of object landmarks through
  conditional image generation}, in: \bibinfo{booktitle}{Proceedings of the
  32nd International Conference on Neural Information Processing Systems}, pp.
  \bibinfo{pages}{4020--4031}.
%Type = Article
\bibitem[{Klein et~al.(2009)Klein, Staring, Murphy, Viergever and
  Pluim}]{klein2009elastix}
\bibinfo{author}{Klein, S.}, \bibinfo{author}{Staring, M.},
  \bibinfo{author}{Murphy, K.}, \bibinfo{author}{Viergever, M.A.},
  \bibinfo{author}{Pluim, J.P.}, \bibinfo{year}{2009}.
\newblock \bibinfo{title}{Elastix: a toolbox for intensity-based medical image
  registration}.
\newblock \bibinfo{journal}{IEEE transactions on medical imaging}
  \bibinfo{volume}{29}, \bibinfo{pages}{196--205}.
%Type = Article
\bibitem[{Kulkarni et~al.(2019)Kulkarni, Gupta, Ionescu, Borgeaud, Reynolds,
  Zisserman and Mnih}]{kulkarni2019unsupervised}
\bibinfo{author}{Kulkarni, T.}, \bibinfo{author}{Gupta, A.},
  \bibinfo{author}{Ionescu, C.}, \bibinfo{author}{Borgeaud, S.},
  \bibinfo{author}{Reynolds, M.}, \bibinfo{author}{Zisserman, A.},
  \bibinfo{author}{Mnih, V.}, \bibinfo{year}{2019}.
\newblock \bibinfo{title}{Unsupervised learning of object keypoints for
  perception and control}.
\newblock \bibinfo{journal}{arXiv preprint arXiv:1906.11883} .
%Type = Article
\bibitem[{Li et~al.(2009)Li, Dawant, Welch, Chakravarthy, Freehardt, Mayer,
  Kelley, Meszoely, Gore and Yankeelov}]{li2009nonrigid}
\bibinfo{author}{Li, X.}, \bibinfo{author}{Dawant, B.M.},
  \bibinfo{author}{Welch, E.B.}, \bibinfo{author}{Chakravarthy, A.B.},
  \bibinfo{author}{Freehardt, D.}, \bibinfo{author}{Mayer, I.},
  \bibinfo{author}{Kelley, M.}, \bibinfo{author}{Meszoely, I.},
  \bibinfo{author}{Gore, J.C.}, \bibinfo{author}{Yankeelov, T.E.},
  \bibinfo{year}{2009}.
\newblock \bibinfo{title}{A nonrigid registration algorithm for longitudinal
  breast mr images and the analysis of breast tumor response}.
\newblock \bibinfo{journal}{Magnetic resonance imaging} \bibinfo{volume}{27},
  \bibinfo{pages}{1258--1270}.
%Type = Inproceedings
\bibitem[{Liu et~al.(2014)Liu, Niethammer, Kwitt, McCormick and
  Aylward}]{liu2014low}
\bibinfo{author}{Liu, X.}, \bibinfo{author}{Niethammer, M.},
  \bibinfo{author}{Kwitt, R.}, \bibinfo{author}{McCormick, M.},
  \bibinfo{author}{Aylward, S.}, \bibinfo{year}{2014}.
\newblock \bibinfo{title}{Low-rank to the rescue--atlas-based analyses in the
  presence of pathologies}, in: \bibinfo{booktitle}{International Conference on
  Medical Image Computing and Computer-Assisted Intervention},
  \bibinfo{organization}{Springer}. pp. \bibinfo{pages}{97--104}.
%Type = Article
\bibitem[{Liu et~al.(2015)Liu, Niethammer, Kwitt, Singh, McCormick and
  Aylward}]{liu2015low}
\bibinfo{author}{Liu, X.}, \bibinfo{author}{Niethammer, M.},
  \bibinfo{author}{Kwitt, R.}, \bibinfo{author}{Singh, N.},
  \bibinfo{author}{McCormick, M.}, \bibinfo{author}{Aylward, S.},
  \bibinfo{year}{2015}.
\newblock \bibinfo{title}{Low-rank atlas image analyses in the presence of
  pathologies}.
\newblock \bibinfo{journal}{IEEE transactions on medical imaging}
  \bibinfo{volume}{34}, \bibinfo{pages}{2583--2591}.
%Type = Article
\bibitem[{Mehrabian et~al.(2018)Mehrabian, Richmond, Lu and
  Martel}]{mehrabian2018deformable}
\bibinfo{author}{Mehrabian, H.}, \bibinfo{author}{Richmond, L.},
  \bibinfo{author}{Lu, Y.}, \bibinfo{author}{Martel, A.L.},
  \bibinfo{year}{2018}.
\newblock \bibinfo{title}{Deformable registration for longitudinal breast mri
  screening}.
\newblock \bibinfo{journal}{Journal of digital imaging} \bibinfo{volume}{31},
  \bibinfo{pages}{718--726}.
%Type = Article
\bibitem[{Modat et~al.(2010)Modat, Ridgway, Taylor, Lehmann, Barnes, Hawkes,
  Fox and Ourselin}]{modat2010fast}
\bibinfo{author}{Modat, M.}, \bibinfo{author}{Ridgway, G.R.},
  \bibinfo{author}{Taylor, Z.A.}, \bibinfo{author}{Lehmann, M.},
  \bibinfo{author}{Barnes, J.}, \bibinfo{author}{Hawkes, D.J.},
  \bibinfo{author}{Fox, N.C.}, \bibinfo{author}{Ourselin, S.},
  \bibinfo{year}{2010}.
\newblock \bibinfo{title}{Fast free-form deformation using graphics processing
  units}.
\newblock \bibinfo{journal}{Computer methods and programs in biomedicine}
  \bibinfo{volume}{98}, \bibinfo{pages}{278--284}.
%Type = Inproceedings
\bibitem[{Mok and Chung(2021)}]{mok2021conditional}
\bibinfo{author}{Mok, T.C.}, \bibinfo{author}{Chung, A.}, \bibinfo{year}{2021}.
\newblock \bibinfo{title}{Conditional deformable image registration with
  convolutional neural network}, in: \bibinfo{booktitle}{International
  Conference on Medical Image Computing and Computer-Assisted Intervention},
  \bibinfo{organization}{Springer}. pp. \bibinfo{pages}{35--45}.
%Type = Article
\bibitem[{Otsu(1979)}]{otsu1979threshold}
\bibinfo{author}{Otsu, N.}, \bibinfo{year}{1979}.
\newblock \bibinfo{title}{A threshold selection method from gray-level
  histograms}.
\newblock \bibinfo{journal}{IEEE transactions on systems, man, and cybernetics}
  \bibinfo{volume}{9}, \bibinfo{pages}{62--66}.
%Type = Article
\bibitem[{Ou et~al.(2011)Ou, Sotiras, Paragios and Davatzikos}]{ou2011dramms}
\bibinfo{author}{Ou, Y.}, \bibinfo{author}{Sotiras, A.},
  \bibinfo{author}{Paragios, N.}, \bibinfo{author}{Davatzikos, C.},
  \bibinfo{year}{2011}.
\newblock \bibinfo{title}{Dramms: Deformable registration via attribute
  matching and mutual-saliency weighting}.
\newblock \bibinfo{journal}{Medical image analysis} \bibinfo{volume}{15},
  \bibinfo{pages}{622--639}.
%Type = Article
\bibitem[{Ou et~al.(2015)Ou, Weinstein, Conant, Englander, Da, Gaonkar, Hsieh,
  Rosen, DeMichele, Davatzikos et~al.}]{ou2015deformable}
\bibinfo{author}{Ou, Y.}, \bibinfo{author}{Weinstein, S.P.},
  \bibinfo{author}{Conant, E.F.}, \bibinfo{author}{Englander, S.},
  \bibinfo{author}{Da, X.}, \bibinfo{author}{Gaonkar, B.},
  \bibinfo{author}{Hsieh, M.K.}, \bibinfo{author}{Rosen, M.},
  \bibinfo{author}{DeMichele, A.}, \bibinfo{author}{Davatzikos, C.}, et~al.,
  \bibinfo{year}{2015}.
\newblock \bibinfo{title}{Deformable registration for quantifying longitudinal
  tumor changes during neoadjuvant chemotherapy}.
\newblock \bibinfo{journal}{Magnetic resonance in medicine}
  \bibinfo{volume}{73}, \bibinfo{pages}{2343--2356}.
%Type = Article
\bibitem[{Rohlfing et~al.(2003)Rohlfing, Maurer, Bluemke and
  Jacobs}]{rohlfing2003volume}
\bibinfo{author}{Rohlfing, T.}, \bibinfo{author}{Maurer, C.R.},
  \bibinfo{author}{Bluemke, D.A.}, \bibinfo{author}{Jacobs, M.A.},
  \bibinfo{year}{2003}.
\newblock \bibinfo{title}{Volume-preserving nonrigid registration of mr breast
  images using free-form deformation with an incompressibility constraint}.
\newblock \bibinfo{journal}{IEEE transactions on medical imaging}
  \bibinfo{volume}{22}, \bibinfo{pages}{730--741}.
%Type = Article
\bibitem[{Sung et~al.(2021)Sung, Ferlay, Siegel, Laversanne, Soerjomataram,
  Jemal and Bray}]{sung2021global}
\bibinfo{author}{Sung, H.}, \bibinfo{author}{Ferlay, J.},
  \bibinfo{author}{Siegel, R.L.}, \bibinfo{author}{Laversanne, M.},
  \bibinfo{author}{Soerjomataram, I.}, \bibinfo{author}{Jemal, A.},
  \bibinfo{author}{Bray, F.}, \bibinfo{year}{2021}.
\newblock \bibinfo{title}{Global cancer statistics 2020: Globocan estimates of
  incidence and mortality worldwide for 36 cancers in 185 countries}.
\newblock \bibinfo{journal}{CA: a cancer journal for clinicians}
  \bibinfo{volume}{71}, \bibinfo{pages}{209--249}.
%Type = Article
\bibitem[{Tang et~al.(2017)Tang, Wu and Fan}]{tang2017groupwise}
\bibinfo{author}{Tang, Z.}, \bibinfo{author}{Wu, Y.}, \bibinfo{author}{Fan,
  Y.}, \bibinfo{year}{2017}.
\newblock \bibinfo{title}{Groupwise registration of mr brain images with
  tumors}.
\newblock \bibinfo{journal}{Physics in Medicine \& Biology}
  \bibinfo{volume}{62}, \bibinfo{pages}{6853}.
%Type = Article
\bibitem[{Tang et~al.(2018)Tang, Yap and Shen}]{tang2018new}
\bibinfo{author}{Tang, Z.}, \bibinfo{author}{Yap, P.T.}, \bibinfo{author}{Shen,
  D.}, \bibinfo{year}{2018}.
\newblock \bibinfo{title}{A new multi-atlas registration framework for
  multimodal pathological images using conventional monomodal normal atlases}.
\newblock \bibinfo{journal}{IEEE Transactions on Image Processing}
  \bibinfo{volume}{28}, \bibinfo{pages}{2293--2304}.
%Type = Article
\bibitem[{Thakran et~al.(2022)Thakran, Cohen, Jahani, Weinstein, Pantalone,
  Hylton, Newitt, DeMichele, Davatzikos and Kontos}]{thakran2022impact}
\bibinfo{author}{Thakran, S.}, \bibinfo{author}{Cohen, E.},
  \bibinfo{author}{Jahani, N.}, \bibinfo{author}{Weinstein, S.P.},
  \bibinfo{author}{Pantalone, L.}, \bibinfo{author}{Hylton, N.},
  \bibinfo{author}{Newitt, D.}, \bibinfo{author}{DeMichele, A.},
  \bibinfo{author}{Davatzikos, C.}, \bibinfo{author}{Kontos, D.},
  \bibinfo{year}{2022}.
\newblock \bibinfo{title}{Impact of deformable registration methods for
  prediction of recurrence free survival response to neoadjuvant chemotherapy
  in breast cancer: Results from the ispy 1/acrin 6657 trial}.
\newblock \bibinfo{journal}{Translational Oncology} \bibinfo{volume}{20},
  \bibinfo{pages}{101411}.
%Type = Article
\bibitem[{Vercauteren et~al.(2009)Vercauteren, Pennec, Perchant and
  Ayache}]{vercauteren2009diffeomorphic}
\bibinfo{author}{Vercauteren, T.}, \bibinfo{author}{Pennec, X.},
  \bibinfo{author}{Perchant, A.}, \bibinfo{author}{Ayache, N.},
  \bibinfo{year}{2009}.
\newblock \bibinfo{title}{Diffeomorphic demons: Efficient non-parametric image
  registration}.
\newblock \bibinfo{journal}{NeuroImage} \bibinfo{volume}{45},
  \bibinfo{pages}{S61--S72}.
%Type = Article
\bibitem[{Wodzinski et~al.(2021)Wodzinski, Ciepiela, Kuszewski, Kedzierawski
  and Skalski}]{wodzinski2021semi}
\bibinfo{author}{Wodzinski, M.}, \bibinfo{author}{Ciepiela, I.},
  \bibinfo{author}{Kuszewski, T.}, \bibinfo{author}{Kedzierawski, P.},
  \bibinfo{author}{Skalski, A.}, \bibinfo{year}{2021}.
\newblock \bibinfo{title}{Semi-supervised deep learning-based image
  registration method with volume penalty for real-time breast tumor bed
  localization}.
\newblock \bibinfo{journal}{Sensors} \bibinfo{volume}{21},
  \bibinfo{pages}{4085}.
%Type = Article
\bibitem[{Yan et~al.(2020)Yan, Cai, Jin, Miao, Harrison, Guo, Tang, Xiao, Lu
  and Lu}]{yan2020self}
\bibinfo{author}{Yan, K.}, \bibinfo{author}{Cai, J.}, \bibinfo{author}{Jin,
  D.}, \bibinfo{author}{Miao, S.}, \bibinfo{author}{Harrison, A.P.},
  \bibinfo{author}{Guo, D.}, \bibinfo{author}{Tang, Y.}, \bibinfo{author}{Xiao,
  J.}, \bibinfo{author}{Lu, J.}, \bibinfo{author}{Lu, L.},
  \bibinfo{year}{2020}.
\newblock \bibinfo{title}{Self-supervised learning of pixel-wise anatomical
  embeddings in radiological images}.
\newblock \bibinfo{journal}{arXiv preprint arXiv:2012.02383} .
%Type = Article
\bibitem[{Yen et~al.(1995)Yen, Chang and Chang}]{yen1995new}
\bibinfo{author}{Yen, J.C.}, \bibinfo{author}{Chang, F.J.},
  \bibinfo{author}{Chang, S.}, \bibinfo{year}{1995}.
\newblock \bibinfo{title}{A new criterion for automatic multilevel
  thresholding}.
\newblock \bibinfo{journal}{IEEE Transactions on Image Processing}
  \bibinfo{volume}{4}, \bibinfo{pages}{370--378}.
%Type = Inproceedings
\bibitem[{Zhang et~al.(2018)Zhang, Guo, Jin, Luo, He and
  Lee}]{zhang2018unsupervised}
\bibinfo{author}{Zhang, Y.}, \bibinfo{author}{Guo, Y.}, \bibinfo{author}{Jin,
  Y.}, \bibinfo{author}{Luo, Y.}, \bibinfo{author}{He, Z.},
  \bibinfo{author}{Lee, H.}, \bibinfo{year}{2018}.
\newblock \bibinfo{title}{Unsupervised discovery of object landmarks as
  structural representations}, in: \bibinfo{booktitle}{Proceedings of the IEEE
  Conference on Computer Vision and Pattern Recognition}, pp.
  \bibinfo{pages}{2694--2703}.

\end{thebibliography}

%\section*{Supplementary Material}

%Supplementary material that may be helpful in the review process should
%be prepared and provided as a separate electronic file. That file can
%then be transformed into PDF format and submitted along with the
%manuscript and graphic files to the appropriate editorial office.

\end{document}